\documentclass[twocolumn,showpacs,preprintnumbers,amsmath,amssymb,aps,pre]{revtex4}

\usepackage{graphicx}
\usepackage{dcolumn}
\usepackage{bm}

\begin{document}

\title{Spectral Characteristics of Network Redundancy}

\author{Ben D. MacArthur}
\thanks{Both authors contributed equally to this work.}
\affiliation{Department of Pharmacology and Systems Therapeutics, Systems Biology Center New York (SBCNY), Mount Sinai School of Medicine, New York, NY, USA.}
\email{ben.macarthur@mssm.edu}
\author{Rub\'en J. S\'anchez-Garc\'\i{}a}
\affiliation{Mathematisches Institut, Heinrich-Heine Universit\"at D\"usseldorf, Universit\"atsstr 1, 40225, D\"usseldorf, Germany.}
\email{sanchez@math.uni-duesseldorf.de}

\date{\today}

\begin{abstract}
Many real-world complex networks contain a significant amount of structural redundancy, in which multiple vertices play identical topological roles. Such redundancy arises naturally from the simple growth processes which form and shape many real-world systems. Since structurally redundant elements may be permuted without altering network structure, redundancy may be formally investigated by examining network automorphism (symmetry) groups. Here, we use a group-theoretic approach to give a complete description of spectral signatures of redundancy in undirected networks. In particular, we describe how a network's automorphism group may be used to directly associate specific eigenvalues and eigenvectors with specific network motifs.   
\end{abstract}

\pacs{89.75.-k 89.75.Fb 05.40.-a 02.20.-a}

\maketitle

\section{\label{sec:intro} Introduction}
Many complex real-world systems -- from chemical reactions inside cells \cite{albert3} to technological systems such as the world wide web \cite{albert2} -- may be represented as networks. Understanding the topological structure of these networks helps understanding the behaviour of the system on which they are based. Thus, there is considerable interest in elucidating the origin and form of common structural features of networks \cite{albertSM,newman,strogatz}. Previous reports have identified a variety of features which are common to a range of disparate networks including: the power-law distribution of vertex degrees \cite{barabasiPA,batty,li}; the `small-world' property \cite{wattsSW}; and network construction from motifs \cite{guimera,milo,shen-orr} amongst others. 

Many common network features derive from common ways in which real-world networks are formed and evolve. So, for instance, growth with preferential attachment naturally leads to a power-law vertex degree distribution \cite{barabasiPA}. As another example, common replicative growth processes, such as growth with duplication \cite{chung2}, naturally endow networks with a certain degree of structural redundancy. Thus, structural redundancy -- in which multiple vertices play an identical topological role -- is common in real-world empirical networks \cite{macarthur}. In terms of system behaviour, structural redundancy can be beneficial since it naturally reinforces against attack by providing structural `backups' should network elements fail \cite{tononi}. Thus, network redundancy is related to system robustness \cite{albertSM,albertAttack}. 

Intuitively, two vertices are topologically equivalent if they may be permuted without altering network structure. A permutation of the vertices of a network which does not affect network adjacency is known as an \emph{automorphism} and the set of network automorphisms forms a group under composition of permutations. Thus, our intuitive notion of structural equivalence may be formally investigated using the mathematical language of permutation groups. Crucially, symmetric networks (those with a nontrivial automorphism group) necessarily contain a certain amount of structural redundancy. In accordance with the observation that common growth processes naturally lead to structural redundancy, many empirical networks have richly structured automorphism groups \cite{macarthur}. In this paper we shall use the automorphism group to investigate the effect of redundancy on network eigenvalue spectra.

Since graph eigenvalues are well-known to be related to a multitude of graph properties \cite{chung} there has been considerable recent interest in studying the spectra of real-world complex networks and their associated models \cite{chung1,dorogovtsevSCN,dorogovtsevES,farkas,goh,kamp,palla,rodgers}. These studies have highlighted the fact that the spectral densities of real-world networks commonly differ significantly from those of the classical ensembles of random matrix theory \cite{farkas, mehta}. For example, in \cite{farkas} the spectral density of Barab{\'a}si-Albert `scale-free' networks \cite{barabasiPA} and Watts-Strogatz `small-world' networks \cite{wattsSW} are considered. Barab{\'a}si-Albert networks are found to have a spectral density which consists of a `triangle-like' bulk with power-law tails; while Watts-Strogatz small world networks are found to have multiple strong local maxima in their spectral densities which are related to the blurring of singularities in the spectral density of the highly ordered $k$-ring structure upon which the Watts-Strogatz model is based. 

\begin{figure}[t!]
\begin{center}
\raisebox{0.0cm}{\includegraphics[width=0.08\textwidth]{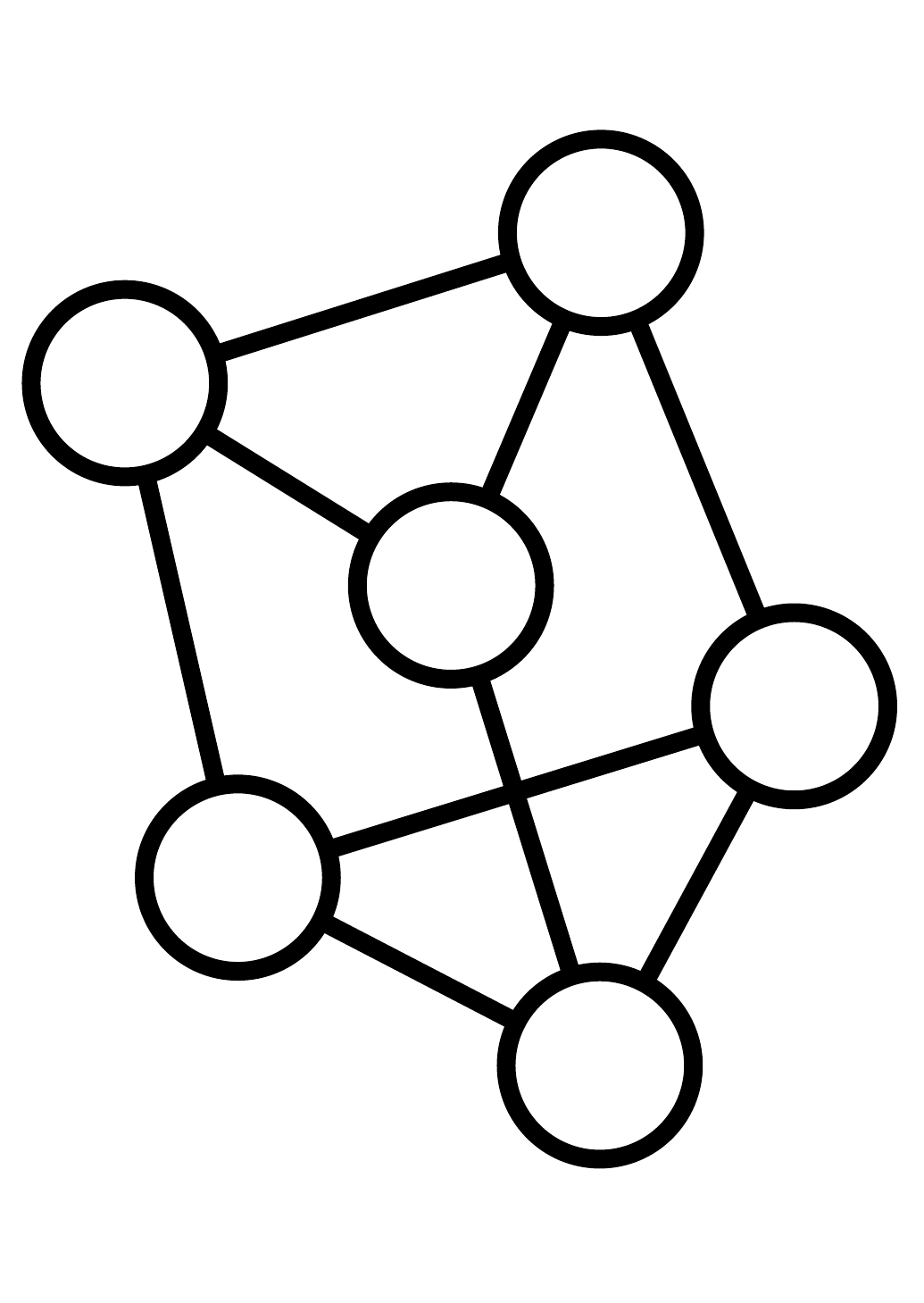}}
\raisebox{0.0cm}{\includegraphics[width=0.15\textwidth]{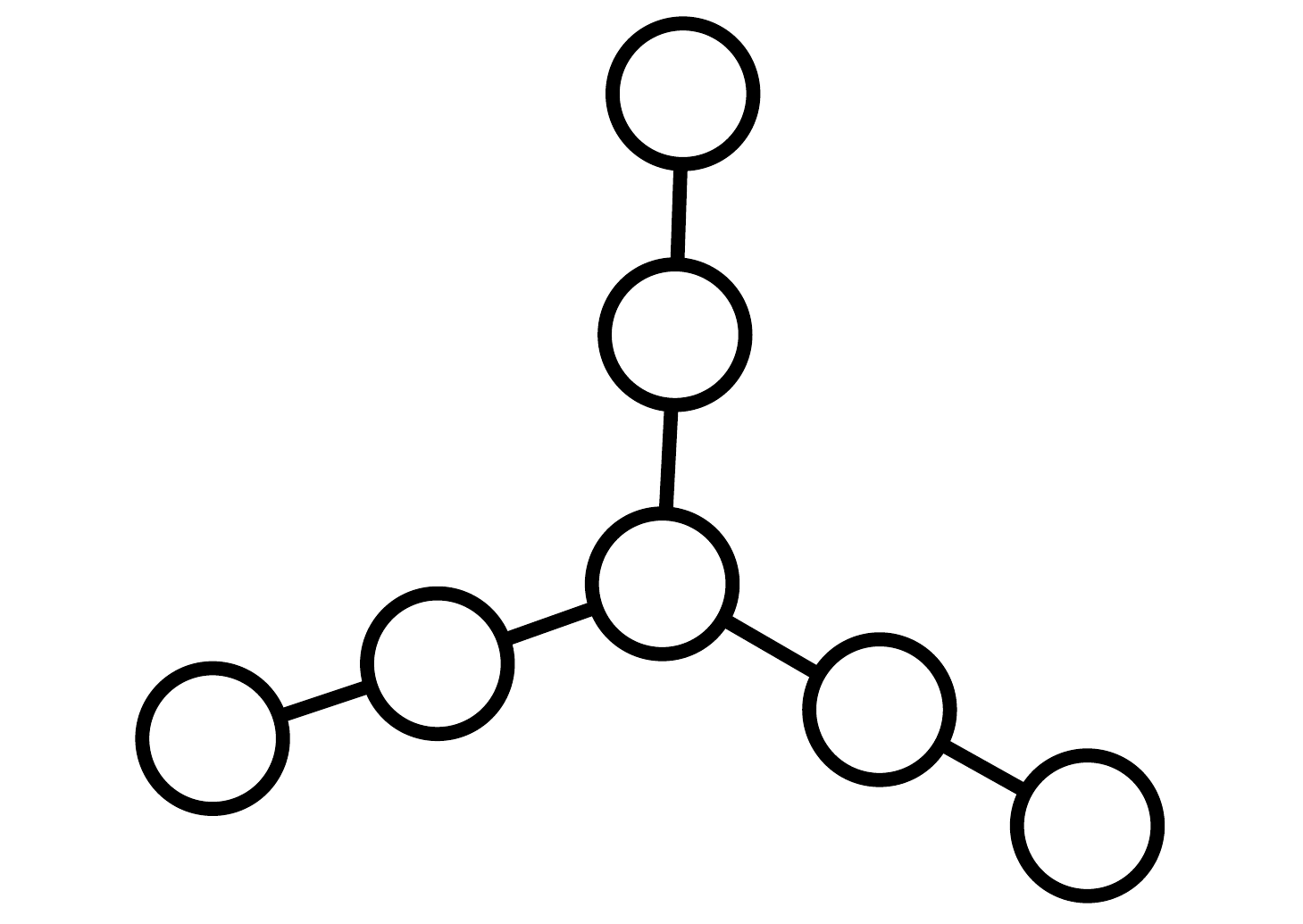}}
\raisebox{-1cm}{\includegraphics[width=0.15\textwidth]{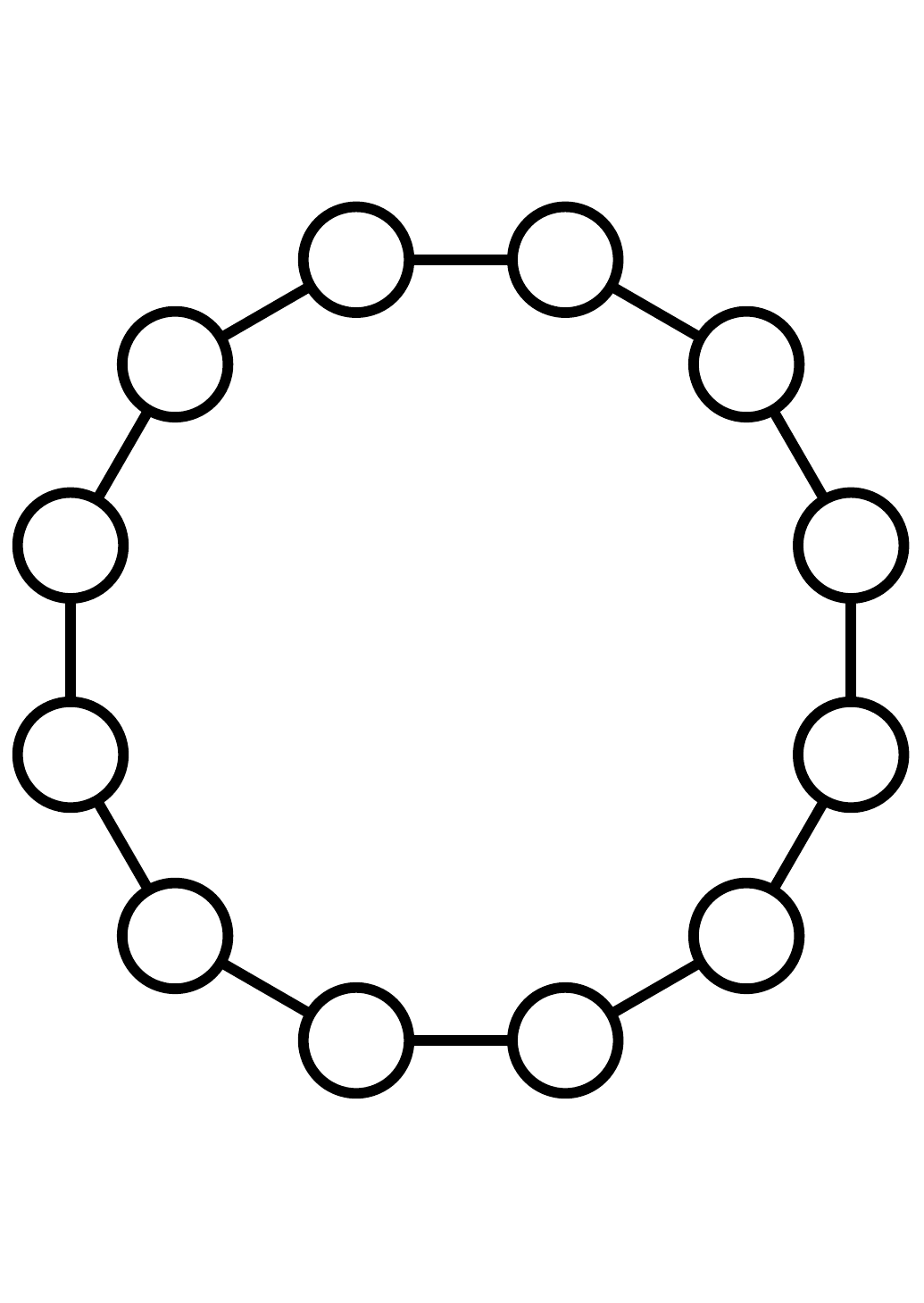}}
\caption{\label{examplegraphs}\small{\textbf{High multiplicity eigenvalues can arise from a variety of network structures}. The first graph has high-multiplicity $0$ eigenvalue but no stars, the second graph has high-multiplicity $-1$ eigenvalue with no cliques, and the third graph has both high-multiplicity $0$ and $-1$ eigenvalues, yet contains neither stars nor cliques. The first two graphs are examples of commonly ocurring symmetric motifs, as we explain later.}}
\end{center}
\end{figure}

Similarly, although they are not usually highly ordered, the spectral densities of real-world networks also often contain singularities. For instance singularities at the 0 and $-1$ eigenvalues are common. Previous discussions have related the singularity at $0$ to local multiplicities in vertices of degree $1$ (stars), and the singularity at $-1$ to complete subgraphs (cliques) \cite{dorogovtsevSCN,dorogovtsevES,goh,kamp} although these explanations are not exhaustive. For example, the graphs in FIG. \ref{examplegraphs} have high multiplicity $0$ and $-1$ eigenvalues which are not due to the presence of stars or cliques respectively.  
In general, since the relationship between a network and its spectrum is nontrivial, determining general conditions for the presence and strength of singularities in the spectral density is an open analytic problem \cite{dorogovtsevSCN}. 

Based upon the observation that high-multiplicity eigenvalues commonly associate with graph symmetries \cite{lauri} we examined the relationship between network symmetry and spectral singularities. Since symmetry can take many forms -- cliques, stars and rings are all symmetric, for example -- symmetry provides a flexible framework for interpreting the effect of a wide variety of redundant network structures on eigenvalue spectra.
  
The structure of the remainder of the paper is as follows: In section \ref{sec:background} we introduce some necessary background material on network symmetry. In particular, we examine the relationship between network topology and automorphism group structure and show how certain subgroups of the automorphism group can be related to specific network motifs. In section \ref{sec:spec} we consider how a network's automorphism group interacts with its spectrum, and discuss how specific eigenvalues and eigenvectors associate with these motifs. We study in detail the most frequent of these motifs and their contribution to the network's spectrum. Finally, we close with some general conclusions. 

\section{Background}\label{sec:background}
\begin{figure}[t!]
\begin{center}
\raisebox{0.0cm}{\includegraphics[width=0.17\textwidth]{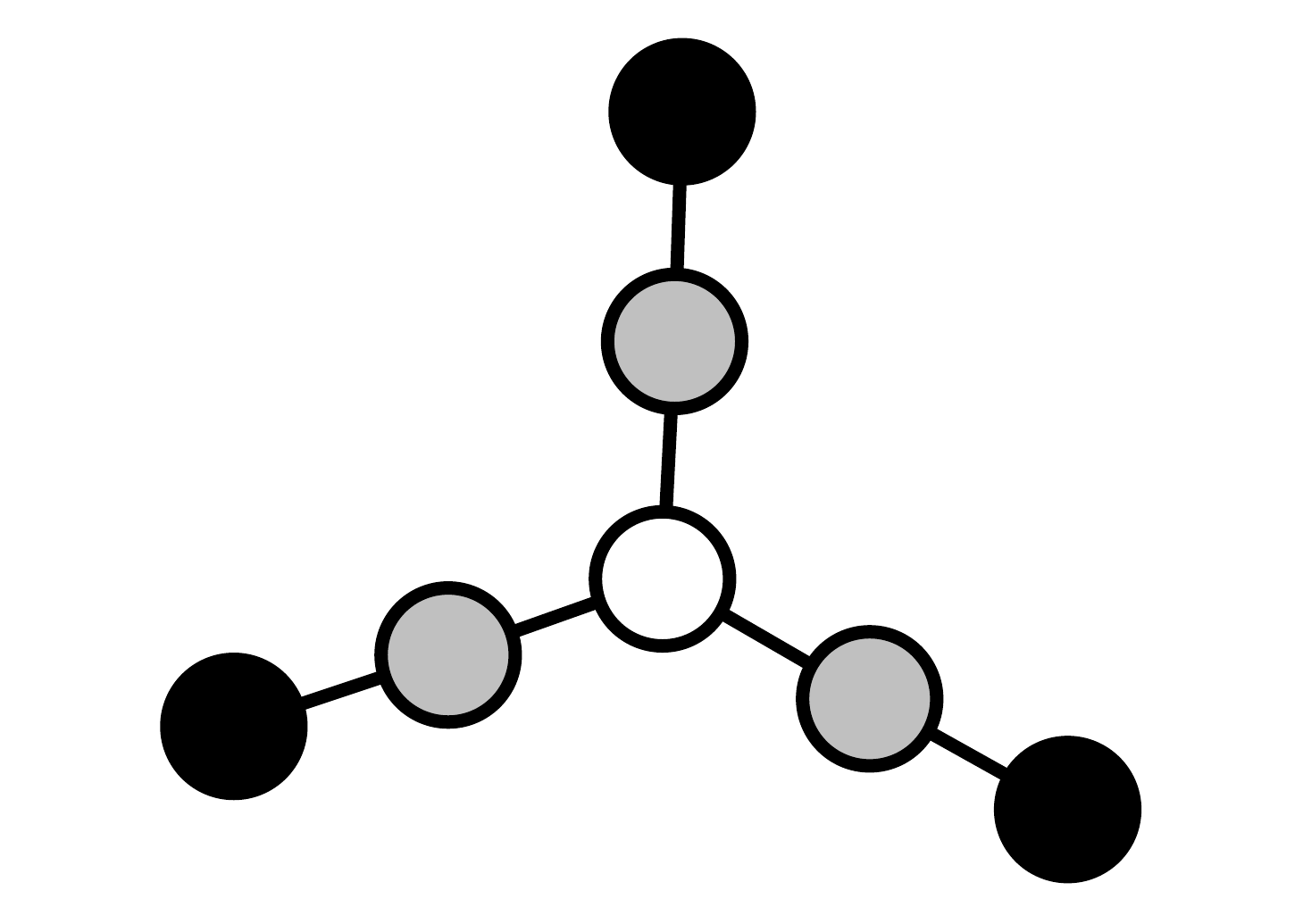}}
\caption{\label{basicexample}\small{\textbf{A symmetric network}. The permutation of the vertices given by the rotation $r$ of $120^{\circ}$ around the central vertex is an automorphism of the graph above, as are the reflections $\sigma_1, \sigma_2$ and $\sigma_3$ through each of the three arms. Together with a rotation of $240^{\circ}$ and the identity (leaving every vertex fixed) they form the automorphism group $G$ of the graph. Multiplication is composition and we have $G = \{1, r, r_2, \sigma_1, \sigma_2=\sigma_1 r, \sigma_3=\sigma_1 r^2\}$. We say that the group $G$ \emph{acts} on the graph, because every element in $G$ correspond to a permutation of the vertex set which preserves adjacency ($G$ is the group of all such permutations). The \emph{orbit} of a vertex $v$ is the set of all vertices to which $v$ can be sent by the action. Vertices are coloured by orbit: there are two orbits of size 3 and one orbit of size 1 (the central vertex is a \emph{fixed point} of the $G$-action).}}
\end{center}
\end{figure}

A network may be thought of as a graph, $\mathcal{G}=\mathcal{G}(V(\mathcal{G}),E(\mathcal{G}))$, with vertex set, $V(\mathcal{G})$ (of size $N$), and edge set, $E(\mathcal{G})$ (of size $M$) where vertices are said to be \emph{adjacent} if there is an edge between them. An \emph{automorphism} is a permutation of the vertices of the network which preserves adjacency. The set of automorphisms under composition forms a group, $G=\textrm{Aut}(\mathcal{G})$, of size $a_{\mathcal{G}}$ \cite{bollobasMGT} --- see FIG. \ref{basicexample} for an example. We say that a network is \emph{symmetric} (respectively \emph{asymmetric}) if it has a nontrivial (respectively trivial) automorphism group. Since automorphisms permute vertices without altering network structure, a network's automorphims group compactly quantifies the degree and nature of the structural redundancy it carries. This correspondence between network symmetry and redundancy forms the basis of the analysis we present in this discussion.

The \emph{support} of an automorphism $p$ is the set of vertices which $p$ moves, $\textup{supp}(p) = \{ v_i \in V(\mathcal{G}) \, | \, p(v_i) \neq v_i \}$. Two sets of automorphisms $P$ and $Q$ are \emph{support-disjoint} if every pair of automorphisms $p \in P$ and $q \in Q$ have disjoint supports. Additionally, we say that the automorphism subgroups $G_P$ and $G_Q$ generated by $P$ and $Q$ are \emph{support-disjoint} if $P$ and $Q$ are. If this is the case, $pq = qp$ for all $p \in P$ and $q \in Q$ and hence $xy = yx$ for all $x \in G_P$ and $y \in G_Q$. Thus, if $G_P$ and $G_Q$ are support-disjoint then we may think of them as acting independently on the network. 

This notion of independent action gives us a useful means to factorize the automorphism groups of complex networks into `irreducible building blocks' \cite{macarthur}. In particular, let $\mathcal{G}$ be a network with automorphism group $G=\textrm{Aut}(\mathcal{G})$ generated by a set $S$ of generators. Partition $S$ into support-disjoint subsets $S = S_1 \cup \ldots \cup S_n$ such that each $S_i$ cannot itself be decomposed into support-disjoint subsets. Call $H_i$ the subgroup generated by $S_i$. Since each $H_i$ commutes with all others, we can construct a direct product decomposition of $G$ from these subgroups:
\begin{equation} \label{decomp}
G = H_1 \times H_2 \times \ldots H_k\,.
\end{equation}
This decomposition splits the automorphism group into smaller pieces, each of which acts independently on the network $\mathcal{G}$. If the set of generators $S$ satisfies two simple conditions, the decomposition of Eq.~\ref{decomp} is unique (up to permutation of the factors) and irreducible \cite{macarthur}. In this case, we call each factor $H_i$ a \emph{geometric factor} and the direct product factorization given in Eq. \ref{decomp} the \emph{geometric decomposition} of $G=\textrm{Aut}(\mathcal{G})$. The motivation for this naming is that this factorization relates strongly to network geometry: each factor $H_i$ may be related to a subgraph of $\mathcal{G}$, as follows. 

The \emph{induced subgraph} on a set of vertices $X \subset V(\mathcal{G})$ is the graph obtained by taking $X$ and any edges whose end points are both in $X$. We call the induced subgraph on the support of a geometric factor $H$ a \emph{symmetric motif}, denoted $\mathcal{M}_H$. Thus $H$ moves the vertices of $\mathcal{M}_H$ while fixing the rest of the vertices of $\mathcal{G}$, and $\mathcal{M}_H$ is the smallest subgraph with this property. 

\begin{figure}[t!]
\begin{center}
\includegraphics[width=0.45\textwidth]{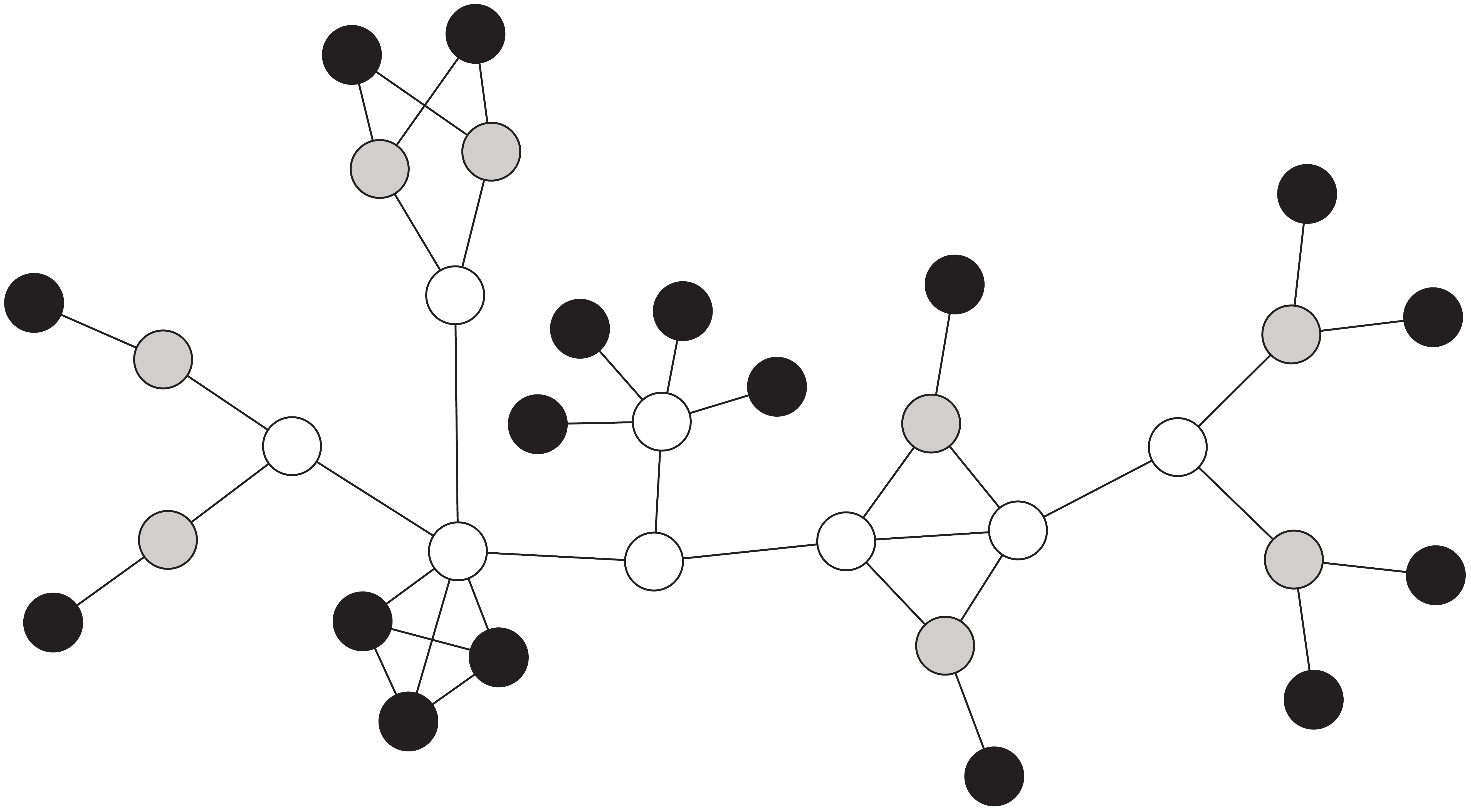}
\includegraphics[width=0.35\textwidth]{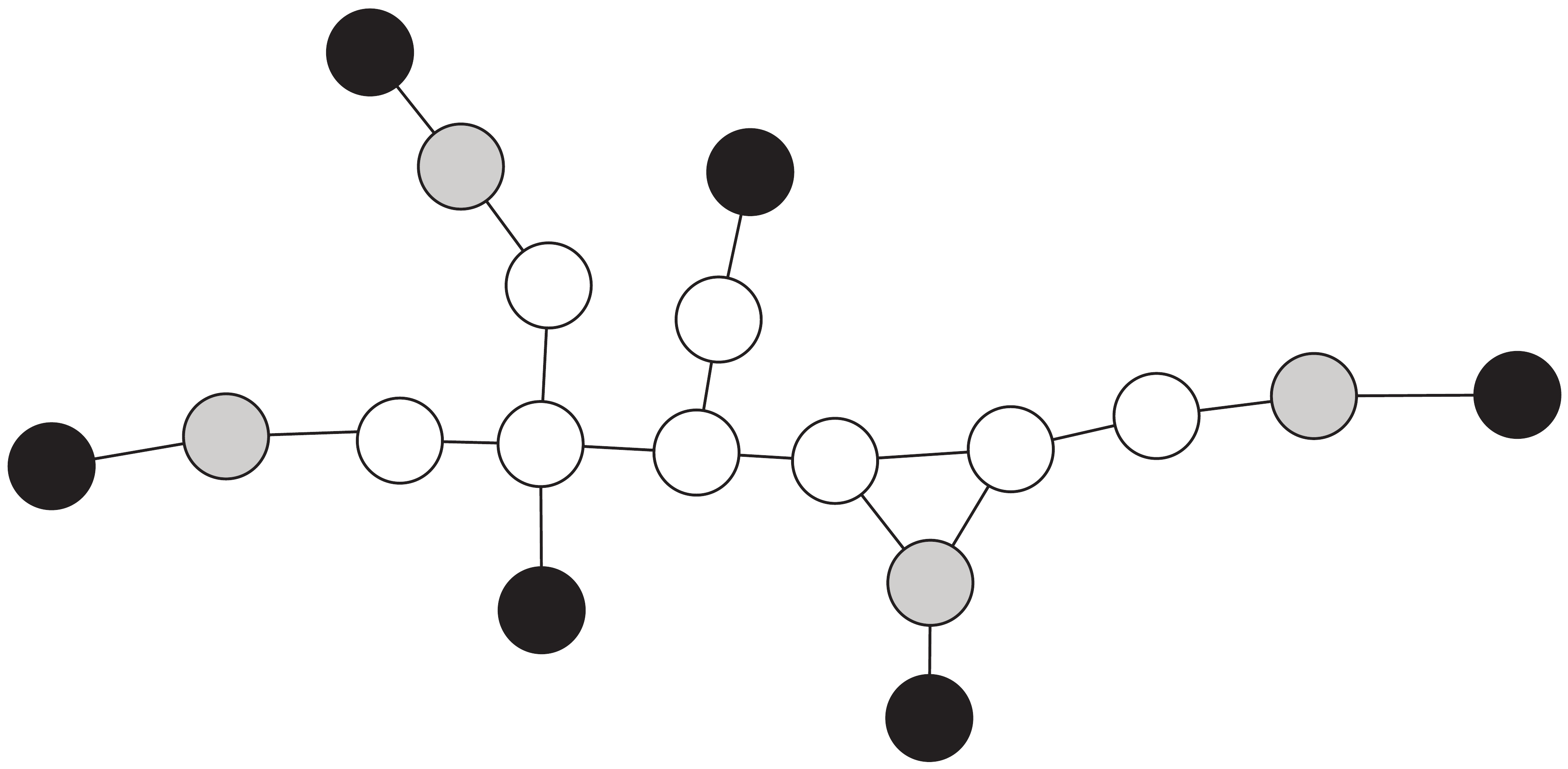}
\caption{\label{example}\small{\textbf{An example network and its quotient}. A network with $\textrm{Aut}(\mathcal{G}) = S_2 \times S_2 \times S_2 \times S_3 \times S_4 \times S_2 \times (S_2 \wr S_2)$ (top), and its quotient (bottom). We write $S_n$ for the group of all permutation of $n$ objects and $\wr$ for the wreath product \cite{rotman}, a mild generalization of the direct product. Vertices are coloured by orbit. Note that the quotient is a multigraph but, for clarity, edge weights and directions have not been represented.}}
\end{center}
\end{figure}

FIG.~\ref{example} shows an example network constructed from a variety of symmetric motifs commonly found in real-world networks, and its associated geometric decomposition. Table \ref{auttable} shows how the factors in the geometric decomposition of this network's automorphism group relate to distinct symmetric motifs in the network. Examples of geometric decompositions of real-world networks can be found elsewhere \cite{macarthur}. Note that for simplicity we consider networks as undirected graphs; a directed version of this decomposition is straightforward.

Since large (Erd{\"o}s-R{\'e}nyi) random graphs are expected to be asymmetric \cite{bollobasRG}, symmetric motifs are commonly over-represented in real-world networks by comparison with random counterparts. Thus, they may (loosely) be thought of as particular kinds of motifs (although undirected) as studied by Milo and co-workers \cite{milo}. However, our definition is much more restrictive than that of Milo and co-workers since we single out motifs preserved by any (global) symmetry of the network. Although this restriction means that we consider only a small subset of possible network motifs, it is useful since the presence of symmetric motifs may be directly linked to network spectra in a way which is not possible for general motifs. 

\begin{table}[!t]
\begin{center}
\begin{ruledtabular}
\begin{tabular}[c]{c  c  c}
Sym. Motif & Geom.~Factor & Eigenvalues \\
\hline\\
\includegraphics[width=0.09\textwidth]{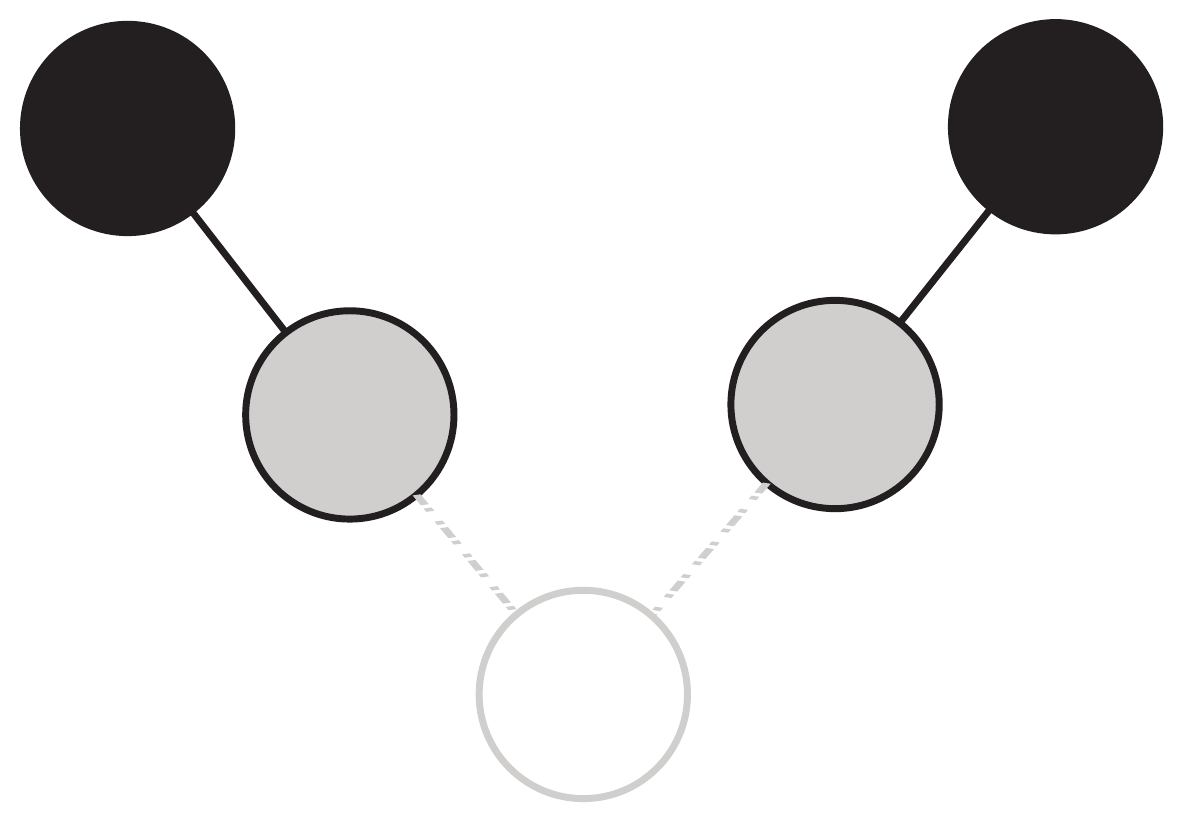} & $S_2$  & [$-1^*,-1,1^*,1$] \\
\includegraphics[width=0.06\textwidth]{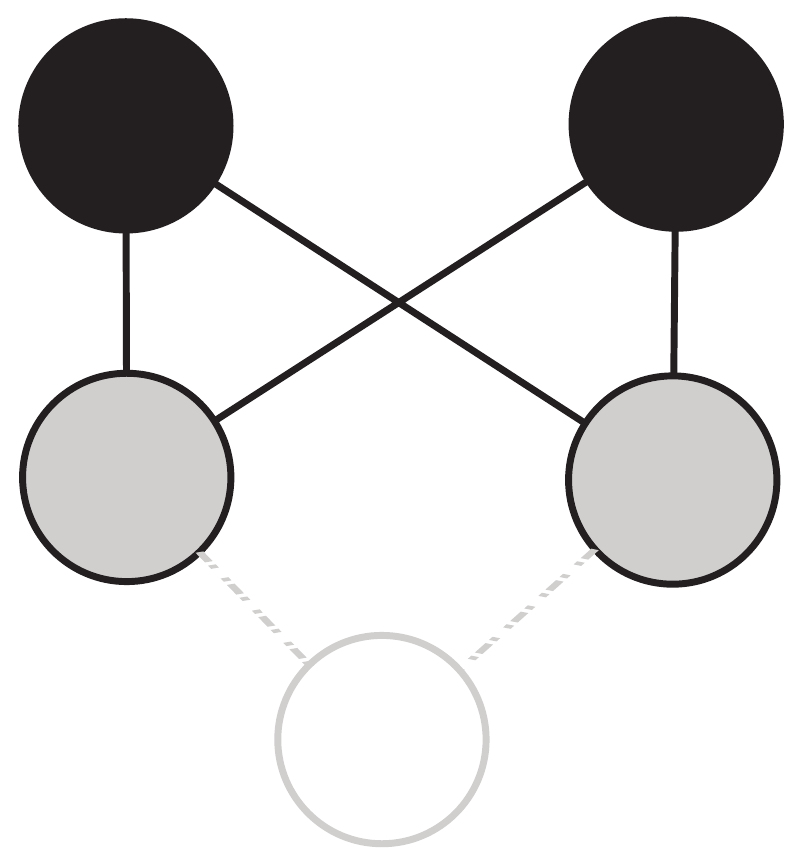} & $S_2$ and $S_2$ & [$-2,0^*,0^*,2$] \\
\includegraphics[width=0.07\textwidth]{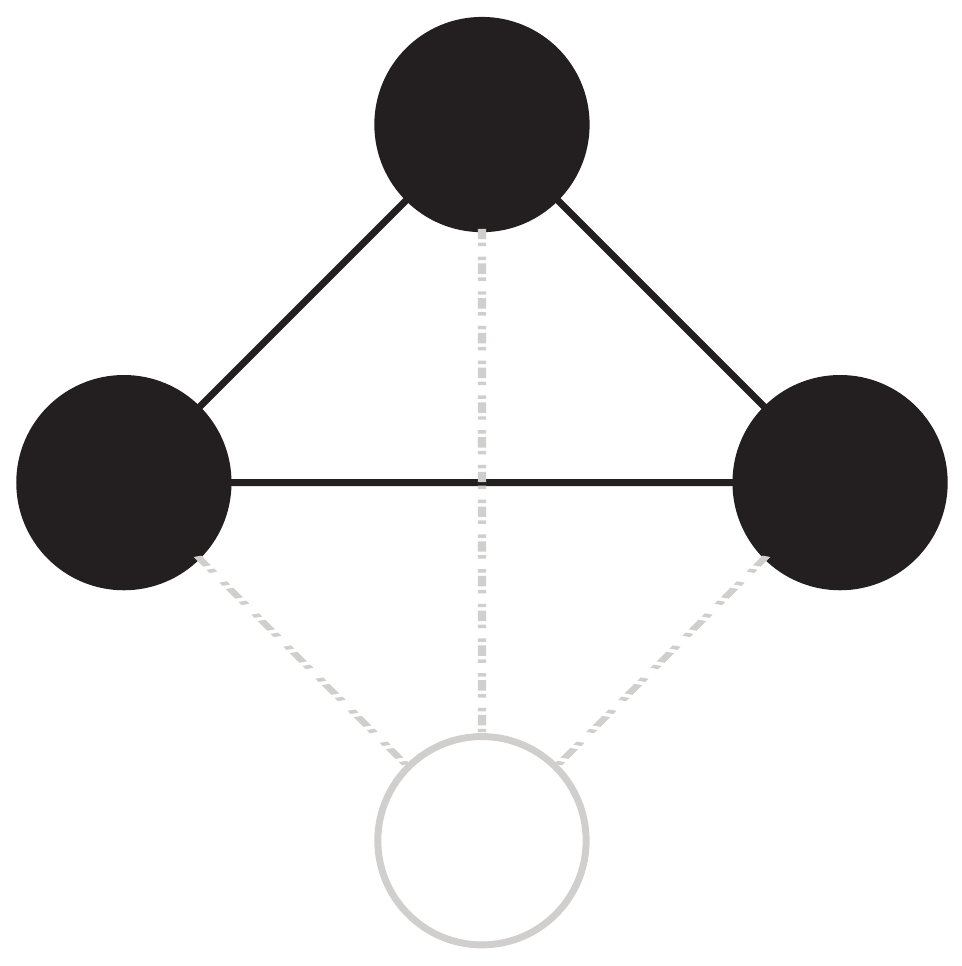} & $S_3$ & [$-1^*,-1^*,2$] \\
\includegraphics[width=0.08\textwidth]{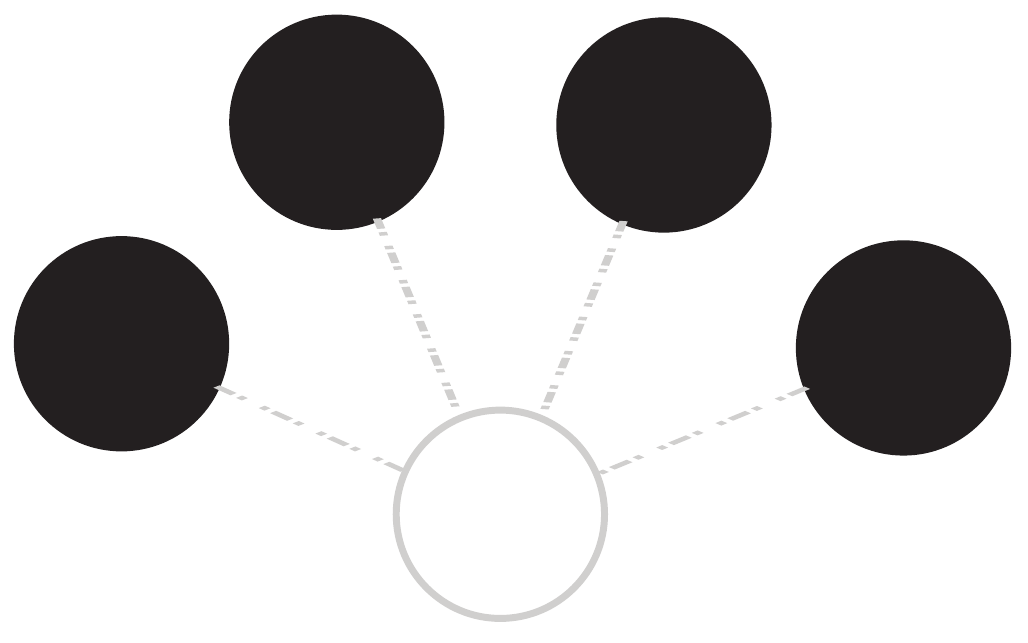} & $S_4$ &  [$0^*,0^*,0^*,0$] \\
\includegraphics[angle=90,width=0.12\textwidth]{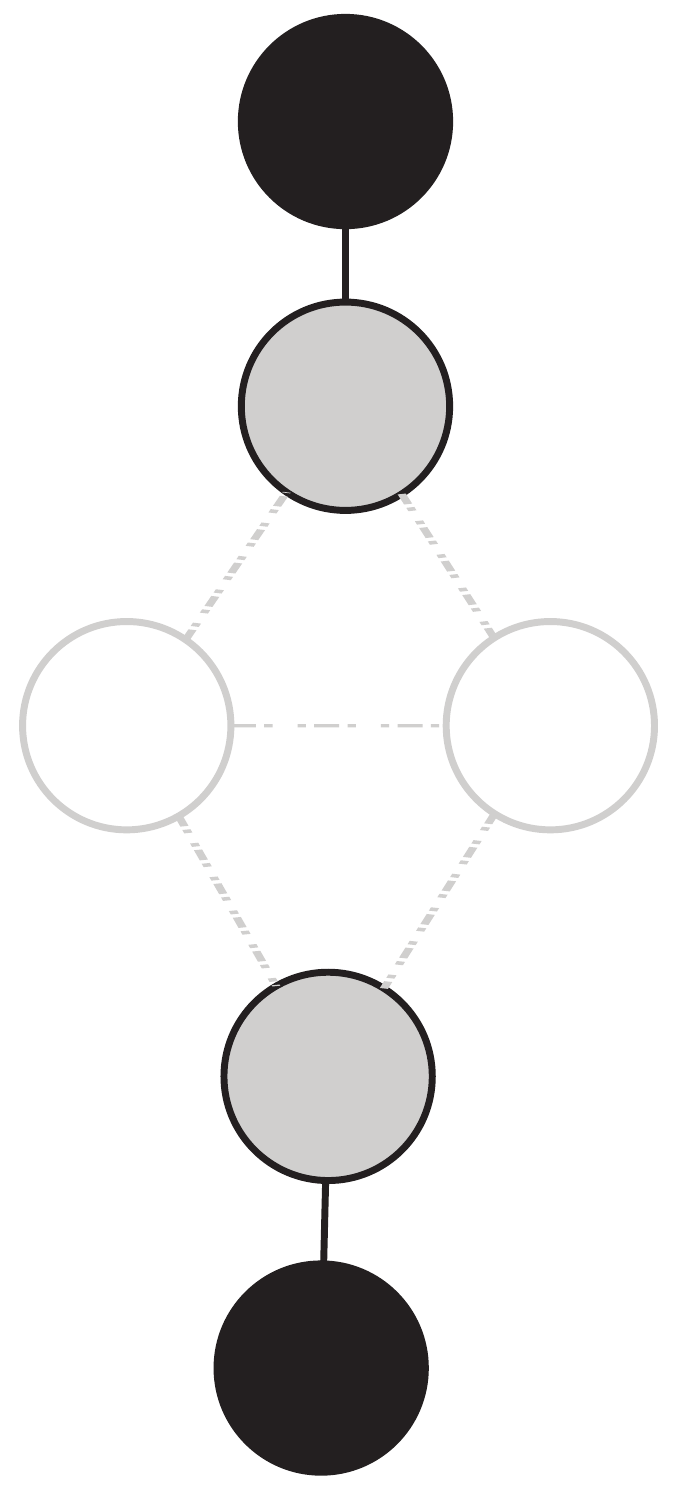} &  $S_2$ &[$-1^*,-1,1^*,1$] \\
\includegraphics[width=0.1\textwidth]{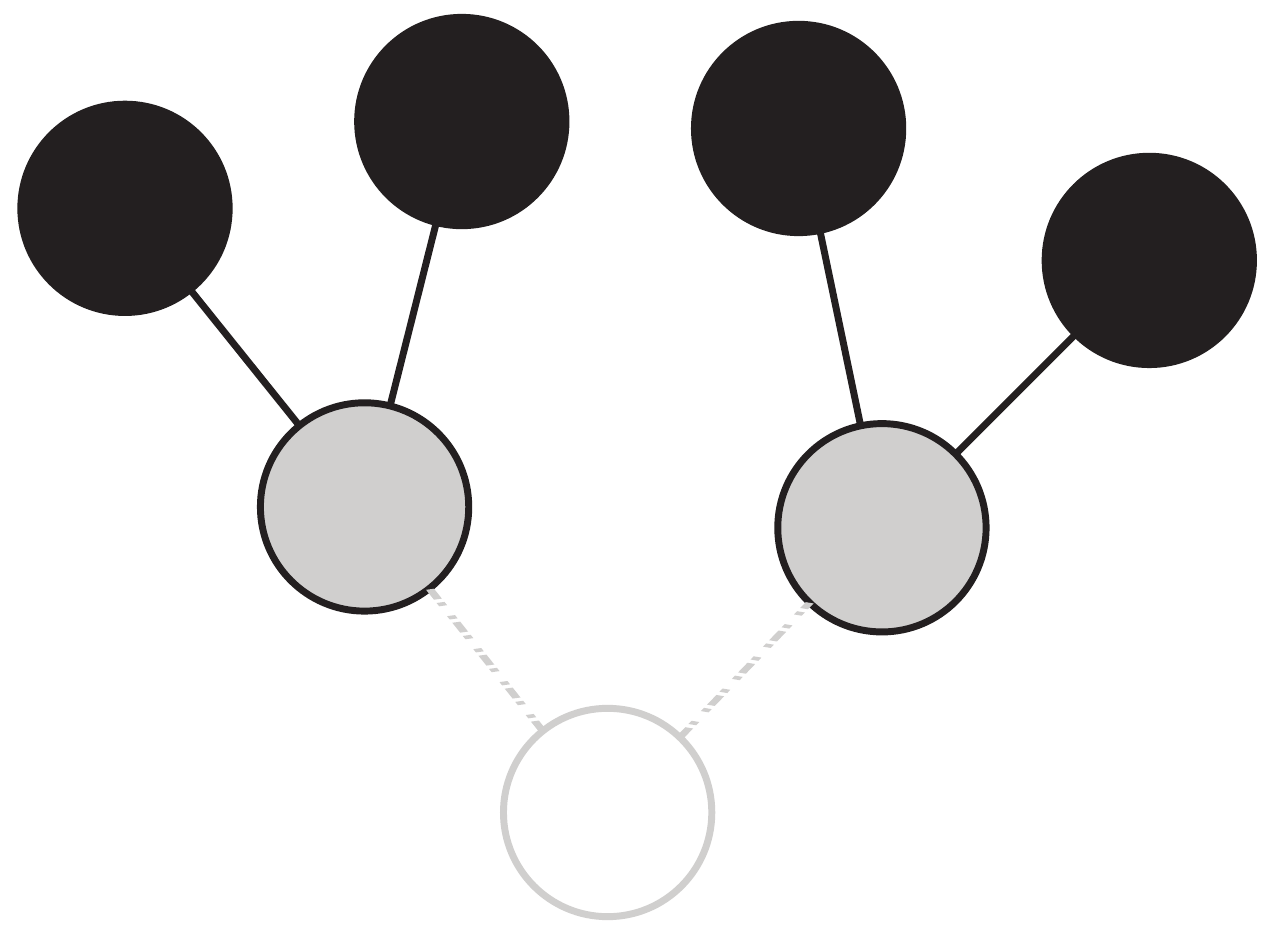} & $S_2 \wr S_2$ & [$-\sqrt{2}^*,-\sqrt{2},0^*,0^*,\sqrt{2}^*,\sqrt{2}$] \\
\end{tabular}
\end{ruledtabular}
\end{center}
\caption{\label{auttable}\small{\textbf{Symmetric motifs, their associated geometric factors and eigenvalues}. Vertices are coloured by orbit, ghost edges and vertices show how the symmetric motif attaches to the network. In the second motif, the action is independent in each orbit, hence is given as two geometric factors. Redundant eigenvalues (see Section \ref{sec:SymmMotifs}) are starred. Notice how different symmetric motifs give rise to the same redundant eigenvalues. The complete spectrum this network is as follows: -2.7337, -2.3923, -2.2758, -2.0291, -1.8546, -1.4181, -1.4142$^*$, -1.1559, -1$^*$, -1$^*$, -1$^*$, -1$^*$, -0.2251, 0$^*$, 0$^*$, 0$^*$, 0$^*$, 0$^*$, 0$^*$, 0$^*$, 0, 0.2712, 0.3812, 0.7218, 1$^*$, 1$^*$, 1.4142$^*$, 1.7570, 1.9740, 2.2323, 2.4236, 2.9431, 3.3804. Fifteen of these eigenvalues are redundant (starred, see section \ref{sec:SymmMotifs}) and the remaining 18 form the spectrum of the quotient graph (see section \ref{sec:quo}). This situation is general and will be explained in Section \ref{sec:SymmMotifs}.}} 
\end{table}

\section{Symmetry and Redundancy in Network Spectra}\label{sec:spec}
The presence of singularities in the eigenvalue spectra of real-world networks has been previously observed and reasons for certain of these peaks has been discussed \cite{dorogovtsevSCN,dorogovtsevES,goh,kamp}. In this section we aim to extend these previous results by outlining a formal framework in which to consider general spectral characteristics of redundancy. We do so by considering interactions between a network's automorphism group and the eigenvalues of its adjacency matrix.

The \emph{adjacency matrix} of a simple network $\mathcal{G}$ is the $ N \times N$ symmetric matrix 
\begin{displaymath}
\mathbf{A}  = A_{ij} =  \left \{ \begin{array}{ll} 1 & \textrm{ if $v_i$ and $v_j$ are adjacent,} \\ 0 & \textrm{otherwise.} \end{array} \right.
\end{displaymath} 
The eigenvalues of $\mathcal{G}$ are the eigenvalues of its adjacency matrix, and the set of eigenvalues is the \emph{network's spectrum}. For undirected networks, the matrix $A$ is symmetric and therefore all eigenvalues are real and there is an orthonormal basis of eigenvectors. For the remainder of this discussion we shall focus on simple undirected networks. 

The spectral density of a simple network $\mathcal{G}$ is the density of its eigenvalues, which can be written as a sum of Dirac delta-functions
\begin{equation}
\rho(\lambda) = \frac{1}{N} \sum_{i=1}^{N} \delta (\lambda - \lambda _i),
\end{equation}
where $\lambda_i$ is the $i$th largest eigenvalue of $\mathcal{G}$. 

Consider $p$, a permutation of the vertices of $\mathcal{G}$, which can be represented by a permutation matrix $\mathbf{P}$ where 
\begin{displaymath}
\mathbf{P} = P_{ij} =  \left \{ \begin{array}{ll} 1 & \textrm{ if $p (v_i) = v_j$,} \\ 0 & \textrm{otherwise.} \end{array} \right.
\end{displaymath}
The relationship between network symmetry and eigenvalue spectra depends centrally upon the fact that $p \in \textrm{Aut}(\mathcal{G})$ if and only if $\mathbf{A}$ and $\mathbf{P}$ commute \cite{lauri}. Thus, if $\mathbf{x}$ is an eigenvector of $\mathbf{A}$ corresponding to the eigenvalue $\lambda$ then $\mathbf{P}\mathbf{x}$ is also an eigenvector of $\mathbf{A}$ corresponding to $\lambda$. Since $\mathbf{Px}$ and $\mathbf{x}$ are generally linearly independent, this means that network symmetry (and thus redundancy) naturally gives rise to eigenvalues with high multiplicity and therefore singularities in the spectral density. In the following sections we shall develop this result a little further and show how certain network eigenvalues may be associated directly with symmetric motifs. First we need to recall the notion of quotient graph.

\subsection{Network Quotients}\label{sec:quo}
Since automorphisms permute vertices without altering network structure, a network's automorphism group may be used to partition its vertex set $V(\mathcal{G})$ into disjoint strutural equivalence classes called \emph{orbits} (see FIG. \ref{basicexample}). For every vertex $v \in V(\mathcal{G})$, the set of vertices to which $v$ maps under the action of the automorphism group $G = \textrm{Aut}(\mathcal{G})$ is called the $G$-orbit of $v$, written $\Delta_G (v)$ or simply $\Delta(v)$. More formally,
\[
\Delta_G(v) = \{g\cdot v \in V : g \in G \}.
\]
Similarly, if $H$ is a subgroup of $G$, the $H$-orbit of a vertex $v$ is the set 
\[
\Delta_H(v) = \{g\cdot v \in V : g \in H \}.
\]
Since vertices in the same orbit may be permuted without altering network structure, they are structurally indistinguishable from each other (that is, they play precisely the same topological role in the network). Thus, a network's orbit structure efficiently quantifies the degree of structural redundancy the network carries. For example, the vertices in FIG. \ref{example} are coloured by orbit. 

Since the vertices in each orbit are structurally equivalent, they may be associated with each other to form the basis of a network coarse-graining known in the context of algebraic graph theory as the \emph{quotient graph}. More specifically, let $\mathbf{\Delta} = \{ \Delta(v_1),\Delta(v_2) \ldots, \Delta(v_s) \}$ be the system of orbits which the vertices of $\mathcal{G}$ are partitioned into under the action of $G$. Let $q_{ij}$ ($i,j = 1,2,\ldots,s$) be the number of edges starting from a vertex in $\Delta_i$ and ending in vertices in $\Delta_j$. Since the orbits partition the vertex set into disjoint equivalence classes, $q_{ij}$ depends on $i$ and $j$ alone. The \emph{quotient} $\mathcal{Q}$ of $\mathcal{G}$ under the action of $G$ is the multi-digraph with vertex set $\mathbf{\Delta}$ and adjacency matrix $(q_{ij})$. We refer to the network $\mathcal{G}$ as the \textit{parent} of $\mathcal{Q}$. Crucially, the quotient of $\mathcal{G}$ retains the unique structural elements of $\mathcal{G}$ yet, by associating structurally equivalent elements, factors out all redundancy. Previous reports have shown that quotients of many empirical networks can be as small as 20\% the size of their parent networks yet preserve precisely key network properties which determine system function \cite{shaw}. Note that we can similarly define the quotient of $\mathcal{G}$ under the action of any subgroup $H$ of $G$ (hence factoring out just a fraction of the redundancy).

\begin{figure}[t]
\begin{center}
\includegraphics[width=0.5\textwidth]{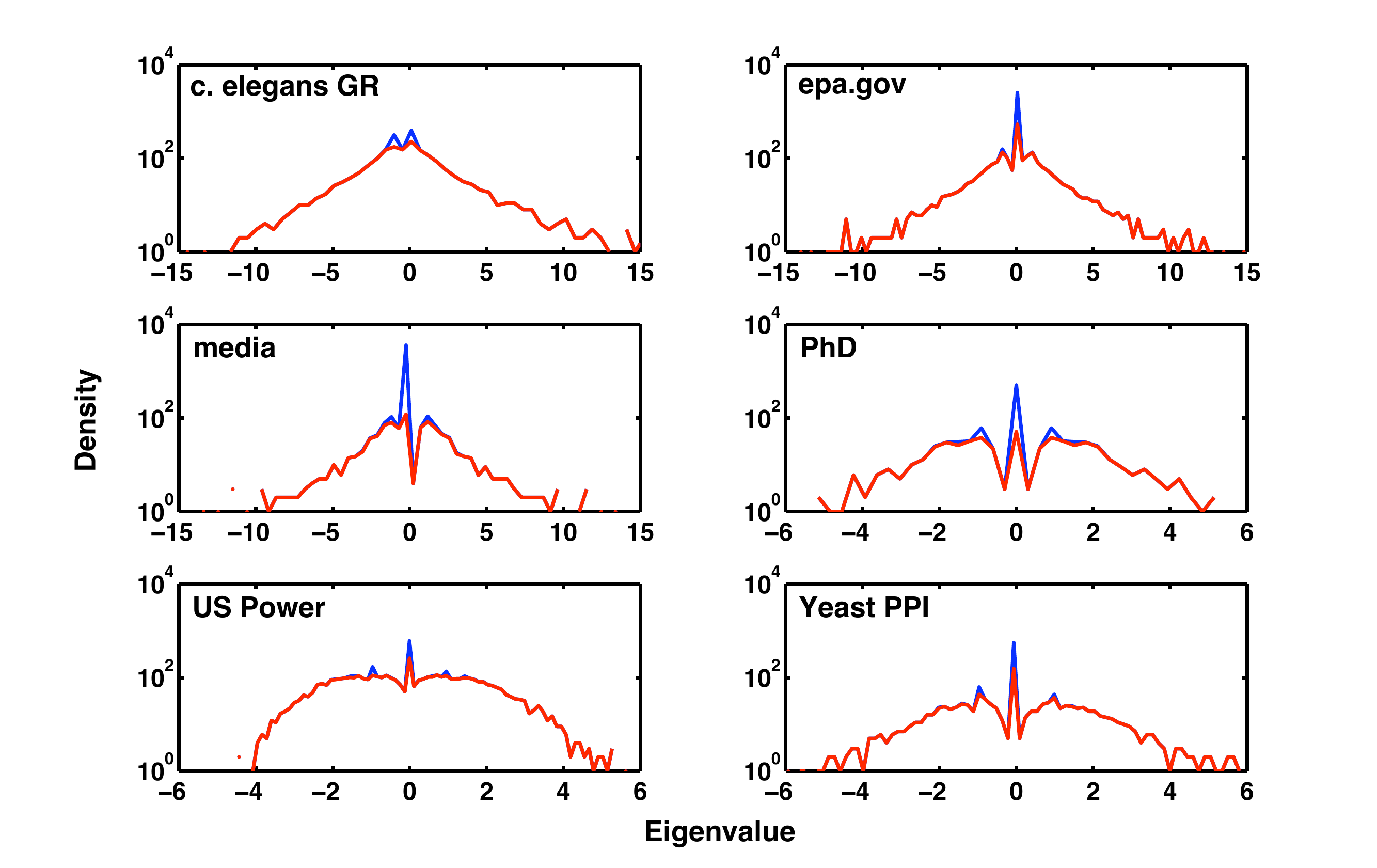}
\caption{\label{spectra}\small{\textbf{Spectral densities of networks and their quotients}} (color online). In all cases, the spectral density of the parent is in dark grey (blue online) while that of the quotient is in light grey (red online). \emph{(a)} the \emph{c. elegans} genetic regulatory network \cite{zhong} \emph{(b)} The www.epa.gov subnetwork \cite{kleinberg2} \emph{(c)} A media ownership network \cite{norlen} \emph{(d)} A network between PhD students and their supervisors \cite{denooy, sigact} \emph{(e)} The US power grid \cite{wattsSW} \emph{(f)} The yeast protein-protein interaction network \cite{jeong}. Note that the y-axis is on a logarithmic scale: in each case, the differences in redundant eigenvalue multiplicities between the parent network and its quotient are significant (see Table \ref{spectratable}).}
\end{center}
\end{figure}

A key result for the present discussion is that, for any given graph $\mathcal{G}$, set of eigenvalues of its quotient are a subset of those of $\mathcal{G}$ \cite{cvetkovic}. Given a graph $\mathcal{G}$ with orbits $\Delta_1, \ldots, \Delta_m$ and an eigenpair $(\lambda, \mathbf{v} = (v_1,\ldots,v_m))$ of the quotient graph $\mathcal{Q}$, then $\lambda$ is also an eigenvalue of the parent network $\mathcal{G}$ with an eigenvector consisting on an identical value $v_i$ on all the vertices of $\Delta_i$. Thus, a network's automorphism group may be used to construct a factorization of its characteristic polynomial, via its quotient. Additionally, since quotients carry less repetition than their parent networks, we find that quotient spectra generally contain less degeneracy than their parent networks. FIG.~\ref{spectra} illustrates this point by giving the spectral densities of 6 representative (biological, social and technological) networks and their quotients. As expected, in each case the spectral density of the parent network contains peaks which are significantly reduced in the spectral density of its quotient. In the following section we will make this relation more explicit by associating specific network eigenvalues with specific symmetric motifs. These, together with the eigenvalues coming from the quotient, describe the entire spectrum of the network.

\subsection{Symmetric Motifs and Network Spectra}\label{sec:SymmMotifs}
There have been some previous attempts to spot the eigenvalues of key subgraphs in network spectra \cite{kamp}. However subgraph eigenvalues are not usually contained in network spectra, and in general they only interlace those of the network \cite{cvetkovic}. Nevertheless, certain eigenvalues associated with symmetric motifs are retained in network spectra. We call them \emph{redundant} eigenvalues, and they are described as follows.

Recall the physical meaning of a eigenvalue-eigenvector pair of an undirected graph. Consider a vector $\mathbf{v}$ on the vertex set of a graph, and write $v_i$ for the value at a vertex $i$. Write on each vertex the sum of the numbers found on the neighbours of that vertex. If the new vector is a multiple of $\mathbf{v}$, say $\lambda \mathbf{v}$, then $\mathbf{v}$ is an eigenvector with eigenvalue $\lambda$. We shall say that an eigenvalue-eigenvector pair $(\lambda,\mathbf{v})$ of a symmetric motif $\mathcal{M} = \mathcal{M}_H$ (considered as an induced subgraph) is \emph{redundant} if, for each $H$-orbit $\Delta_H \in \mathcal{M}$, the sum $\sum_{i \in \Delta_H} v_i =0$. For example, in Table \ref{starstable} the redundant eigenvectors are starred: the coordinates are separeted by orbits and the sum over each orbit is zero. Indeed, it can be shown that if $\mathcal{M}$ has $n$ vertices and $m$ $H$-orbits, there is an orthonormal basis of eigenvectors of $\mathcal{M}$ such that $n-m$ of them are redundant, and the remaining $m$ are constant on each orbit (see Appendix \ref{appendixA}).

\begin{table}[!t]
\begin{center}
\begin{ruledtabular}
\begin{tabular}[c]{c  c  c}
Sym. Motif & Eigenvalues & Eigenvectors \\
\hline 

\raisebox{-0.6cm}{\includegraphics[width=0.15\textwidth]{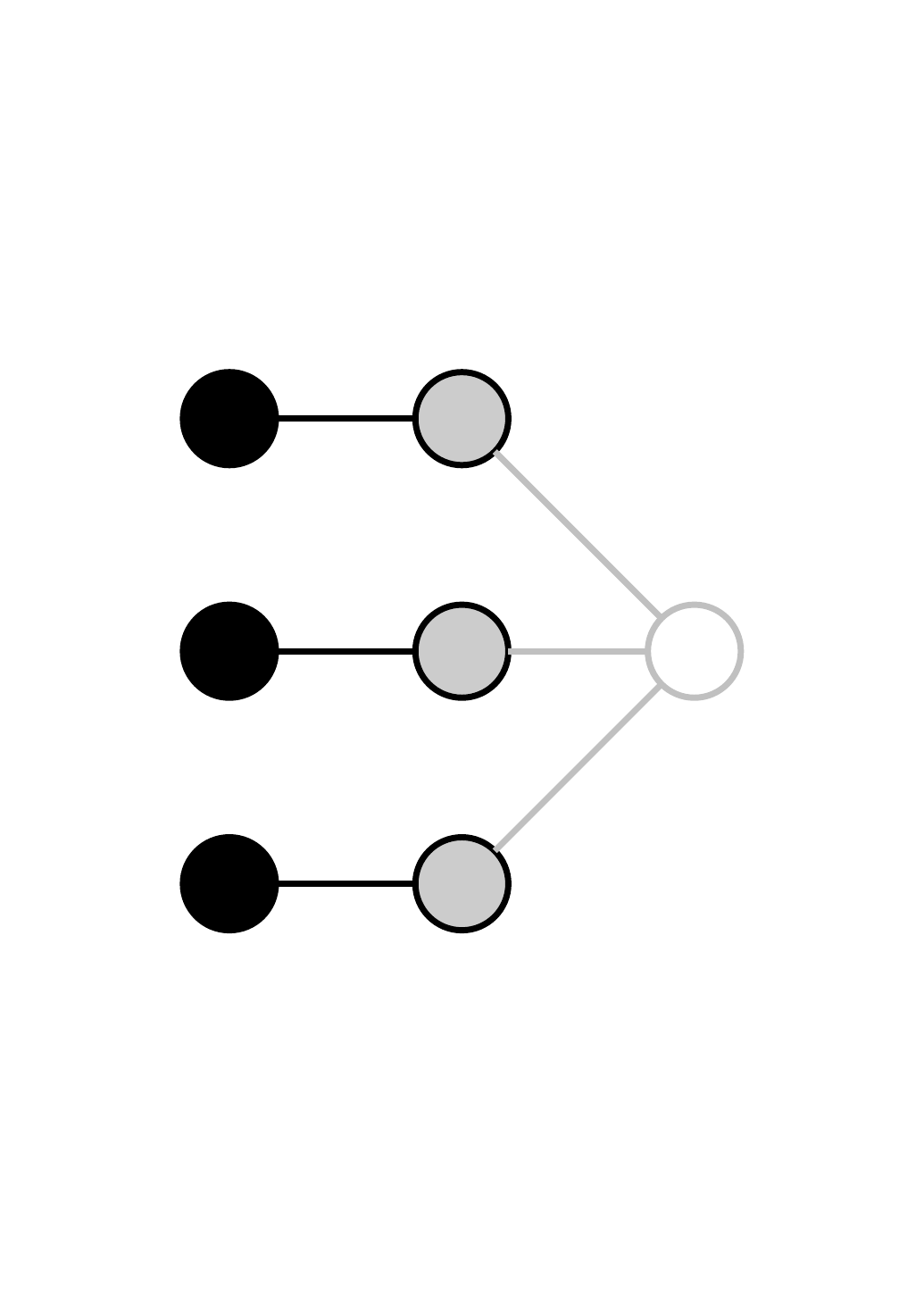}} & \begin{minipage}[b]{0.5cm} \footnotesize{$1^*$ $1^*$ $-1^*$ $-1^*$ $2$ $-2$ $0$} \end{minipage} & \raisebox{0.0cm}{\begin{minipage}[b]{4.0cm} \footnotesize{$(1, -1, 0, | 1, -1, 0, | 0)^*$ $(1, 0, -1, | 1, 0, -1, | 0)^*$ $(1, -1, 0, | -1, 1, 0, | 0)^*$ $(1, 0, -1, | -1, 0, 1, | 0)^*$ $(1, 1, 1, | 2, 2, 2, | 3)$ $(1, 1, 1, | -2, -2,  -2, | 3)$ $(1, 1, 1, | 0, 0, 0, | -1)$} \end{minipage}}\\

\hline 

\raisebox{-0.4cm}{\includegraphics[width=0.15\textwidth]{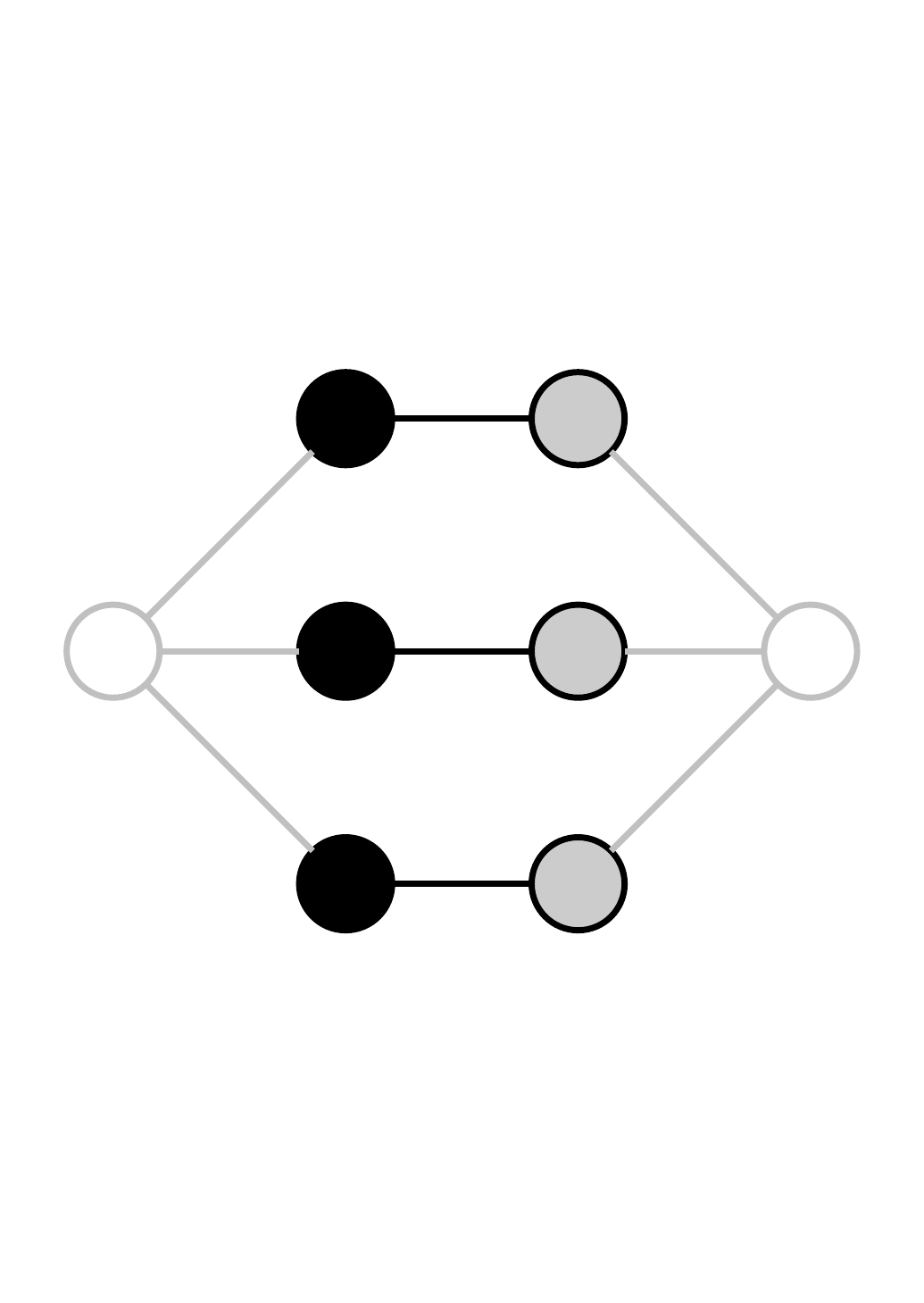}} & \begin{minipage}[b]{0.5cm} \footnotesize{$1^*$\\ $1^*$\\ $-1^*$\\ $-1^*$\\ $\lambda_1$} \\ $\lambda_2$ \\ $-\lambda_1$ \\ $-\lambda_2$
\end{minipage} & \raisebox{0.0cm}{\begin{minipage}[b]{4.0cm} \footnotesize{$(0, | 1, -1, 0,| 1, -1, 0, | 0)^*$\\ $(0, | 1, 0, -1,| 1, 0, -1 | 0)^*$\\ $(0, | 1, -1, 0,| -1, 1, 0, | 0)^*$\\ $(0, | 1, 0, -1,| -1, 0, 1,| 0)^*$\\ $(-\lambda_2,|1, 1, 1, | 1, 1, 1,|-\lambda_2)$\\ $(-\lambda_1,|1, 1, 1, | 1, 1, 1,|-\lambda_1)$\\ $(\lambda_2,|1, 1, 1, | -1, -1, -1,|-\lambda_2)$\\ $(\lambda_1,|1, 1, 1, | -1, -1, -1,|-\lambda_1)$}\\\end{minipage}}\\

\hline 

\raisebox{-0.17cm}{\includegraphics[width=0.10\textwidth]{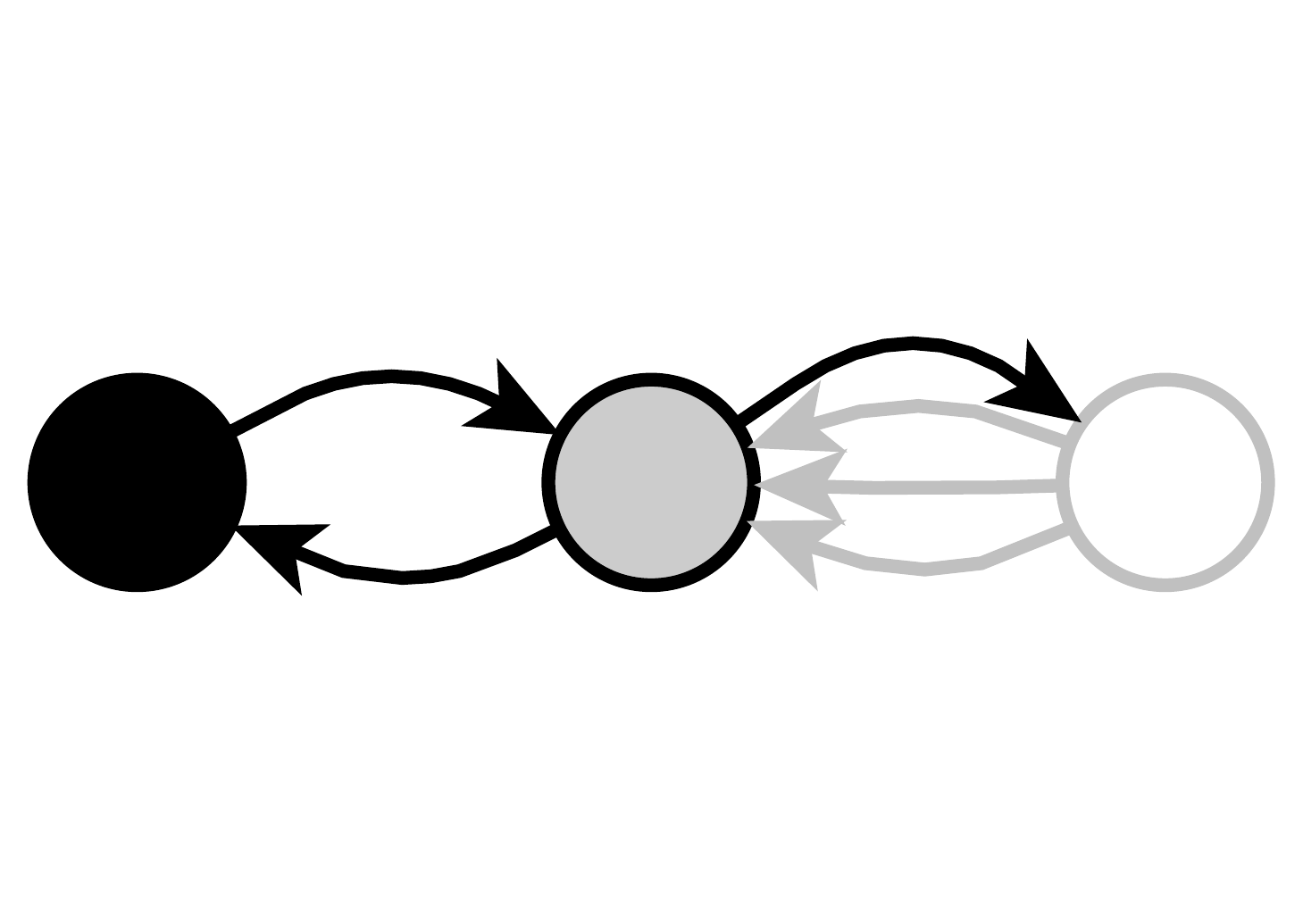}} & \begin{minipage}[b]{0.5cm} \footnotesize{$2$ $-2$ $0$} \end{minipage} & \raisebox{0.0cm}{\begin{minipage}[b]{4.0cm} \footnotesize{$(1, 2, 3)$\\ $(1, -2, 3)$\\ $(1, 0, -1)$} \end{minipage}}\\

\hline 

\raisebox{-0.3cm}{\includegraphics[width=0.15\textwidth]{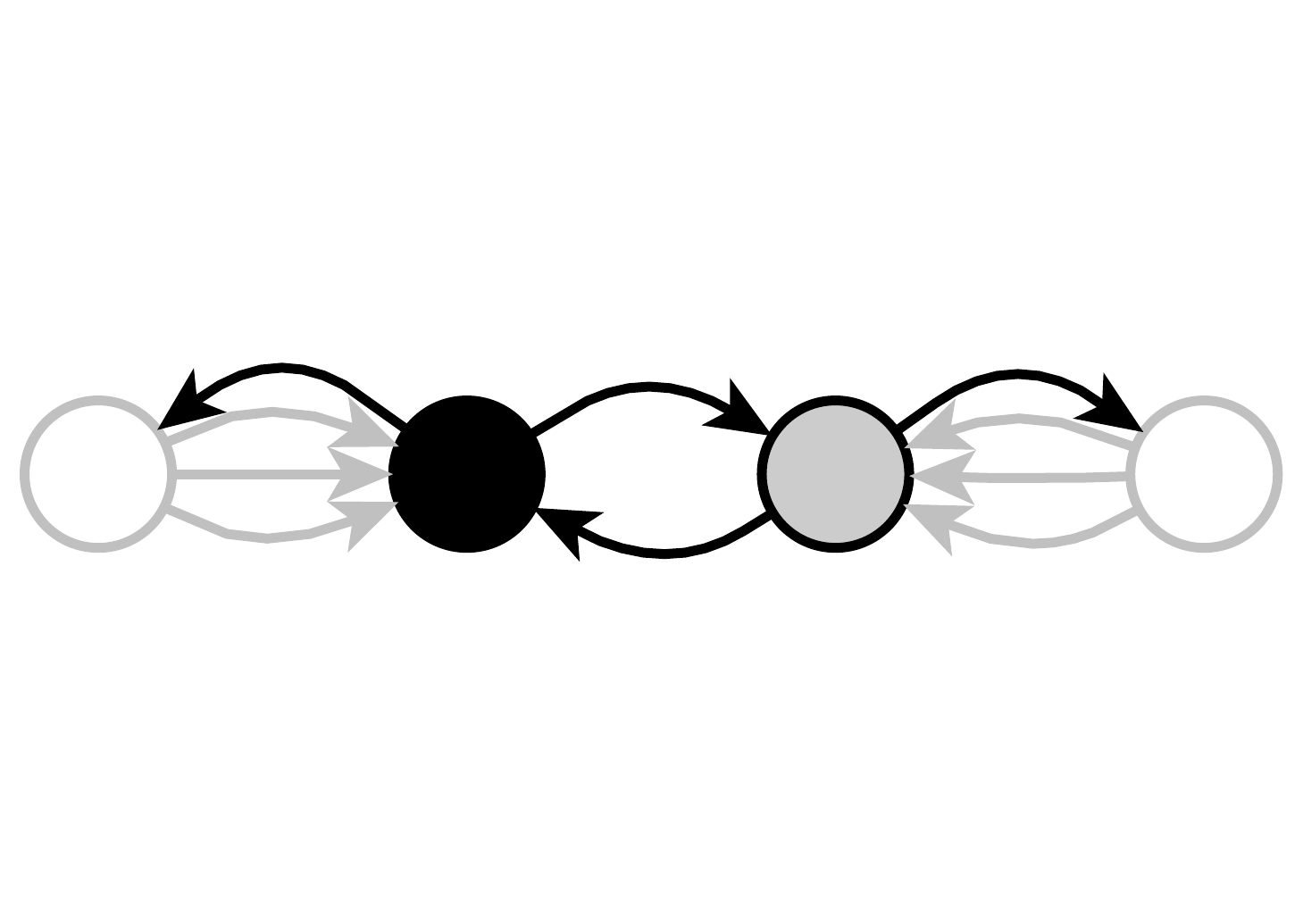}} & \begin{minipage}[b]{0.5cm} \footnotesize{$\lambda_1$ $\lambda_2$ $-\lambda_1$ $-\lambda_2$} \end{minipage} & \raisebox{0.0cm}{\begin{minipage}[b]{4.0cm} \footnotesize{$(-\lambda_2, 1, 1,-\lambda_2)$\\ $(-\lambda_1, 1, 1, -\lambda_1)$\\ $(\lambda_2, 1, -1, -\lambda_2)$\\ $(\lambda_1, 1, -1, -\lambda_1)$} \end{minipage}}\\

\end{tabular}
\end{ruledtabular}
\end{center}
\caption{\label{starstable}\small{\textbf{Examples of redundant spectra}. Two symmetric motifs (top two rows) and their quotients (bottom two rows) are shown. Vertices are coloured by orbit and ghost vertices and edges show how each motif attaches to a (hypothetical) network. In both cases the underlying symmetric motif is $V_3 \equiv V_3$ (see Section \ref{section:BSMs} and compare with Table \ref{BSM2}) with geometric factor $S_3$, permuting each orbit simultaneously. Eigenvector entries are separated by orbit. The eigenvalues $\lambda_1 \approx 2.30$, and $\lambda_2 \approx -1.30$ are the solutions of $\lambda^2-\lambda-3=0$. Redundant eigenvectors (the sum of the entries on each orbit is zero) and their eigenvalues are starred. The redundant eigenvalues are the same in both cases, while the non-redundant ones vary. Note that the non-redundant eigenvectors are constant in each orbit. Observe also that the number of redundant eigenvectors equals the number of vertices minus the number of orbits (see Appendix \ref{appendixA}) and the quotient graphs retain exactly the non-redundant part of the spectrum.}} 
\end{table}

We say that an eigenvalue $\lambda$ is \emph{redundant with multiplicity $m$} if there are up to $m$ linearly independent eigenvectors $v_1, \ldots, v_m$ such that all the pairs $(\lambda, \mathbf{v}_i)$ are redundant. For example, the eigenvalue $0$ in the fourth motif of Table \ref{auttable} has multiplicity 4 but redundant multiplicity 3. The crucial property is that redundant eigenvalues are retained, with their \emph{redundant} multiplicity, in the network spectrum: if $(\lambda, \mathbf{v})$ is a redundant eigenvalue-eigenvector pair of a symmetric motif $\mathcal{M}$ then $(\lambda,\overline{\mathbf{v}})$ is an eigenvalue-eigenvector pair for the whole network, where $\overline{\mathbf{v}}$ is formed by setting $\overline{v}_i = v_i$ for all $i \in \mathcal{M}$ and setting $\overline{v}_i=0$ for all $i \notin \mathcal{M}$ \footnote{Every vertex $j \in \mathcal{G} - \mathcal{M}$ is adjacent to either all or none of the vertices of each $H$-orbit $\Delta$, since the action of $H$ permutes transitively the vertices of $\Delta$ while fixing $j$, hence the value at $j$ remains zero and satisfies the eigenvector-eigenvalue relation $\lambda \cdot 0 = 0$.}. 
We call such an eigenvector \emph{$\mathcal{M}$-local}: it is constructed from a redundant eigenvector of a symmetric motif $\mathcal{M}$ by setting entries to zero on the vertices outside $\mathcal{M}$.

The non-redundant eigenvalues of $\mathcal{M}$ will not, in general, be retained in the network spectrum but rather will change depending on how the motif is embedded in the network (more precisely, on the topology of the quotient graph) -- for instance, see the examples in Tables \ref{auttable} and \ref{starstable}.

\textit{Remark:} The argument above applies naturally to symmetric motifs but not necessarily to single orbits: a redundant eigenvector of an orbit will not necessarily give an eigenvector of the whole network (see for instance the closing remark on Table \ref{BSM2}). The reason is that it may not be possible to treat one orbit on its own if the action is not `independent' on this orbit. The smallest independent actions (and their associated subgraphs) are precisely given by the geometric factorization of Eq.~\ref{decomp}. The symmetric motifs are the smallest subgraphs whose redundant eigenvalues survive to the spectrum of the whole network.

On the other hand, consider the quotient graph of a network $\mathcal{G}$.  Recall that if $(\lambda, \mathbf{v})$ is an eigenpair of the quotient then $(\lambda, \widehat{\mathbf{v}})$ is an eigenpair of $\mathcal{G}$, where $\widehat{\mathbf{v}}$ is obtained setting the identical value $v_i$ on all the vertices of the $i$th orbit. We say that the eigenvector of the parent network $\widehat{\mathbf{v}}$ is \emph{lifted} from the eigenvector $\mathbf{v}$ of the quotient. 

The key result is that these two procedures explain completely the whole spectrum of $\mathcal{G}$: if $\mathcal{G}$ has $n$ vertices and $m$ orbits, we can find a basis of eigenvectors $\widehat{\mathbf{v}}_1, \ldots, \widehat{\mathbf{v}}_m, \overline{\mathbf{v}}_{m+1}, \ldots \overline{\mathbf{v}}_n$ such that the first $m$ are lifted from a basis of eigenvectors of the quotient $\mathbf{v}_1, \ldots, \mathbf{v}_m$ and the remaining come from the redundant eigenvectors of the symmetric motifs of $\mathcal{G}$. See Appendix \ref{appendixB} for full details and Table \ref{starstable} for examples. Finally, note that the $\widehat{\mathbf{v}}_i$'s are constant on each orbit and the $\overline{\mathbf{v}}_i$'s are redundant on each orbit (the sum of the coordinates is zero).

Recall that the spectrum of the quotient graph is a subset of the spectrum of the parent network. The redundant eigenvalues are exactly the ones `lost' in the spectrum of the quotient graph (Appendix \ref{appendixA}). Hence the proportion of a network's spectrum due to redundancy is obtained by comparing the size of the parent graph to the size of its quotient. This varies from network to network but can be as small as 20\% \cite{macarthur}. Thus this phenomena is non-trivial and can account for up to 80\% of the network spectrum.

Until now we have been counting repeated eigenvalues separately (that is, we have considered eigenpairs after fixing an appropriate basis of eigenvectors).  What can we say about the multiplicity of these redundant eigenvalues? There is no general principle beyond the general rule of thumb that the multiplicity is directly correlated to the size of the automorphism group. For example, if a network has an orbit of $n$ vertices such that all the permutations of these vertices are allowed (ie.~$S_n$ acts naturally on the orbit) then there will be a redundant $\mathcal{M}$-local eigenvalue with multiplicity at least $n-1$, (see Appendix \ref{appendixC}). Conversely, a graph with only simple eigenvalues has an automorphism which is a subgroup of $S_2 \times \ldots \times S_2$ \cite{cvetkovic}.

One obvious question remains: what are the possible redundant eigenvalues associated with symmetric motifs? In principle, there is no restriction so we should rephrase the question as: what are the most commonly ocurring redundant eigenvalues in `real-world' networks? We now address this question by focussing on the most commonly ocurring symmetric motifs.

\subsection{Basic symmetric motifs}\label{section:BSMs}
Most symmetric motifs (typically more than 90\%) found in real-world networks conform to the following pattern \cite{macarthur}: they consist of one or more orbits of $n$ vertices ($n \ge 2$) with a natural symmetric action, that is, the geometric factor $H$ (the subgroup of symmetries permuting only vertices of the motif) consists of \emph{all} the permutations of the $n$ vertices of each orbit and hence $H = S_n$. Therefore, each $H$-orbit is either the empty graph $V_n$, or the complete graph $K_n$, on $n$ vertices. Every vertex not in the motif is a fixed point with respect to $H$ and hence is joined to either all or none of the vertices of each orbit. Moreover, two orbits may be joined in one of only four possible ways  shown in Table \ref{tablejoints} (for a proof see \cite{liebeck}). For example, the graphs in Table \ref{auttable} would be, in this notation, $V_2 \equiv V_2$, $V_2 \ast V_2$, $K_3$, $V_4$ and $V_2 \equiv V_2$, while the last graph does not follow this pattern.

\begin{table}[!t]
\begin{center}
\begin{ruledtabular}
\begin{tabular}[c]{c  c c}
Orbits & Graphic notation & Written notat. \\
\hline \\[-1.0cm]
\raisebox{-1.2cm}{\includegraphics[angle=90,width=0.2\textwidth]{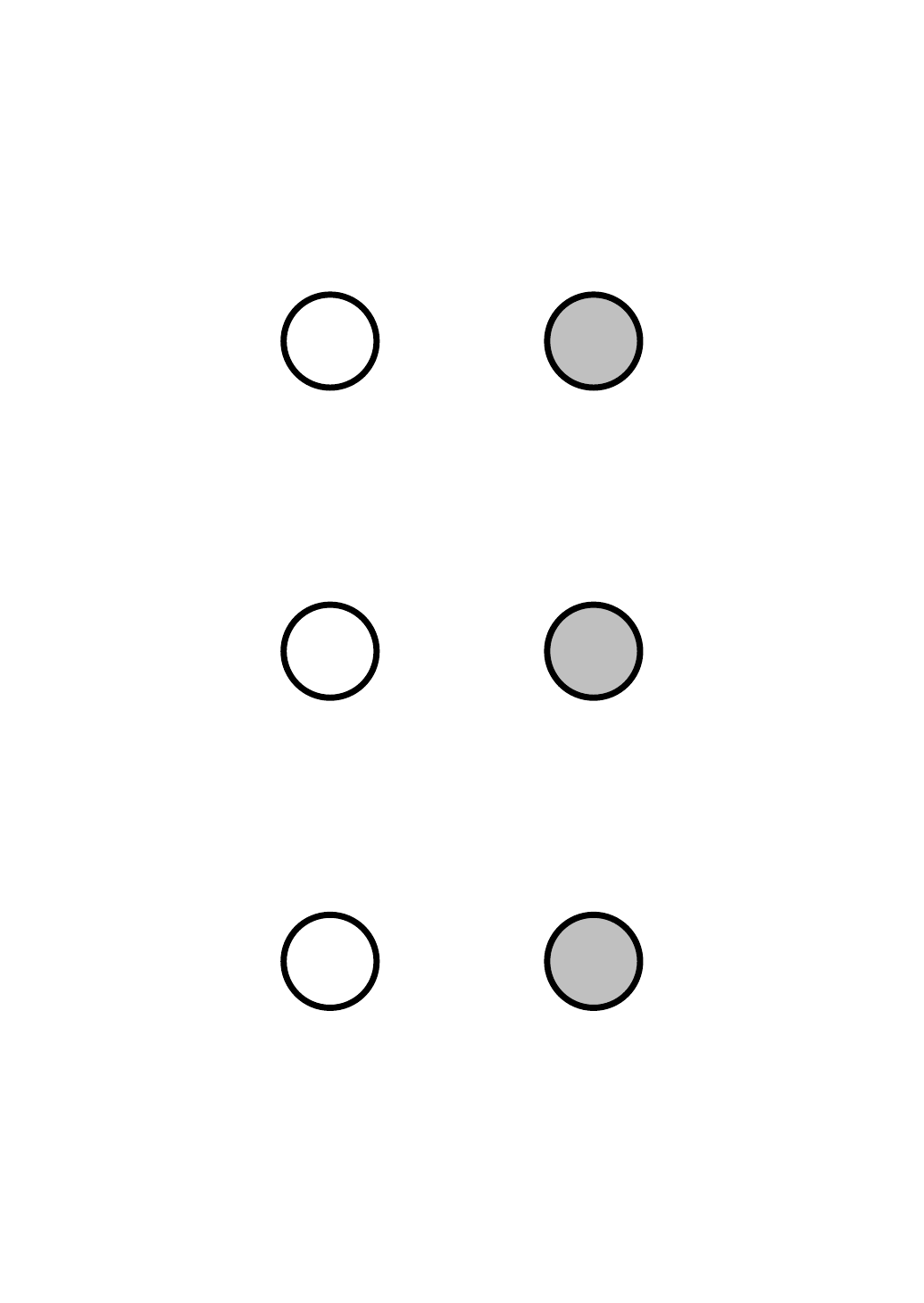}} & 
\raisebox{-1.6cm}{\includegraphics[width=0.135\textwidth]{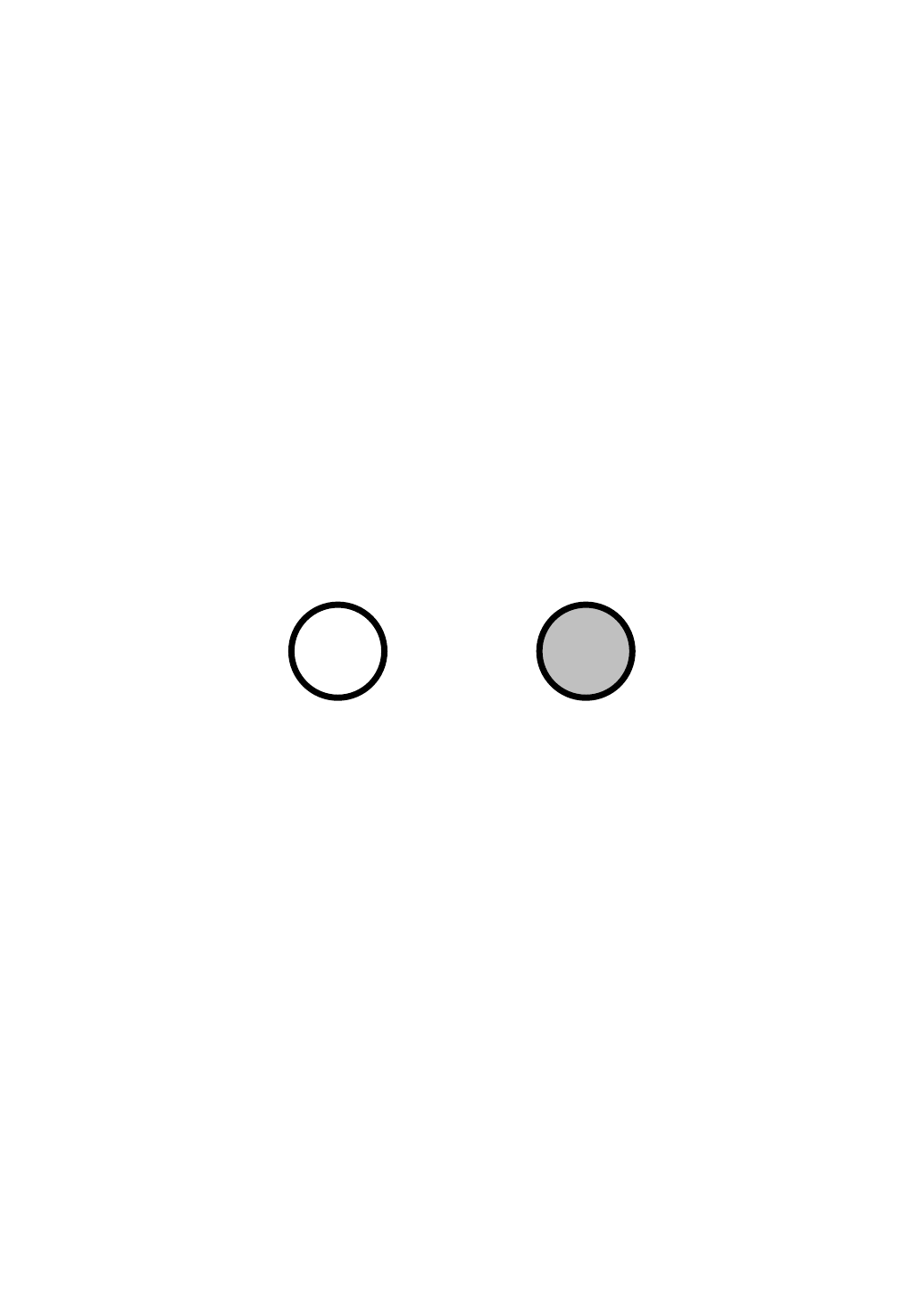}} & 
\begin{minipage}[t]{2cm}{$\Delta_1 \circ \Delta_2$}\end{minipage}\\[-0.6cm] 

\hline\\[-1.0cm]

\raisebox{-1.2cm}{\includegraphics[angle=90,width=0.2\textwidth]{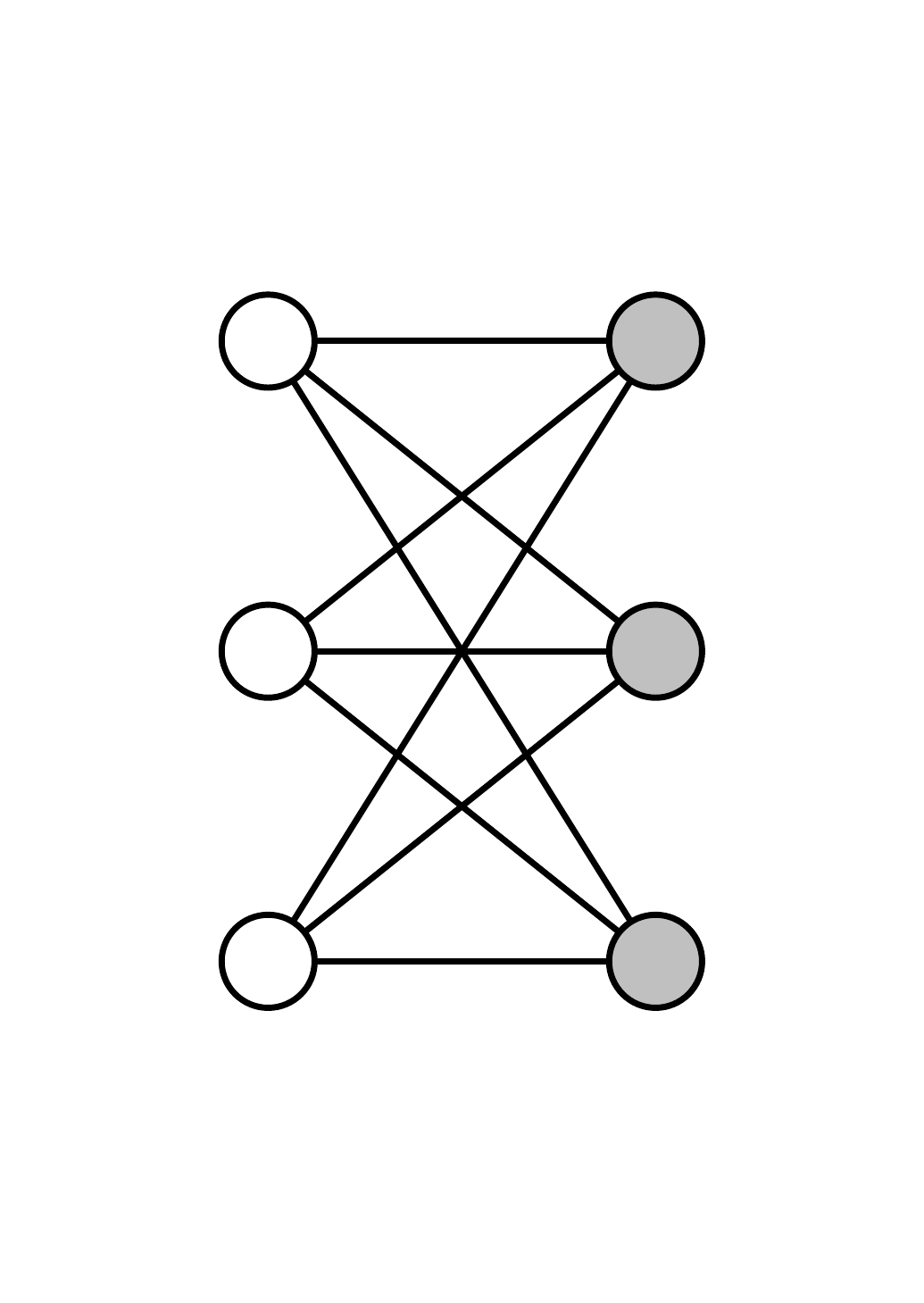}} & 
\raisebox{-1.5cm}{\includegraphics[width=0.135\textwidth]{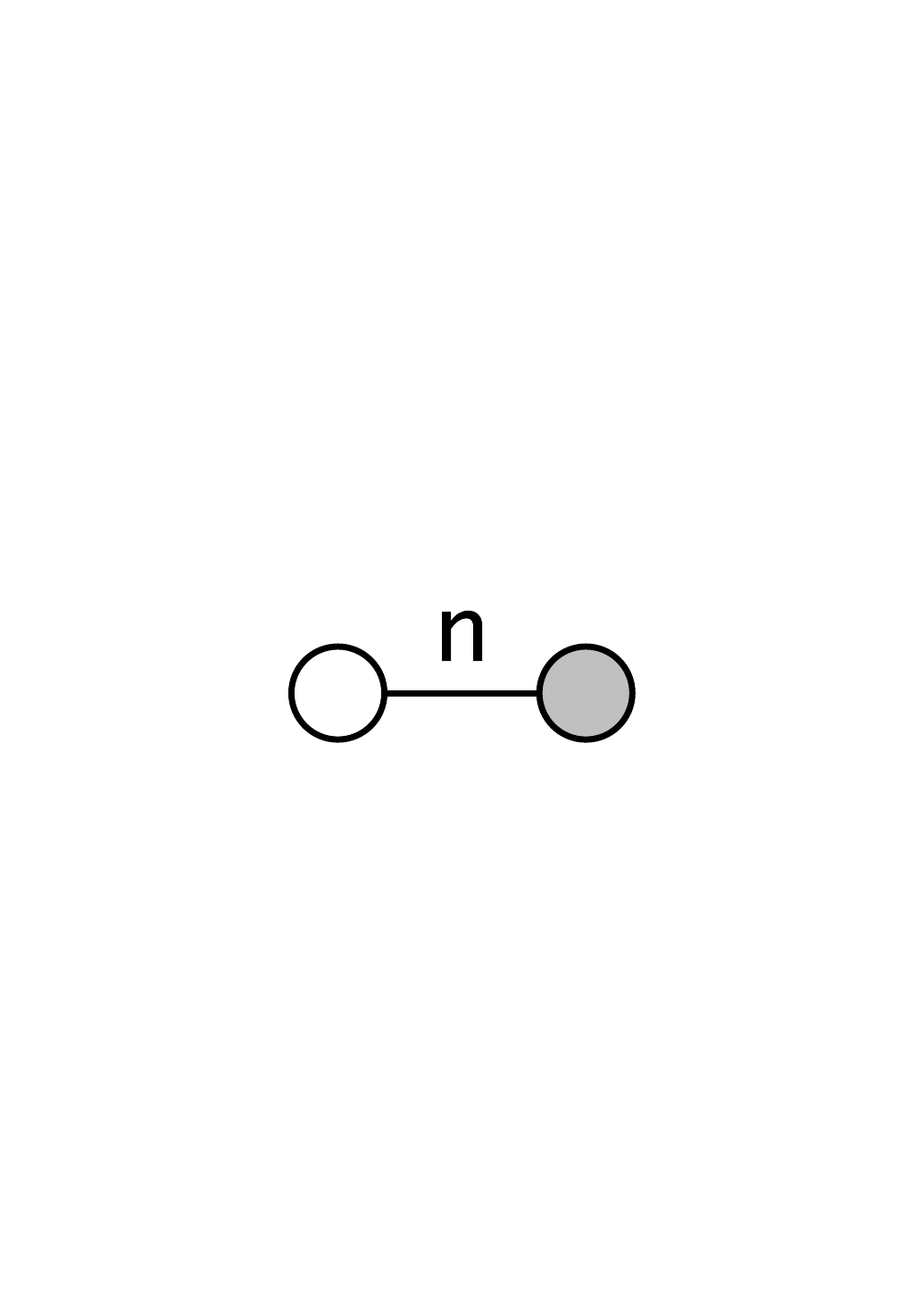}} & 
\begin{minipage}[t]{2cm}{$\Delta_1 \ast \Delta_2$}\end{minipage}\\[-0.4cm]

\hline\\[-1.0cm]

\raisebox{-1.2cm}{\includegraphics[angle=90,width=0.2\textwidth]{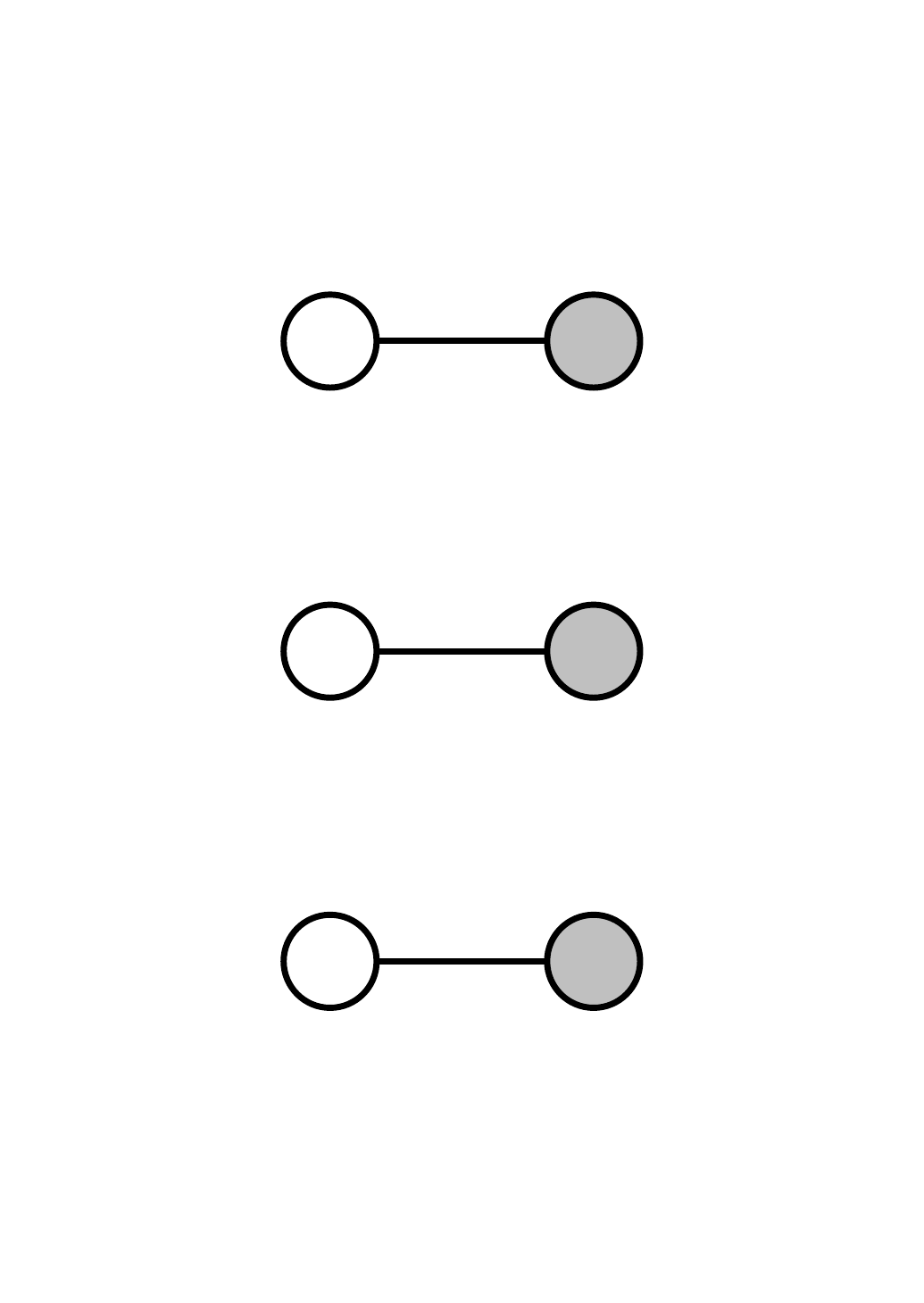}} & 
\raisebox{-1.5cm}{\includegraphics[width=0.135\textwidth]{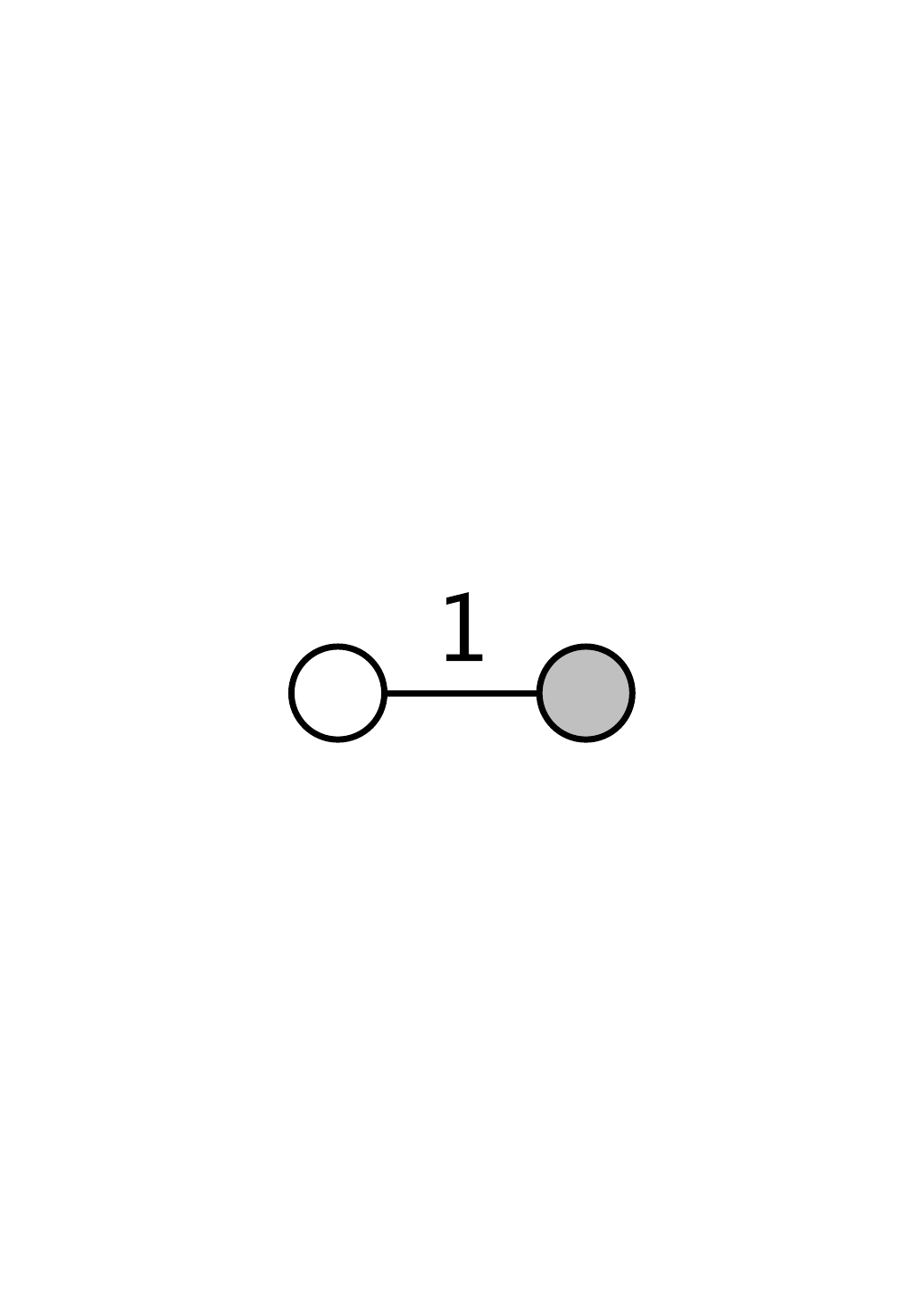}} & 
\begin{minipage}[t]{2cm}{$\Delta_1 \equiv \Delta_2$}\end{minipage}\\[-0.6cm] 

\hline\\[-1.0cm]

\raisebox{-1.2cm}{\includegraphics[angle=90,width=0.2\textwidth]{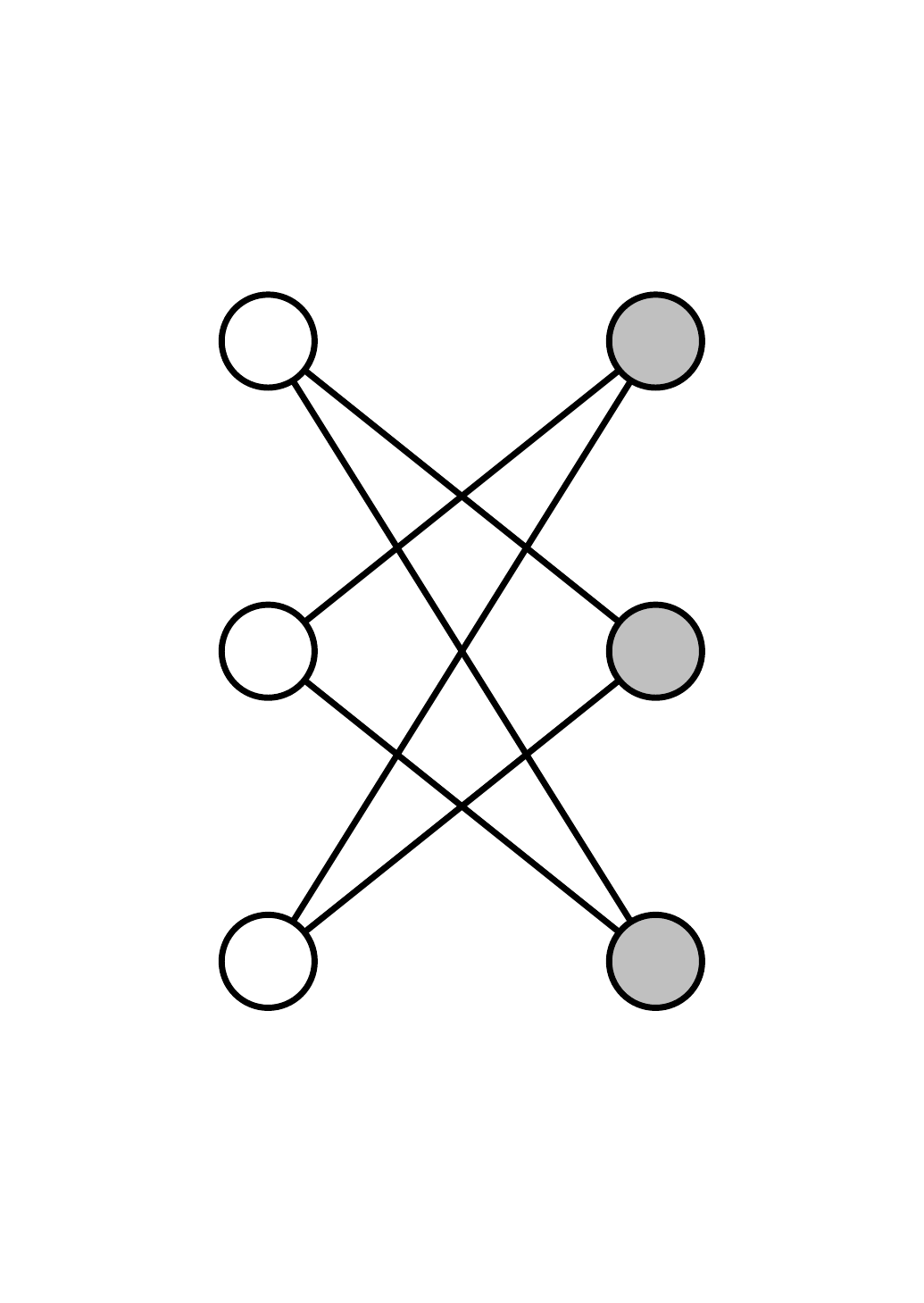}} & 
\raisebox{-1.5cm}{\includegraphics[width=0.135\textwidth]{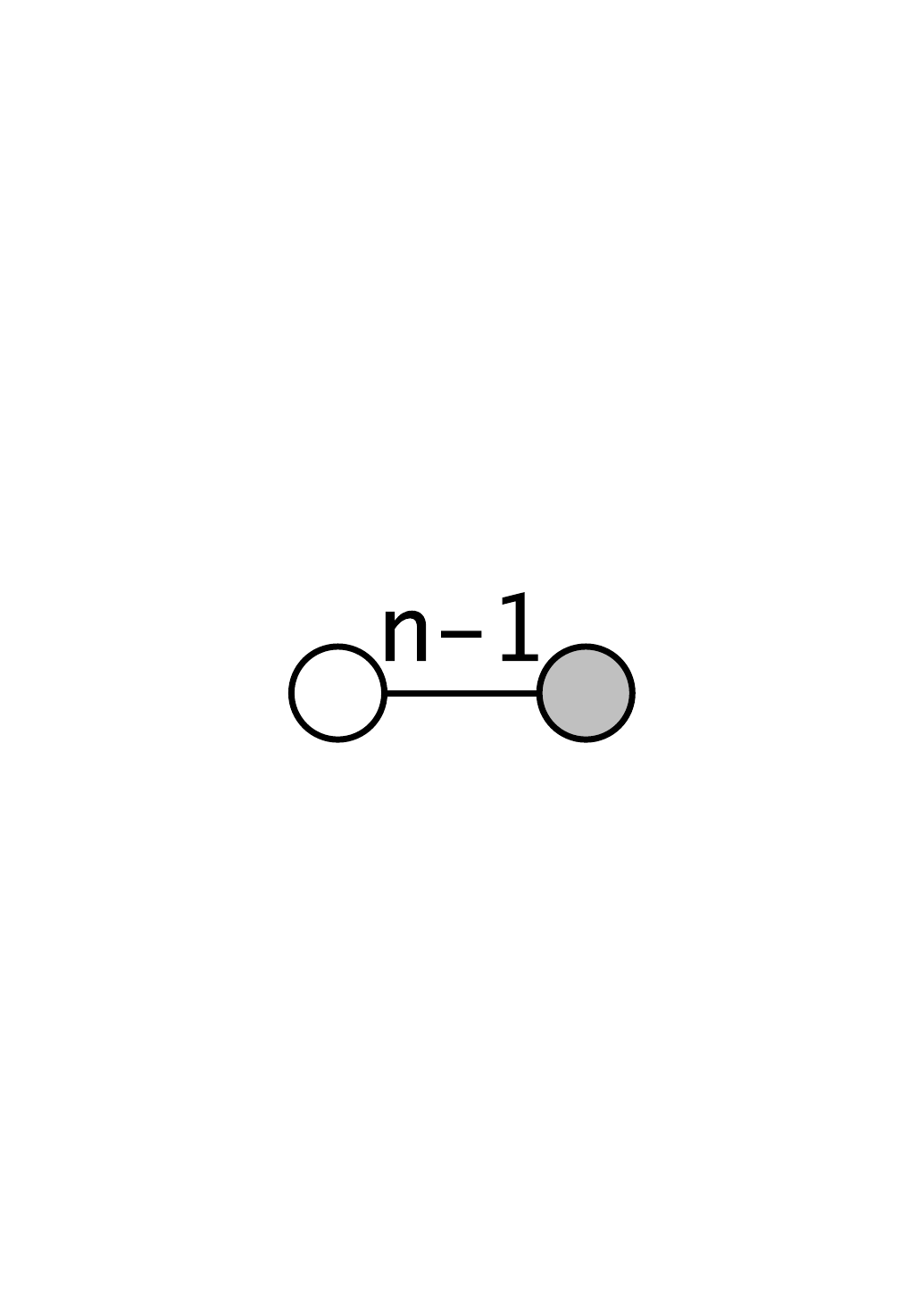}} & 
\begin{minipage}[t]{2cm}{$\Delta_1 \bowtie \Delta_2$}\end{minipage}\\[-0.6cm]
\end{tabular}
\end{ruledtabular}
\end{center}
\caption{\label{tablejoints}\small{\textbf{Joints of two orbits in a basic symmetric motif}. Two orbits of $n$ vertices can be joined in one of these four ways (possibly after a suitable permutation of the vertices): each vertex is joined to either none, all, exactly one or exactly $n-1$ of the vertices of the other orbit. Each orbit can be either a complete or an empty graph on $n$ vertices. To illustrate these joints we have taken both orbits to be $V_3$ (hence the graphs are bipartite) although the same argument holds in the more general case.}} 
\end{table}

We call a symmetric motif as above a \emph{basic symmetric motif} (BSM), while all others which do not conform to this pattern we call \emph{complex}. Complex motifs are rare \cite{macarthur} and their spectrum can be studied separately. However, since they have a constrained shape, it is possible to systematically analyze all the possible contributions that BSMs make to the spectra of the whole network. In particular, specific network eigenvalues may be directly associated with BSMs. We have carried this analysis out for BSMs up to 3 orbits. In all cases, each redundant eigenvalue of a BSM will have multiplicity a multiple of $n-1$ (Appendix \ref{appendixC}).

There are two symmetric motifs with one orbit, $K_n$ and $V_n$, and they are both basic. Their spectrum is shown in Table \ref{BSM1}. We use the notation $\mathbf{e}_i$ for the (redundant) vector with non-zero entries 1 in the first position and $-1$ on the $i$th position ($2\le i\le n$), and \textbf{1} for the vector with constant entries 1.

As predicted, each motif has a redundant eigenvalue of multiplicity $n-1$, which survives as an eigenvalue of the same multiplicity in the spectrum of \emph{any} network containing such a subgraph as a symmetric motif. This amounts to the usual association of the $-1$ and $0$ eigenvalues to cliques and stars respectively, as discussed in previous publications. However, our general setting now allows us to go further.

\begin{table}[!t]
\begin{center}
\begin{ruledtabular}
\begin{tabular}[c]{c c c c c}
Notation & Sym.~motif & Eigenvalue & Multiplicity & Eigenvectors \\
\hline \\

\includegraphics[width=0.05\textwidth]{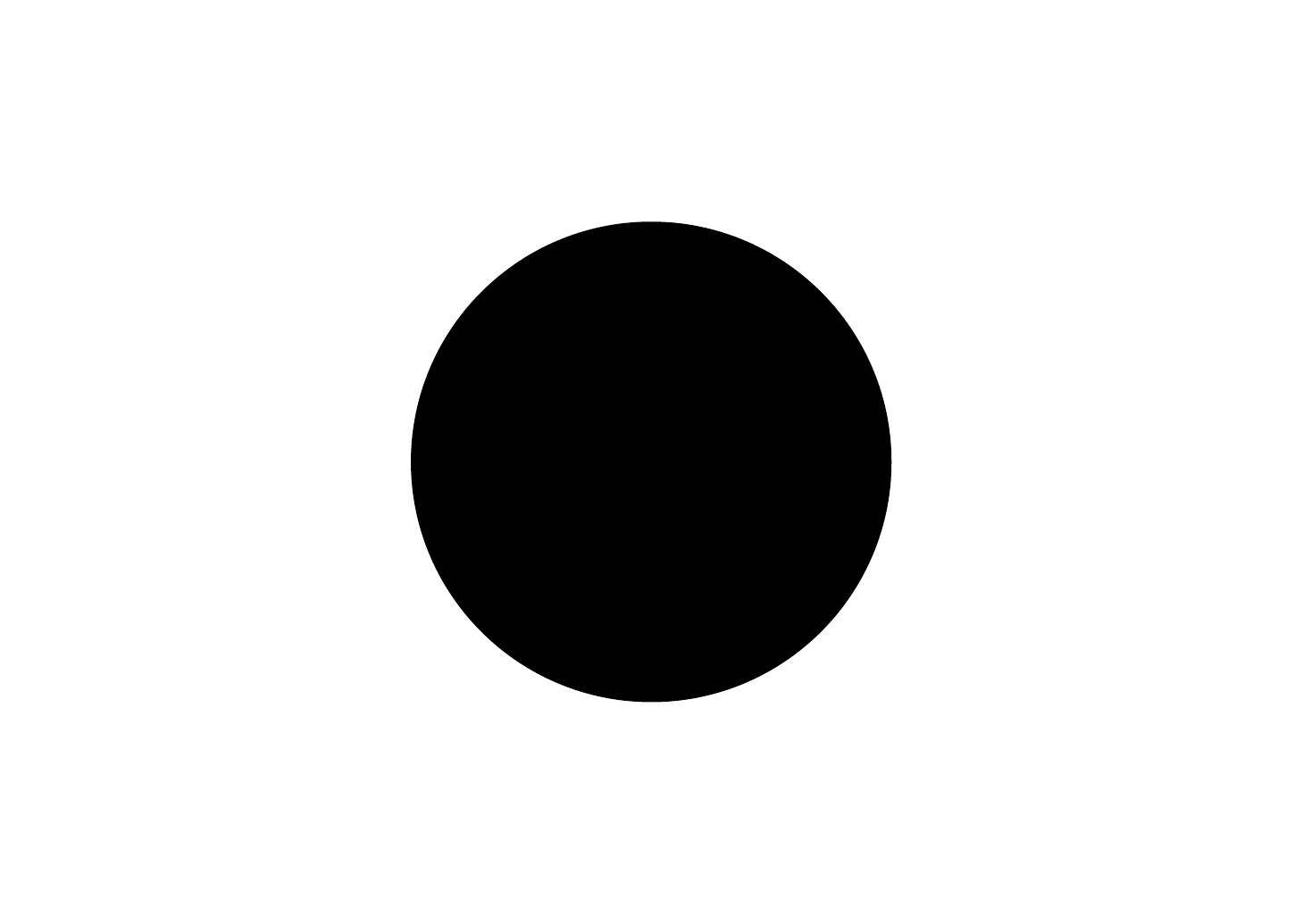}  & \raisebox{0.2cm}{$K_n$} & \begin{minipage}[b]{1cm} $-1^*$ $n-1$ \end{minipage} & \begin{minipage}[b]{0.7cm} $n-1$  $1$ \end{minipage} & \begin{minipage}[b]{0.7cm} $\{\mathbf{e}_i\}$ $\textbf{1}$ \end{minipage} \\
\hline \\
\includegraphics[width=0.05\textwidth]{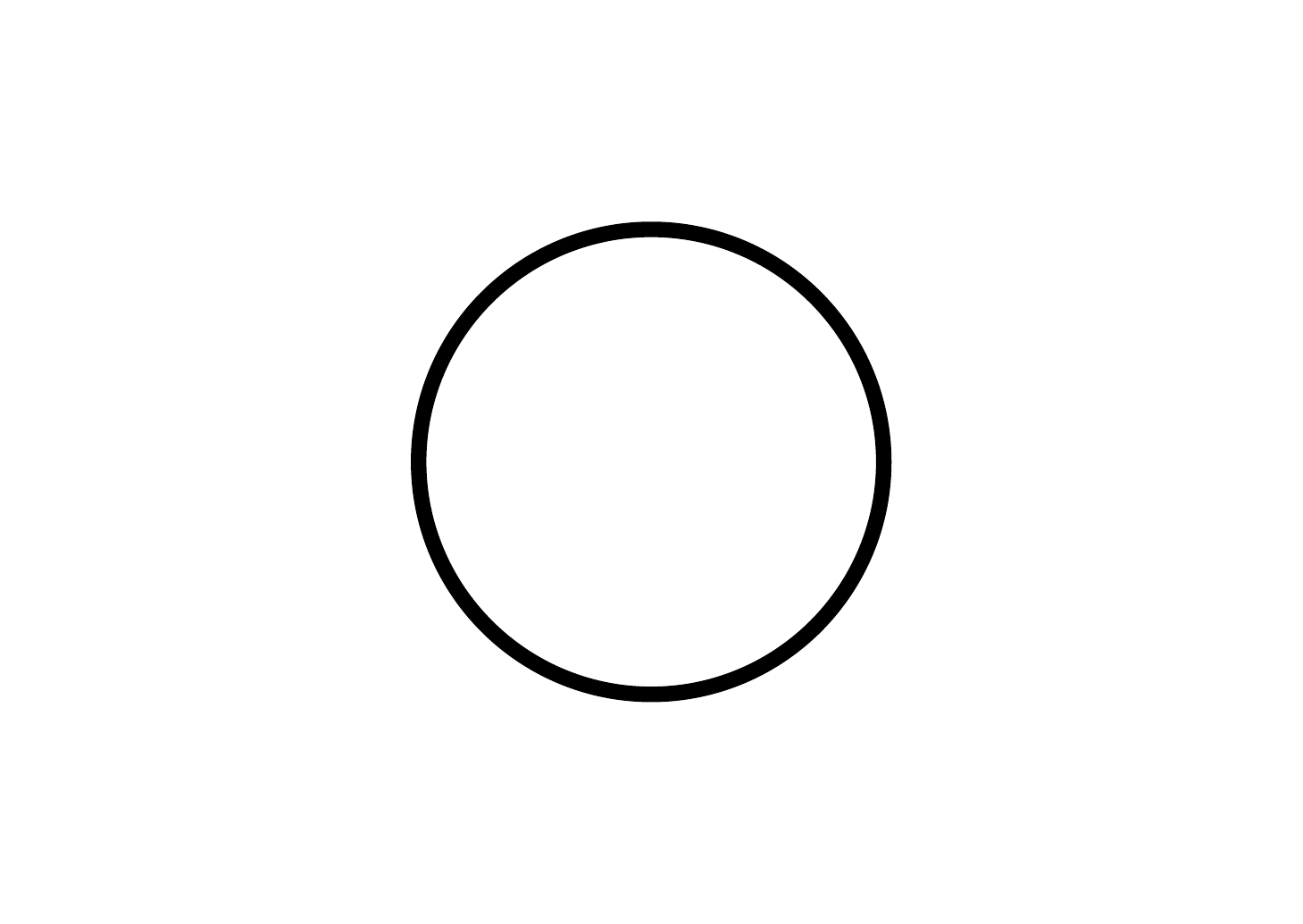}  & \raisebox{0.1cm}{$V_n$} & \begin{minipage}[b]{0.3cm} $\, 0^*$ $0$  \end{minipage} & \begin{minipage}[b]{0.7cm} $n-1$  $1$ \end{minipage} & \begin{minipage}[b]{0.7cm} $\{\mathbf{e}_i\}$ $\textbf{1}$ \end{minipage} \\
\end{tabular}
\end{ruledtabular}
\end{center}
\caption{\label{BSM1}\small{\textbf{Spectra of symmetric motifs with 1 orbit}. The geometric factor is always $S_n$. Redundant eigenvalues are starred. Notice that 0 is an eigenvalue of $V_n$ with multiplicy $n$ but redundant multiplicity $n-1$.}} 
\end{table}

Before moving on we make two brief observations.

Firstly, note that the following BSMs cannot appear in practice. Call a BSM \emph{reducible} if it has an $H$-orbit $\Delta$ joined to all other $H$-orbits $\Delta_j$ by joints of type `$\ast$' or `$\circ$' (Table \ref{tablejoints}), that is, $\Delta \ast \Delta_j$ or $\Delta \circ \Delta_j$ for all $j$. In this case we would obtain an independent geometric factor of type $S_n$ just permuting the vertices of $\Delta$. For example, the second motif of Table \ref{auttable} (a bifan) has $S_2$ and $S_2$ as geometric factors. Such motifs are included in our analysis as two separate symmetric motifs.
Secondly, consider the \emph{complement} $\overline{\mathcal{G}}$ of a graph $\mathcal{G}$, that is, the graph with same vertex set and complement edge set (two vertices are joined in $\overline{\mathcal{G}}$ if and only if they are not joined in $\mathcal{G}$). Note that the complement of a BSM is also a BSM, replacing $K_n$ by $V_n$, $\ast$ by $\circ$, $\equiv$ by $\bowtie$, and viceversa. If $\lambda$ is an eigenvalue of a BSM with multiplicity $p>1$ then $-\lambda - 1$ is an eigenvalue of the complement BSM with the same multiplicity \footnote{See the proof of 8.5.1 in \cite{godsil}.}.

There are 12 BSMs with two orbits of $n$ vertices: 6 of these are non-reducible and it is sufficient to compute 3 cases, since the other 3 are their complement. Table \ref{BSM2} summarizes the results. The first two motifs have complementary spectra while the third is self-complementary. Observe that $-1$ and $0$ arise again as redundant eigenvalues (and hence survive in the network's spectrum), however, this time they are associated with motifs other than stars or cliques.

\begin{table}[!t]
\begin{center}
\begin{ruledtabular}
\begin{tabular}[c]{c c c c c}
Notation & Sym.~motif & Eigenvalue & Red.~Mult. & Eigenvectors \\
\hline \\
\raisebox{0.4cm}{\includegraphics[width=0.07\textwidth]{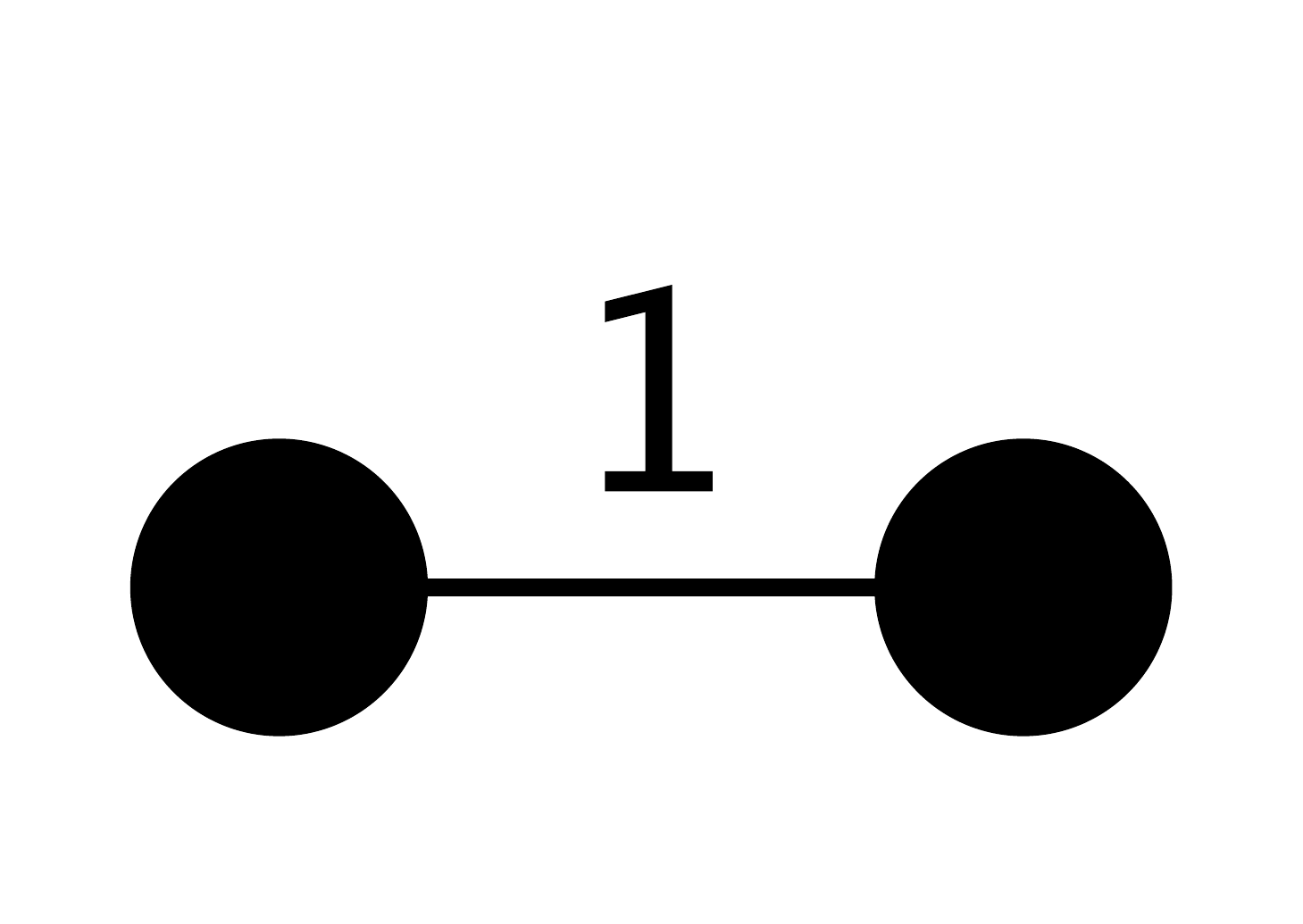}}  & \raisebox{0.6cm}{$K_n \equiv K_n$} & \begin{minipage}[b]{0.8cm} $\,\,\,\,\,0^*$ $-2^*$ $n$ $n-2$ \end{minipage} &  \begin{minipage}[b]{0.8cm} $n-1$ $n-1$ $1$ \\ $1$ \end{minipage} &  \begin{minipage}[b]{1.5cm} $\{(\mathbf{e}_i| \mathbf{e}_i)\}$ $\{(\mathbf{e}_i| -\mathbf{e}_i)\}$ $(\textbf{1}|\textbf{1})$ $(\textbf{1}|\textbf{-1})$ \end{minipage} \\
\hline \\
\raisebox{0.4cm}{\includegraphics[width=0.07\textwidth]{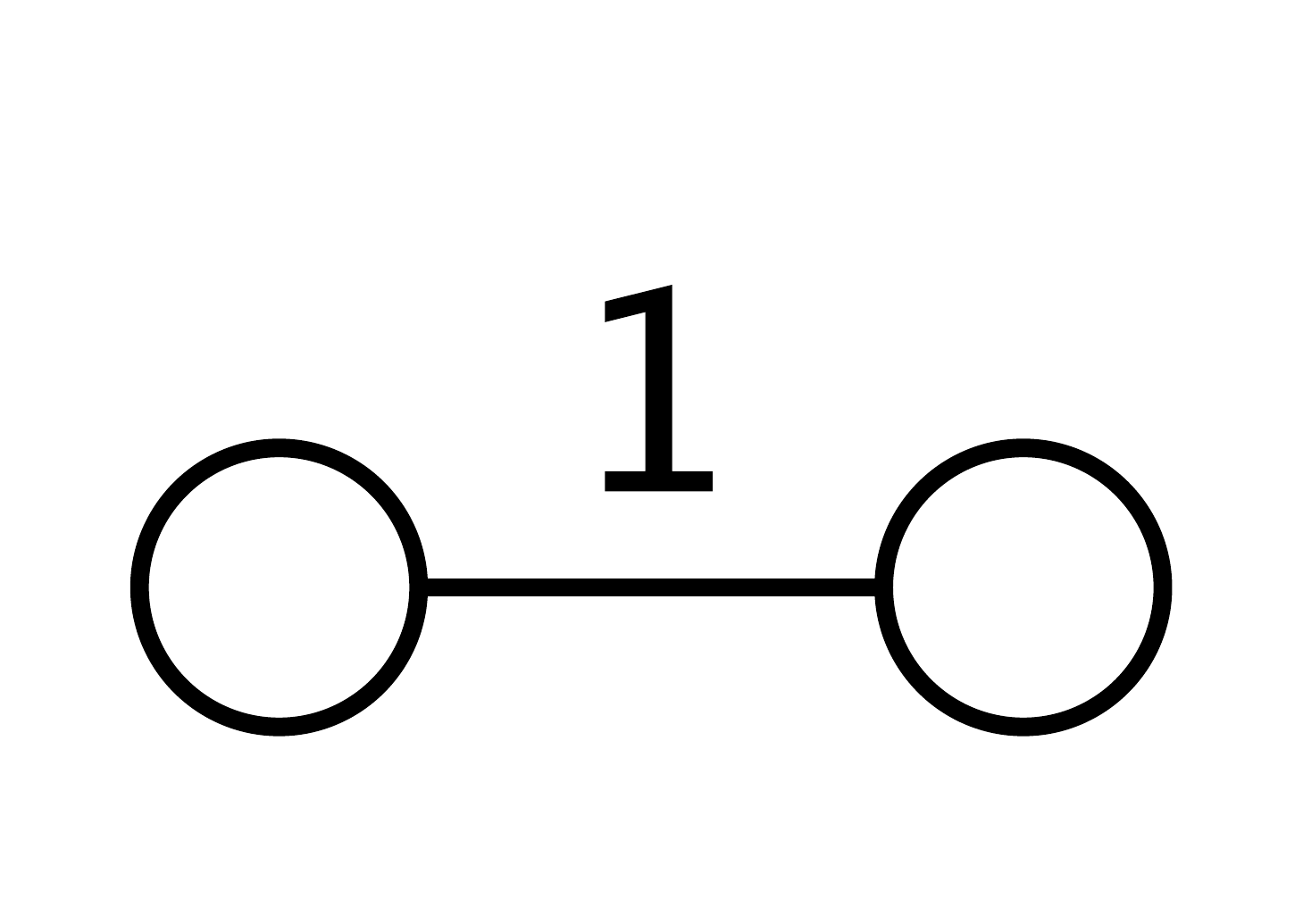}}  & \raisebox{0.6cm}{$V_n \equiv V_n$} & \begin{minipage}[b]{0.6cm} $\,\,\,\,\,1^*$ $-1^*$ $\,\,\,\,\,\,1$ $-1$ \end{minipage} &  \begin{minipage}[b]{0.7cm} $n-1$ $n-1$ $1$ \\ $1$ \end{minipage} &  \begin{minipage}[b]{1.5cm} $\{(\mathbf{e}_i| \mathbf{e}_i)\}$ $\{(\mathbf{e}_i| -\mathbf{e}_i)\}$ $(\textbf{1}|\textbf{1})$ $(\textbf{1}|\textbf{-1})$ \end{minipage} \\
\hline \\
\raisebox{0.4cm}{\includegraphics[width=0.07\textwidth]{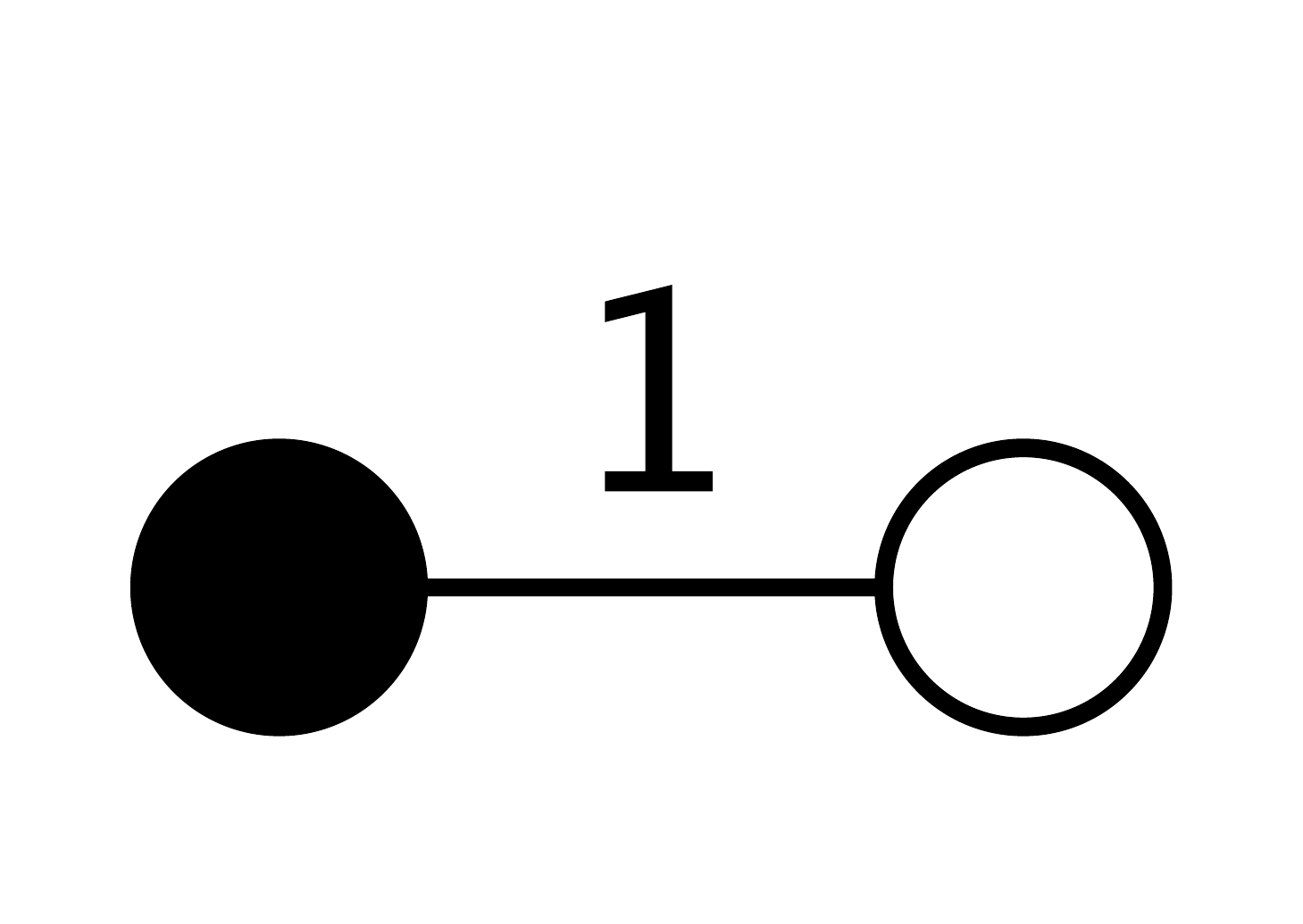}}  & \raisebox{0.6cm}{$K_n \equiv V_n$} & \begin{minipage}[b]{0.6cm} $\lambda_{1}^*$ $\lambda_2^*$ $a_n$ $b_n$ \end{minipage} &  \begin{minipage}[b]{0.7cm} $n-1$ $n-1$ $1$ \\ $1$ \end{minipage} &  \begin{minipage}[b]{1.9cm} $\{(\lambda_1\mathbf{e}_i| \mathbf{e}_i)\}$ $\{(\lambda_2\mathbf{e}_i| -\mathbf{e}_i)\}$\\ $(a_n \textbf{1}|\textbf{1})$ $(b_n \textbf{1}|\textbf{-1})$ \end{minipage} \\
\end{tabular}
\end{ruledtabular}
\end{center}
\caption{\label{BSM2}\small{\textbf{Spectra of basic symmetric motifs with 2 orbits}. Redundant eigenvalues are starred. Eigenvector coordinates are separated by orbit $(x|y)$. The eigenvalues $\lambda_i$ are the roots of $\lambda^2+\lambda-1$, that is, $\lambda_1=\varphi - 1=\frac{-1+\sqrt{5}}{2} \approx 0.6180$ and $\lambda_2=-\varphi=\frac{-1-\sqrt{5}}{2}\approx -1.6180$, where $\varphi$ is the golden ratio.  For completeness, we also give $a_n = \frac{n-1}{2}+\frac{\sqrt{n^2-2n+5}}{2}$ and $b_n = \frac{n-1}{2}-\frac{\sqrt{n^2-2n+5}}{2}$, although they are not redundant. Observe that the redundant eigenvalues of each single orbit ($-1$ and $0$ for $K_n$ respectively $V_n$) do \emph{not} ocurr in the spectrum of the  corresponding BSM.}} 
\end{table}

Define $\textup{RSpec}_m$ as the set of redundant eigenvalues of basic symmetric motifs up to $m$ orbits. We have shown so far that
\begin{eqnarray*}
	\textup{RSpec}_1 &=& \{ -1, 0\} \quad \textup{and}\\ \textup{RSpec}_2 &=& \{-2,-\varphi,-1,0,\varphi-1, 1\}\,,
\end{eqnarray*}
where $\varphi$ is the golden ratio.

For $m \ge 3$ orbits, exactly the same analysis may be conducted. However, the number of possible different BSMs with $m$ orbits increases dramatically with $m$. We have nevertheless computed the redundant eigenvalues of most BSMs with 3 orbits, as shown in Table \ref{BSM3}\footnote{The missing 3-orbit BSMs consist of the $(n-1)!$ permutations of the third orbit for the triangle-shaped BSMs, where $n$ is the number of vertices of each orbit.}. Observe that the 20 non-complementary BSMs organise themselves into 7 different redundant spectrum types. We have therefore shown that:
\begin{align*}
\{ &-3, -2, -1, 0, 1, \pm \sqrt{2}, \pm \sqrt{3}, -1\pm \sqrt{2}, -1\pm \sqrt{3},\\ &\mu_1, \mu_2, \mu_3, \nu_1, \nu_2, \nu_3 \} \subset \textup{RSpec}_3\,.
\end{align*}

It would be interesting to find out all the possible eigenvalues of BSMs of any number of orbits, if there is a pattern. However this is a purely mathematical problem since their relevance (i.e. frequency) in real-world networks decays rapidly with the number of orbits. 

\begin{table}[!h]
\begin{center}
\begin{ruledtabular}
\begin{tabular}[c]{c c c c c c}
Notation & $\lambda$ & Red.~Mult. & $-\lambda - 1$ \\
\hline 
 & & & \\
\begin{minipage}[t]{4cm}
\raisebox{-0.5cm}{\includegraphics[width=0.5\textwidth]{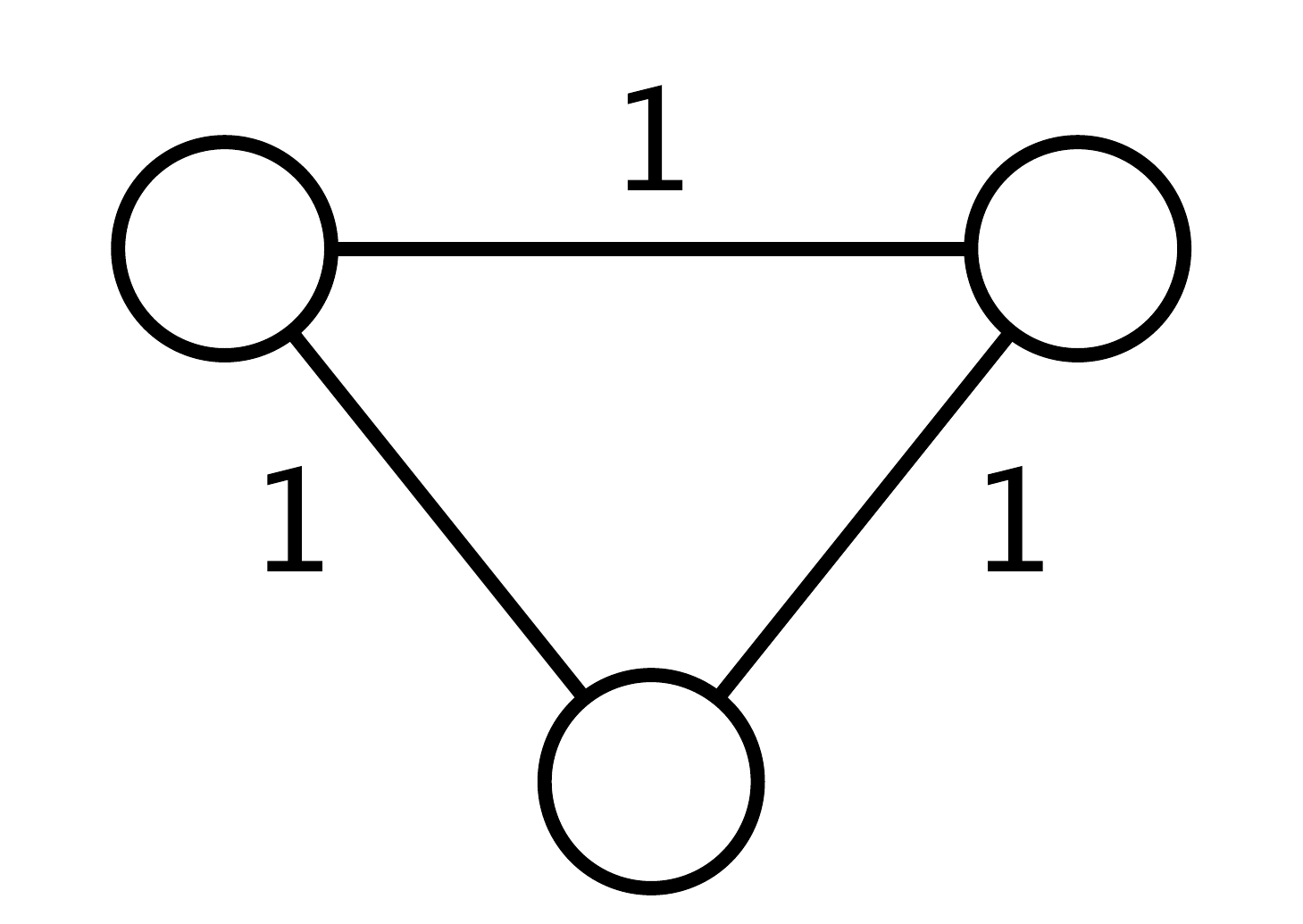}
\includegraphics[width=0.5\textwidth]{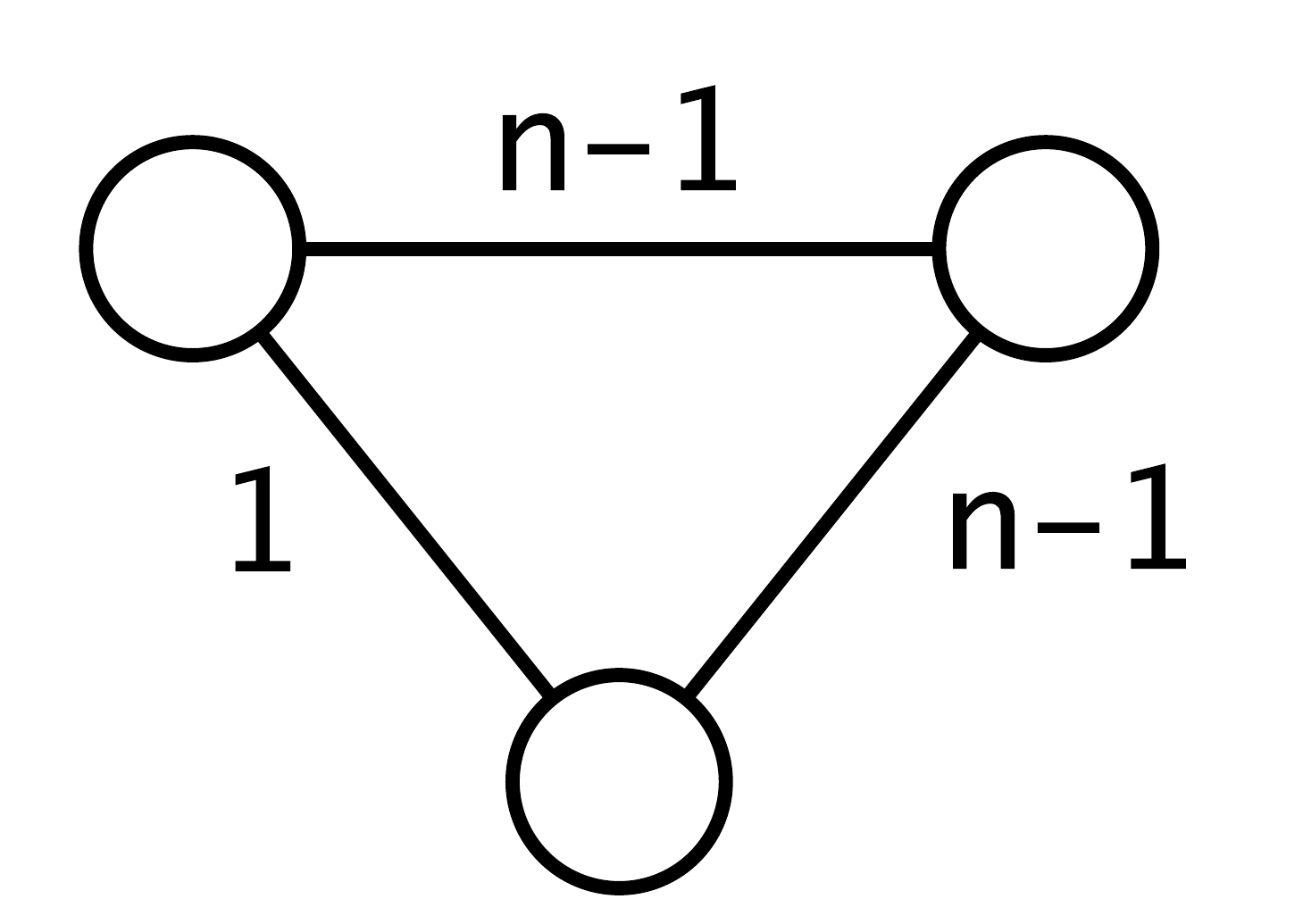}}
\end{minipage} & 
\begin{minipage}{1cm} $-1$\\  $2$ \end{minipage} & 
\begin{minipage}{1.1cm} $2(n-1)$\\ $n-1$ \end{minipage} &
\begin{minipage}{0.5cm} $0$\\  $-3$ \end{minipage} \\[0.8cm]

\hline

\begin{minipage}[t]{4cm}
   \raisebox{-0.2cm}{\includegraphics[width=0.5\textwidth]{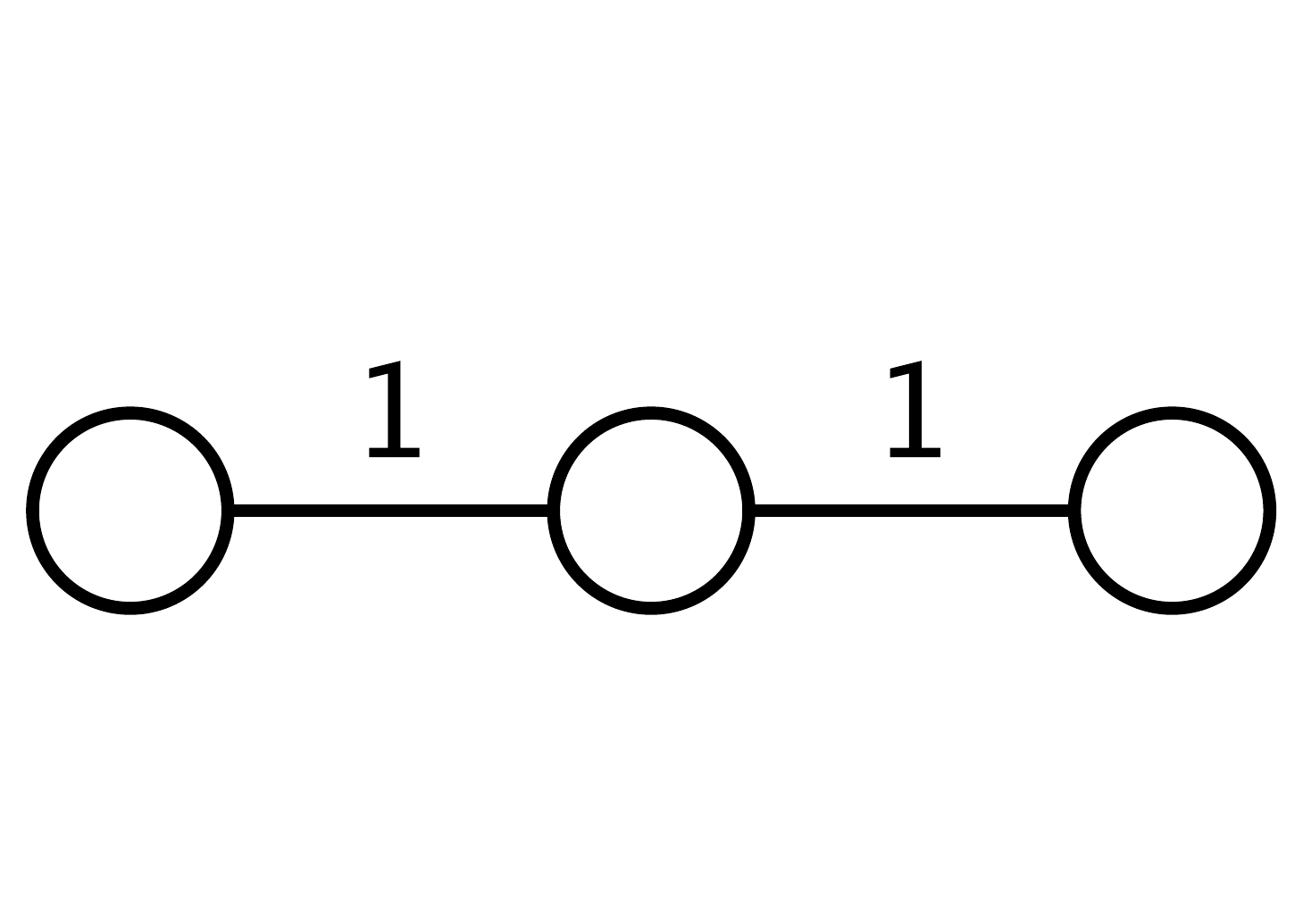}
   \includegraphics[width=0.5\textwidth]{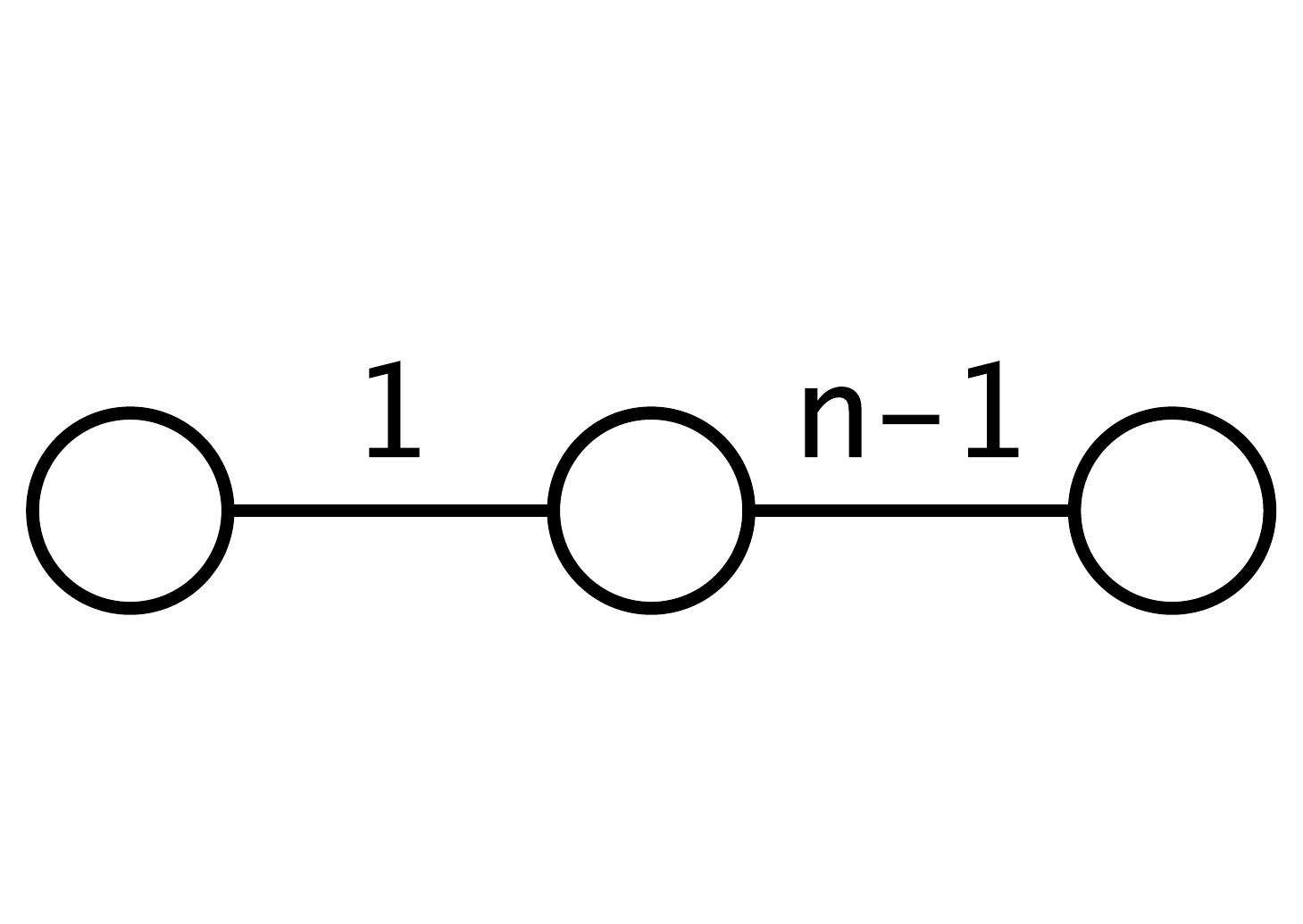}}\\[-0.6cm]
   \raisebox{-0.2cm}{\includegraphics[width=0.5\textwidth]{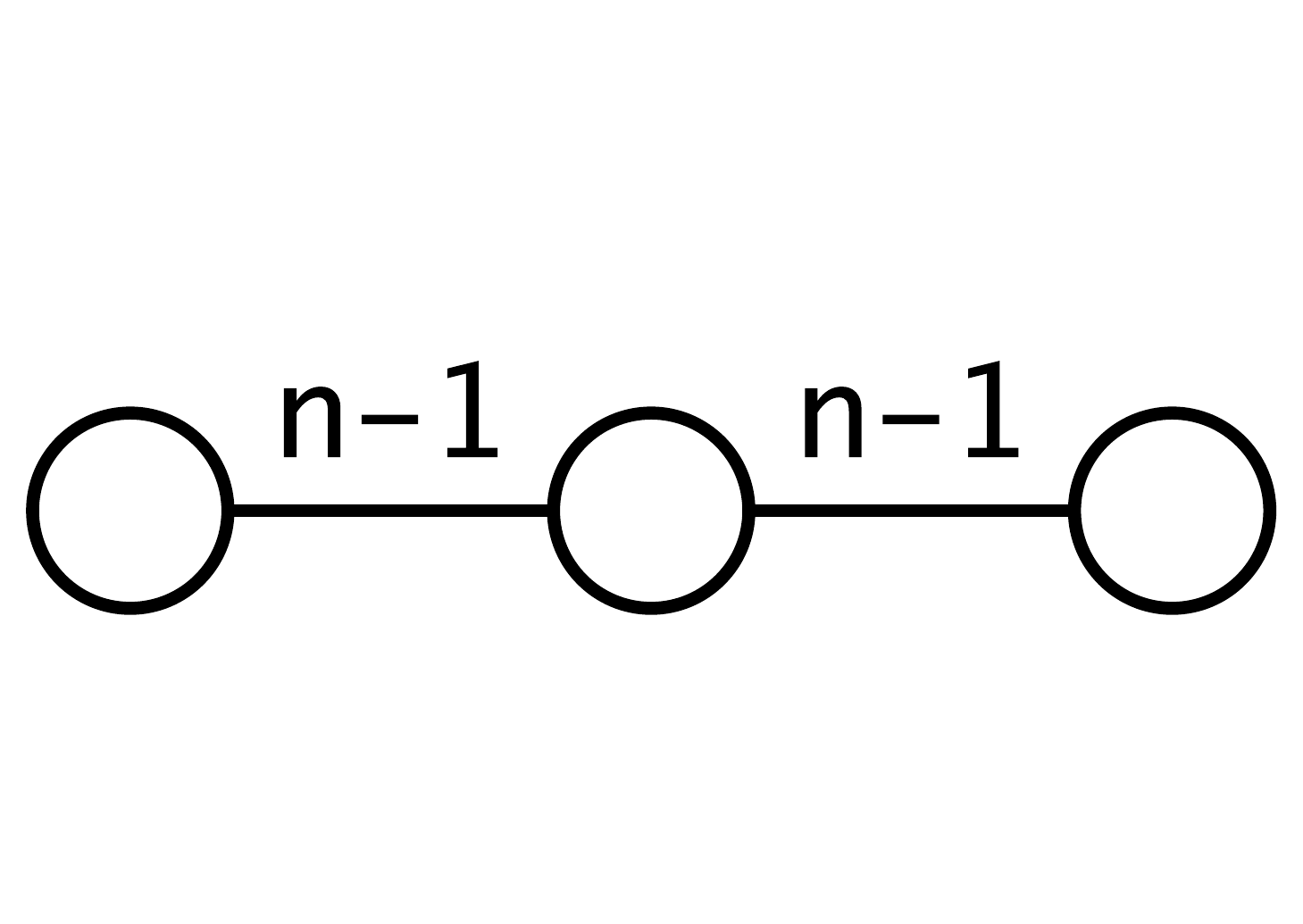}}
\end{minipage}
   &
\begin{minipage}[c]{1cm} $0$\\  $\sqrt{2}$\\ $-\sqrt{2}$ \end{minipage} & 
\begin{minipage}[c]{1.1cm} $n-1$\\ $n-1$\\ $n-1$ \end{minipage} &
\begin{minipage}[c]{1.2cm} $-1$\\  $-1-\sqrt{2}$\\ $-1+\sqrt{2}$ \end{minipage} \\

\hline
 & & & \\
\begin{minipage}[t]{4cm}
\raisebox{-0.7cm}{\includegraphics[width=0.5\textwidth]{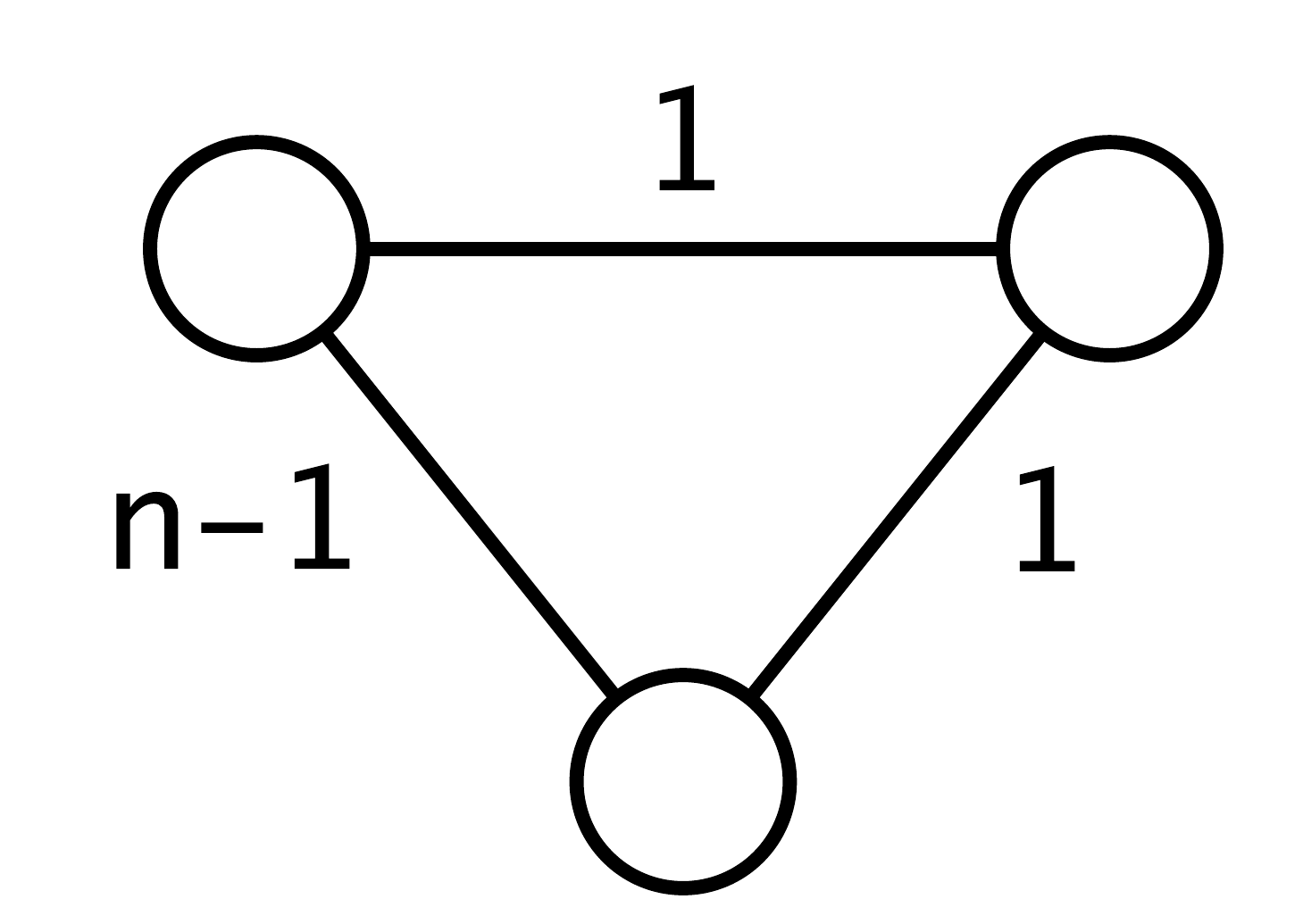}
\includegraphics[width=0.5\textwidth]{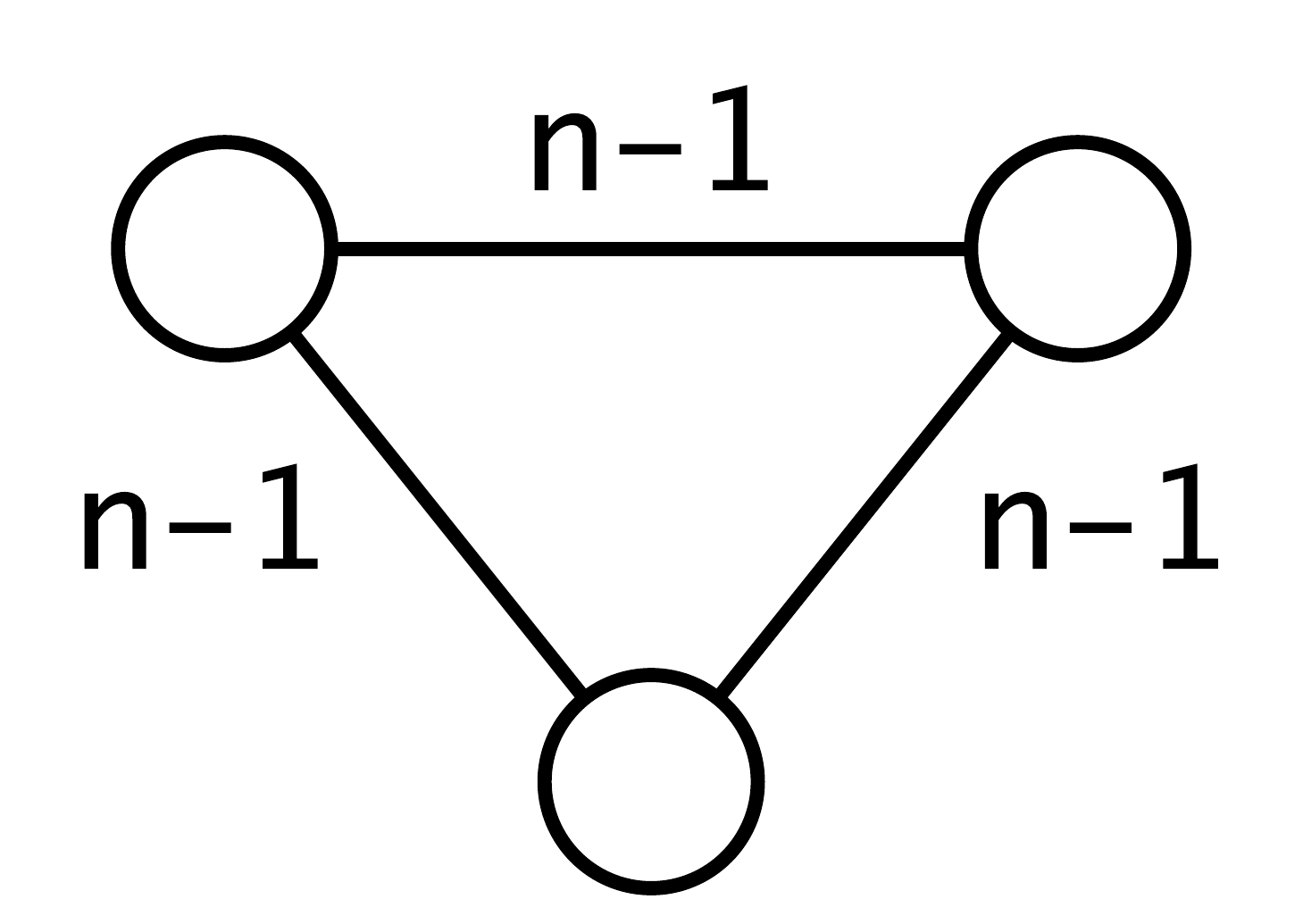}}
\end{minipage} &
\begin{minipage}{1cm} $1$\\  $-2$ \end{minipage} & 
\begin{minipage}{1.1cm} $2(n-1)$\\ $n-1$ \end{minipage} &
\begin{minipage}{0.5cm} $-2$\\  $1$ \end{minipage} \\[0.8cm]

\hline
 & & & \\
\begin{minipage}[t]{4cm}
   \raisebox{-0.2cm}{\includegraphics[width=0.5\textwidth]{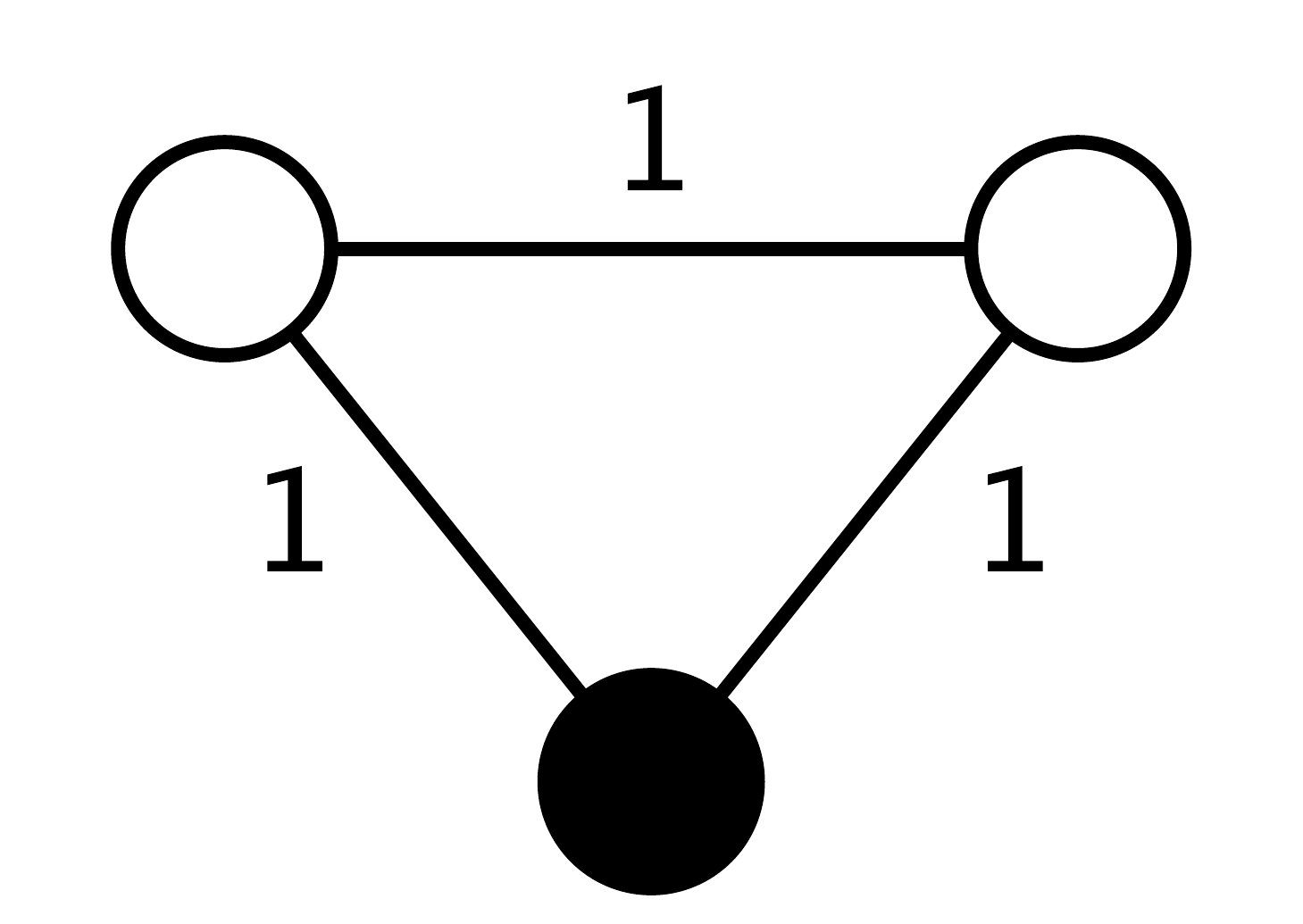}
   \includegraphics[width=0.5\textwidth]{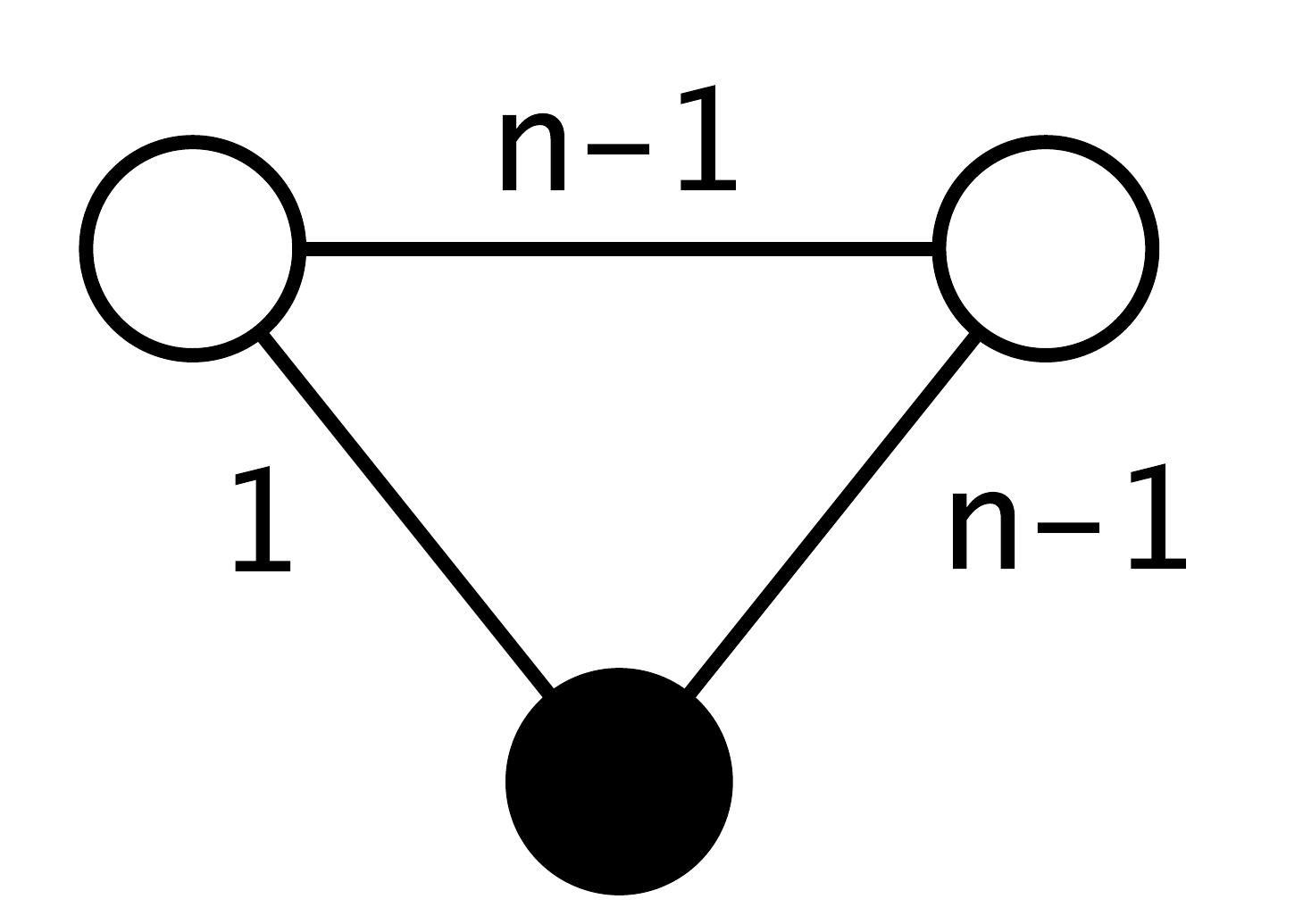}}\\[-0.0cm]
   \raisebox{-0.2cm}{\includegraphics[width=0.5\textwidth]{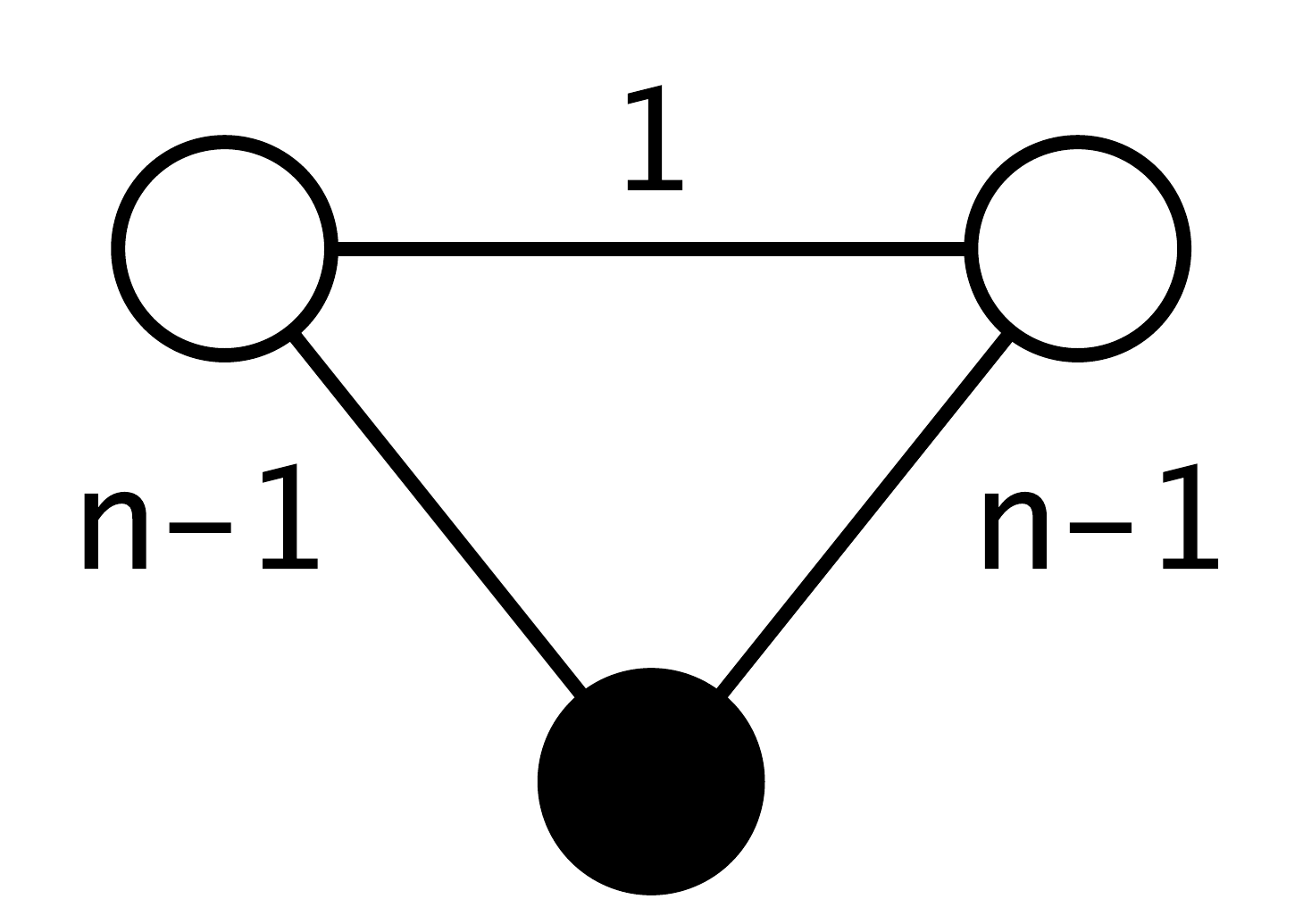}}
\end{minipage}
   &
\begin{minipage}[c]{1cm} $-1$\\  $\sqrt{3}$\\ $-\sqrt{3}$ \end{minipage} & 
\begin{minipage}[c]{1.1cm} $n-1$\\ $n-1$\\ $n-1$ \end{minipage} &
\begin{minipage}[c]{1.2cm} $0$\\  $-1-\sqrt{3}$\\ $-1+\sqrt{3}$ \end{minipage} \\
 & & & \\
\hline        
 & & & \\
\begin{minipage}[t]{4cm}
   \raisebox{-0.2cm}{\includegraphics[width=0.5\textwidth]{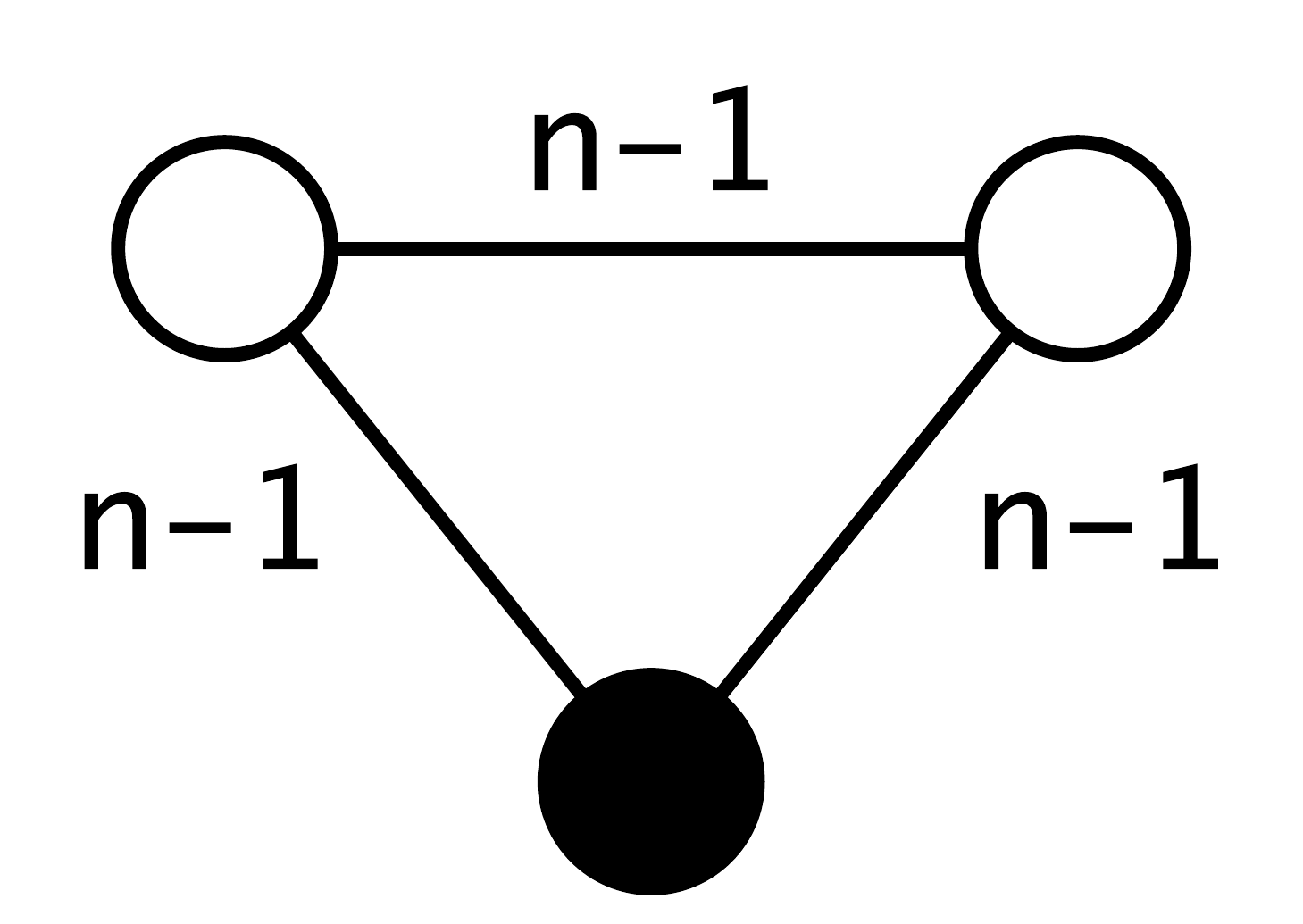}
   \includegraphics[width=0.5\textwidth]{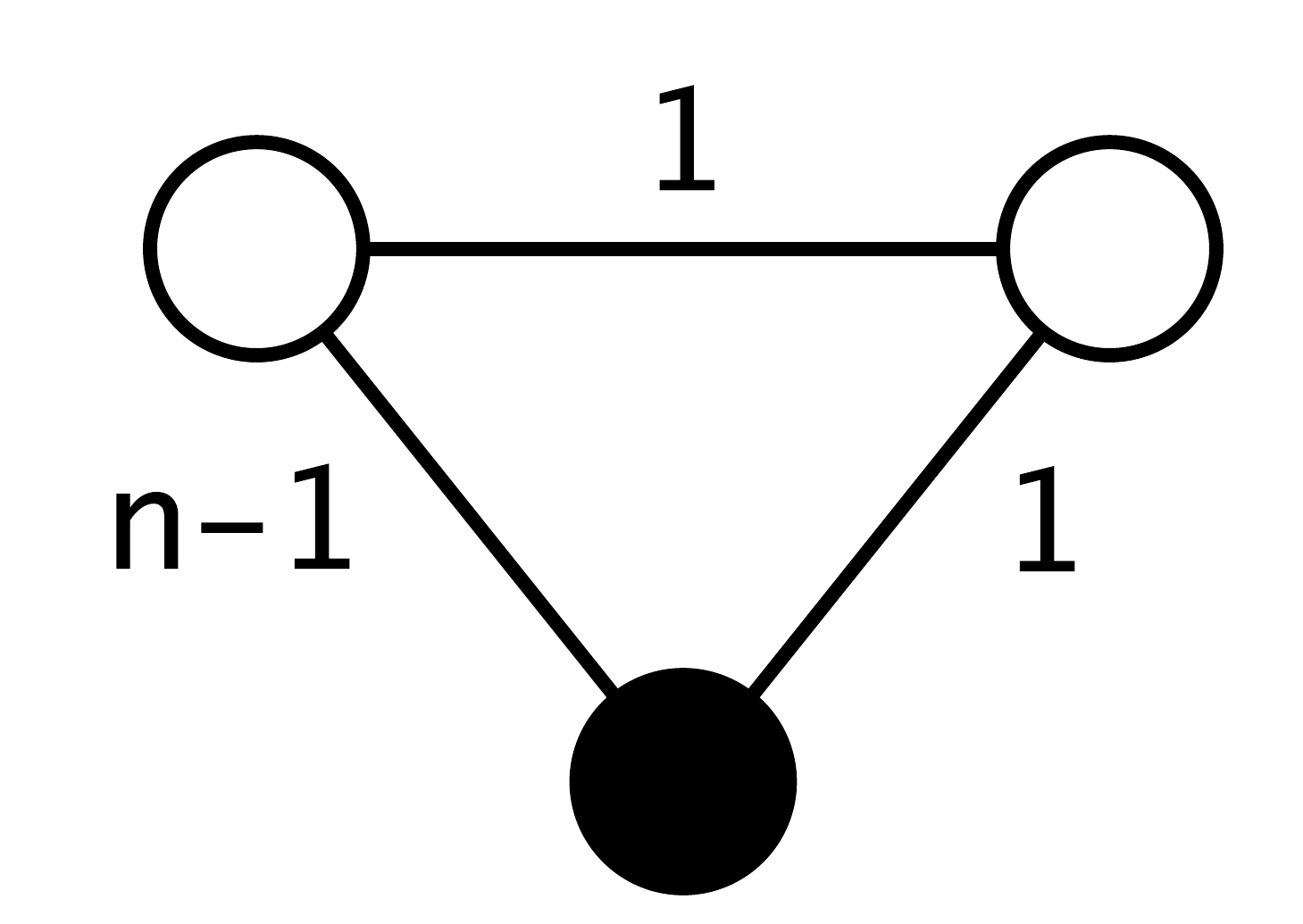}}\\[-0.0cm]
   \raisebox{-0.2cm}{\includegraphics[width=0.5\textwidth]{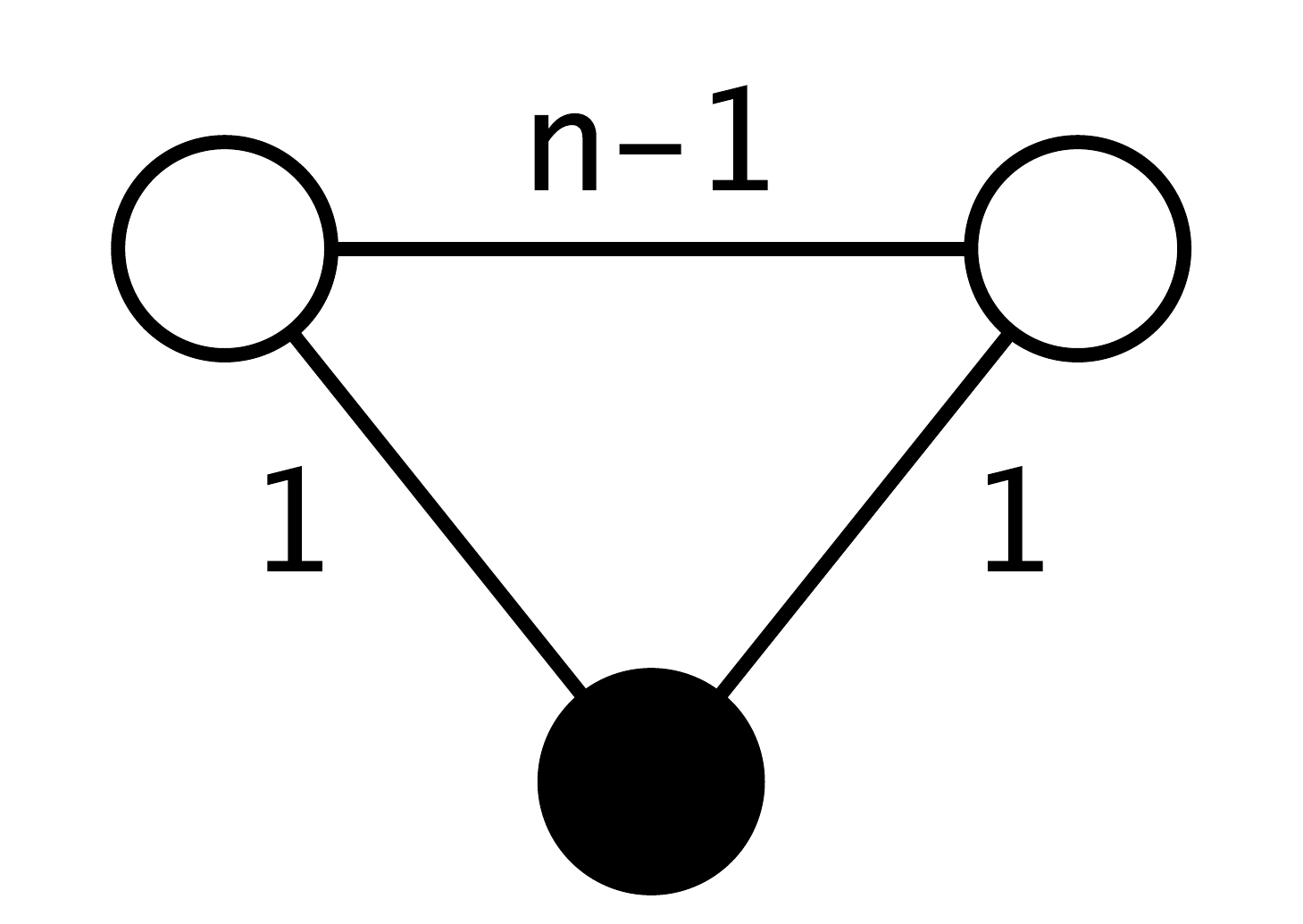}}
\end{minipage}
   &
\begin{minipage}[c]{1.2cm} $-1$\\  $-1+\sqrt{2}$\\ $-1-\sqrt{2}$ \end{minipage} & 
\begin{minipage}[c]{1.1cm} $n-1$\\ $n-1$\\ $n-1$ \end{minipage} &
\begin{minipage}[c]{1.2cm} $-2$\\  $-\sqrt{2}$\\ $\sqrt{2}$ \end{minipage} \\
 & & & \\
\hline

\begin{minipage}[t]{4cm}
   \raisebox{-0.2cm}{\includegraphics[width=0.5\textwidth]{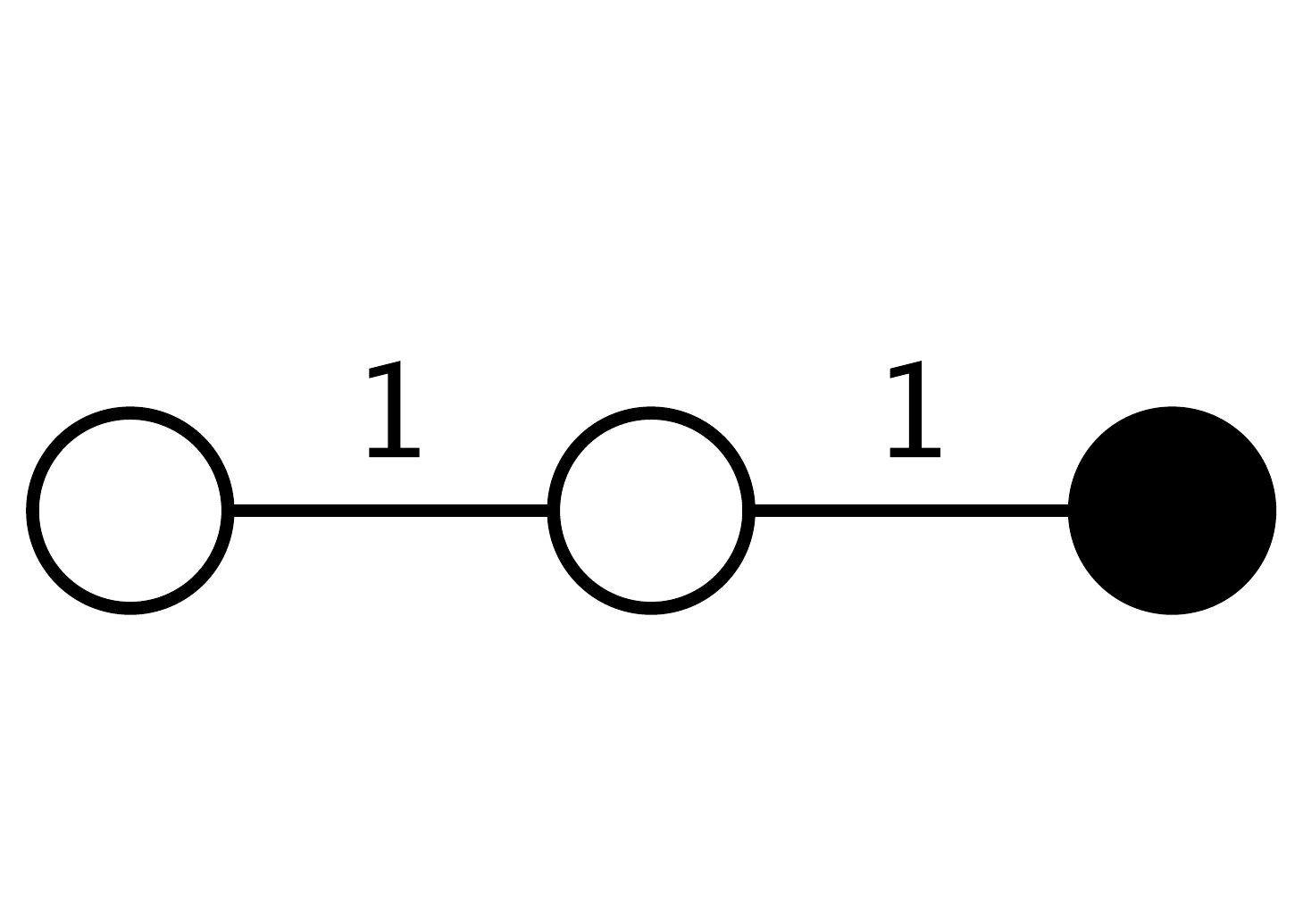}
   \includegraphics[width=0.5\textwidth]{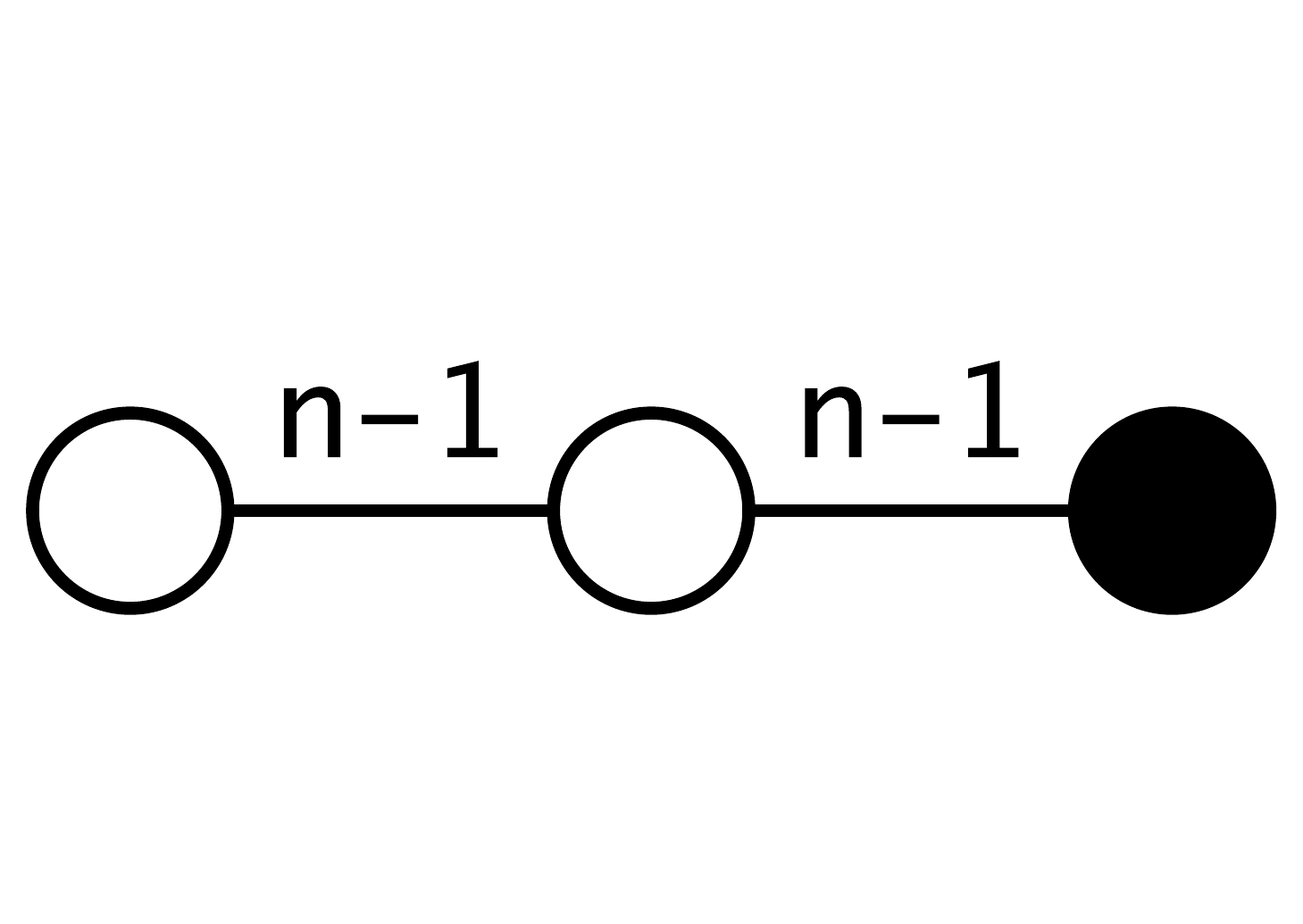}}\\[-0.0cm]
   \raisebox{-0.2cm}{\includegraphics[width=0.5\textwidth]{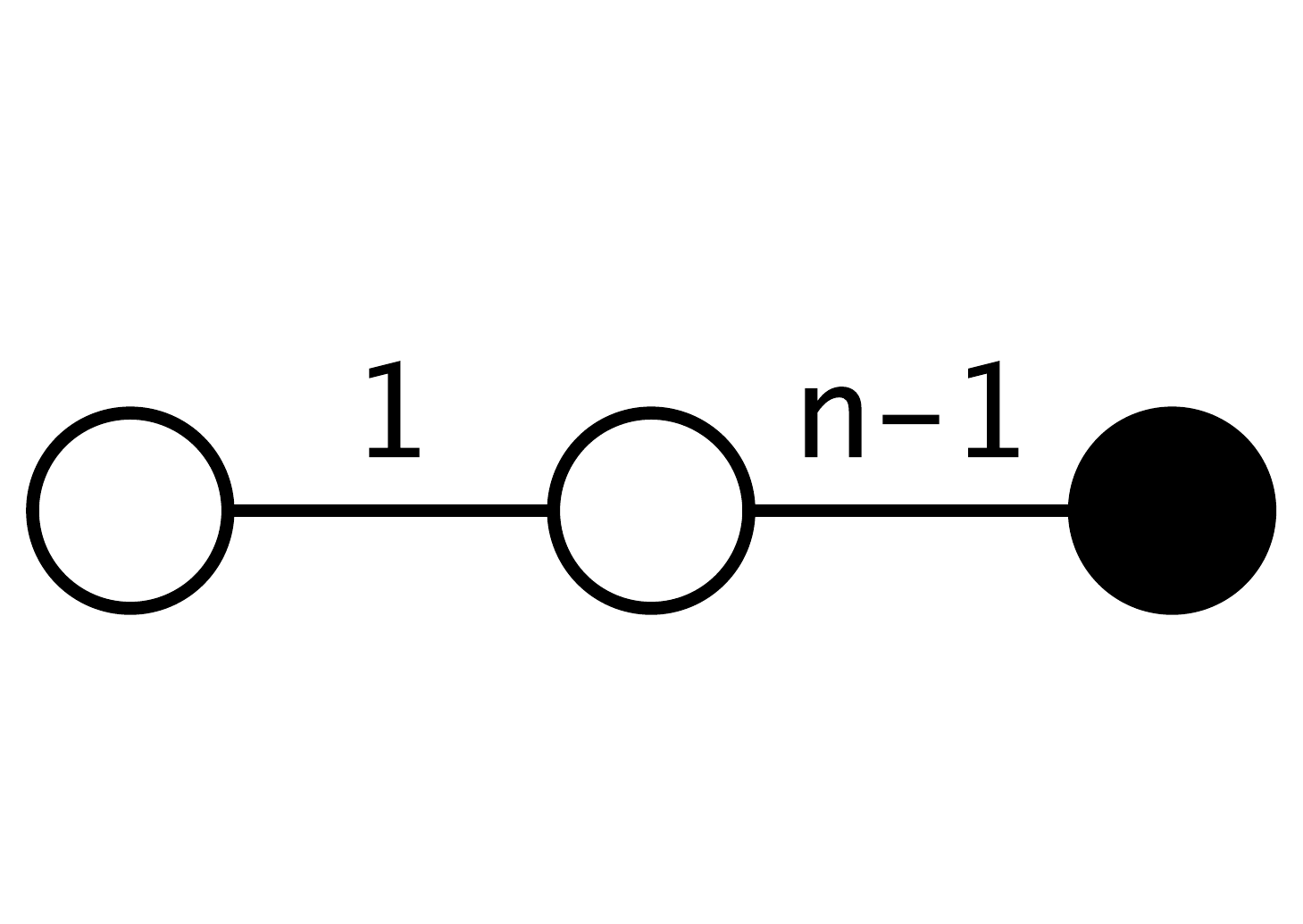}   
   \includegraphics[width=0.5\textwidth]{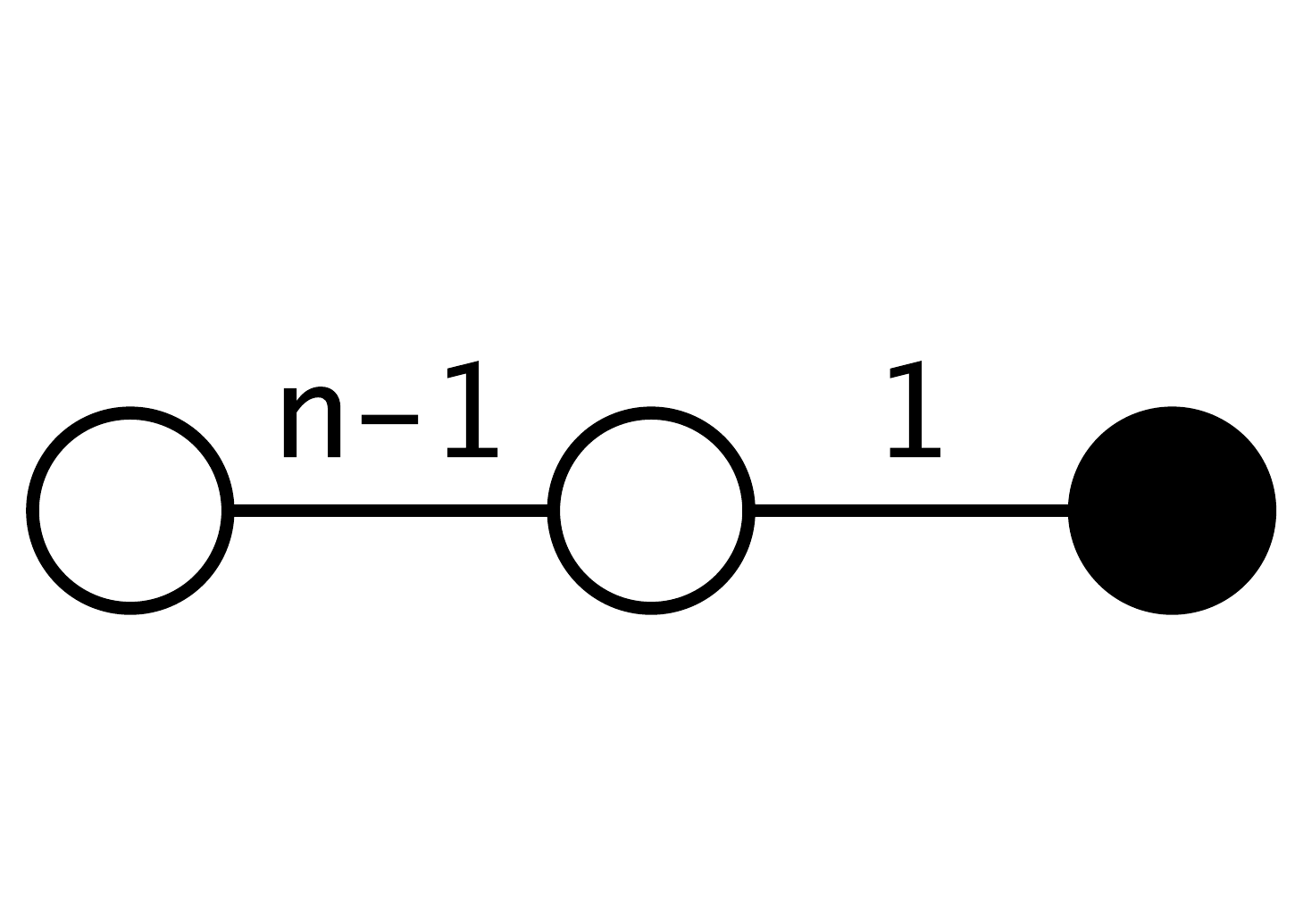}}
\end{minipage}
   &
\begin{minipage}[c]{1.2cm} $\mu_1$\\  $\mu_2$\\ $\mu_3$ \end{minipage} & 
\begin{minipage}[c]{1.1cm} $n-1$\\ $n-1$\\ $n-1$ \end{minipage} &
\begin{minipage}[c]{1.2cm} $\nu_1$\\  $\nu_2$\\ $\nu_3$ \end{minipage} \\
 & & & \\
\hline

\begin{minipage}[t]{4cm}
   \raisebox{-0.2cm}{\includegraphics[width=0.5\textwidth]{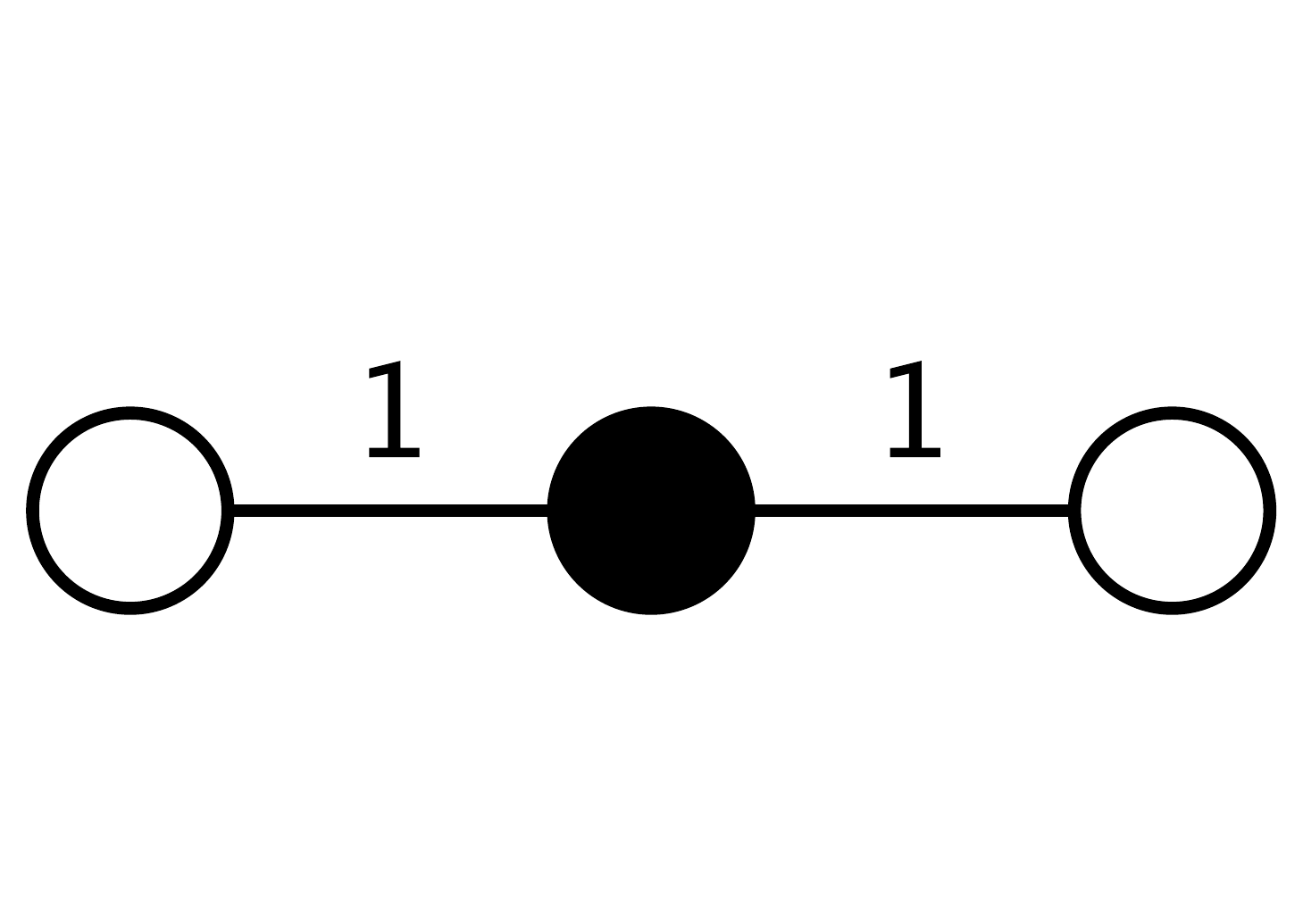}
   \includegraphics[width=0.5\textwidth]{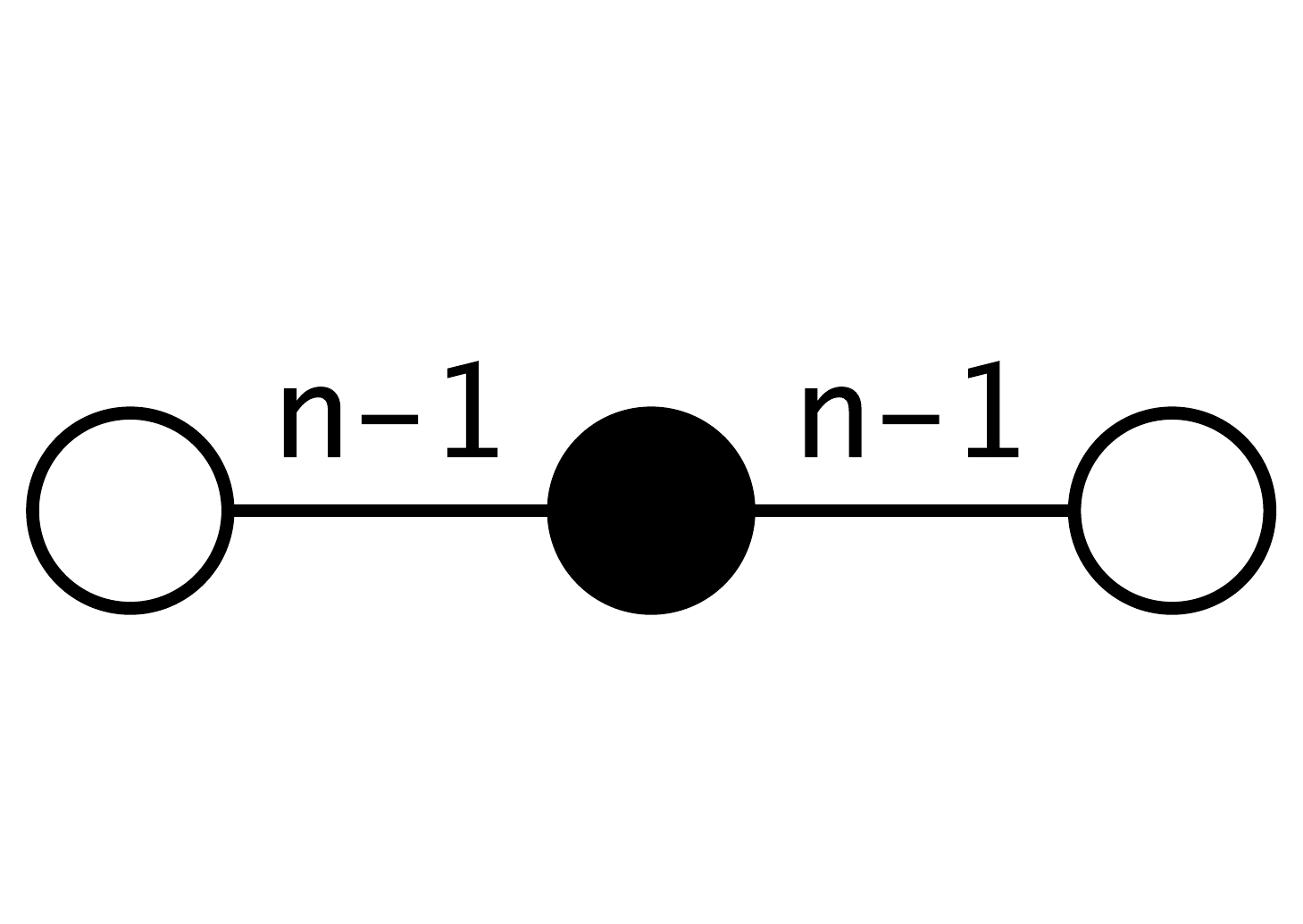}}\\[-0.6cm]
   \raisebox{-0.2cm}{\includegraphics[width=0.5\textwidth]{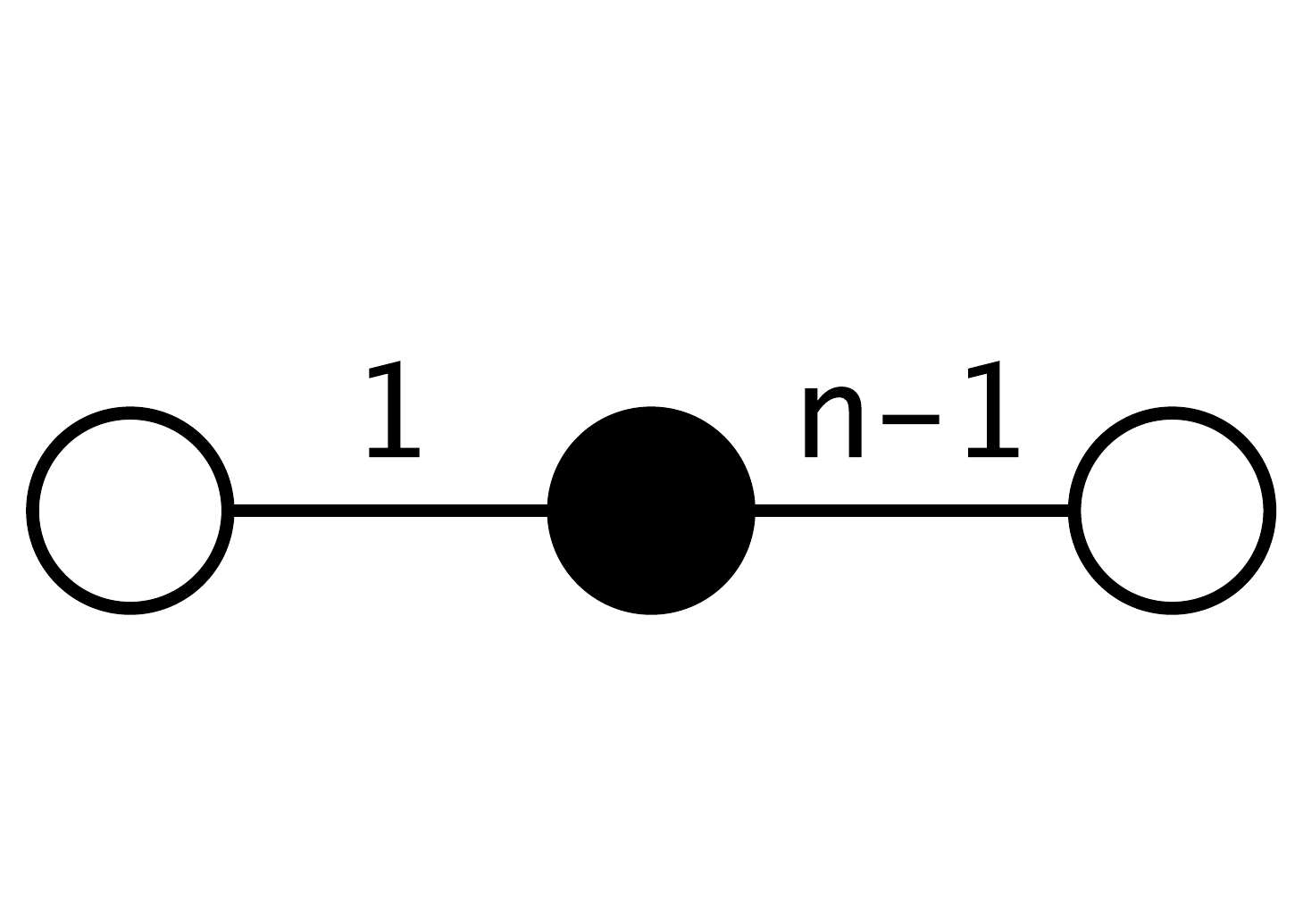}}
\end{minipage}
   &
\begin{minipage}{1cm} $0$\\  $1$\\$-2$ \end{minipage} & 
\begin{minipage}{1.1cm} $n-1$\\ $n-1$\\$n-1$ \end{minipage} &
\begin{minipage}{0.5cm} $-1$\\  $-2$\\ $1$ \end{minipage}                                                        

\end{tabular}
\end{ruledtabular}
\end{center}
\caption{\label{BSM3}\small{\textbf{Spectra of basic symmetric motifs with 3 orbits}. Only redundant eigenvalues $\lambda$ are shown, and their complements $-\lambda-1$, corresponding to complement motifs. Here $\mu_i$ are the roots of the polynomial $p(\lambda)=\lambda^3+\lambda^2-2\lambda-1$, that is, $\mu_1 \approx -1.8019$, $\mu_2 \approx -0.4450$ and $\mu_3 \approx 1.2470$. The complements $\nu_i = -1-\mu_i$ are the roots of the polynomial $p(-1-\lambda)=-\lambda^3-\lambda^2+\lambda+1$: $\nu_1 \approx -2.2470$, $\nu_2 \approx 0.5550$ and $\nu_3 \approx 0.8019$.}}
\end{table}

\begin{figure}[t]
\begin{center}
\includegraphics[scale=0.25]{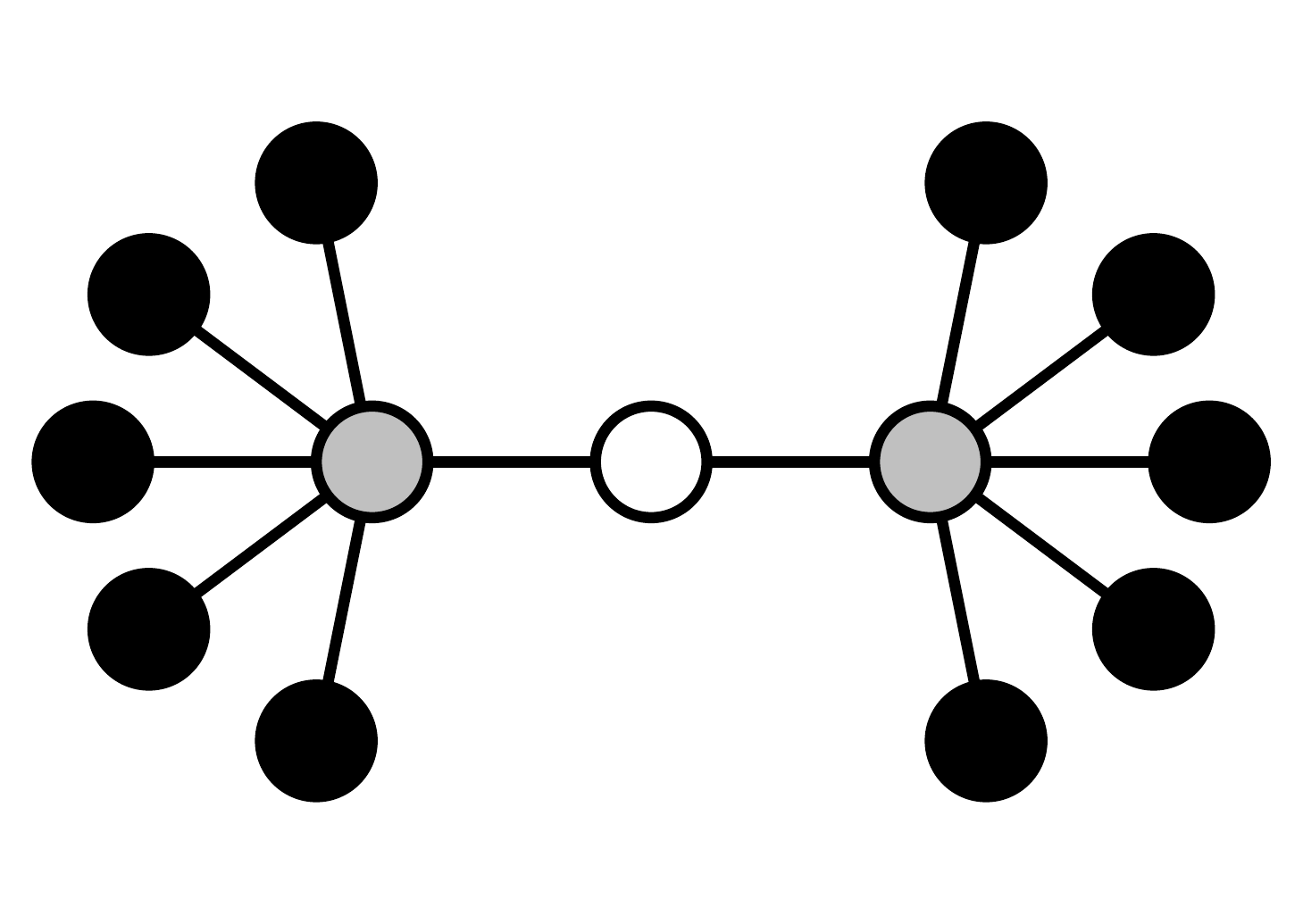}\\
$[-\sqrt{5}^\ast, -\sqrt{5}, 0^\ast, \ldots, 0^\ast,0, \sqrt{5}^\ast,\sqrt{5}] $
\caption{\label{motifsqrt5}\small{\textbf{A complex symmetric motif and its spectrum.}} This complex motif appears in the network of ties between PhD students and their supervisors \cite{denooy, sigact}. The redundant eigenvalues of this motif (starred) survive in the spectrum of the network as a whole.}
\end{center}
\end{figure}

In order to place these abstract results in a more concrete setting, we have computed spectral characteristics of redundancy in the real-world empirical networks of whose spectra are given in FIG \ref{spectra}. All high-multiplicity eigenvalues of these networks are listed in Table \ref{spectratable}. Note that, with the exeception of $\pm\sqrt{5}$ in the spectrum of the network of ties between PhD students and their supervisors -- which comes from the complex motif shown in FIG \ref{motifsqrt5} -- each redundant eigenvalue is in our set $\textup{RSpec}_3$.

\begin{table}[!t]
\begin{center}
\begin{ruledtabular}
\begin{tabular}[c]{l | c r r}
Network & $\lambda$ & $m_\mathcal{P}$ & $m_\mathcal{Q}$ \\
\hline\\[-2ex]
c.elegans GR & $-1$ & 147 & 6 \\ 
			 & 0 & 212 & 45 \\ 
\hline			 
epa.gov & $-1$ & 23 & 0 \\
  & 0 & 2532 & 518 \\
   & 1 & 8 & 4\\
\hline   
media & $-2$& 2& 0\\
 & $-\sqrt{2}$ & 13&6 \\
  & $-1$ & 32& 6\\
   & 0 & 3621& 119\\
    & 1 & 33& 7\\
 &$\sqrt{2}$ & 13 &6 \\
 \hline
PhD & $-\sqrt{5}$	& 2	& 1	\\ 
 	& 	$-\sqrt{3}$	& 3	& 	3\\
 	& 	$-\sqrt{2}$	& 6	& 0	\\
 	& 	$-1$	& 27	& 4	\\ 	
 	& 	0	& 507	& 51	\\
 	& 	1	& 27	& 4	\\
 	& 	$\sqrt{2}$	& 6	& 0	\\ 
 	& 	$\sqrt{3}$	& 3	& 3	\\
 	& 	$\sqrt{5}$	& 2	& 	1\\ 	 
\hline 	 		 	
US Power  & $-2.9150$	& 2	& 2	\\ 
 	& $-\varphi$		& 5	& 	0\\ 
 	& $-\sqrt{2}$	& 13 & 3	\\
 	& $-1$		& 73	& 15	\\
 	& 	0	& 593	& 241	\\ 	
 	& 	$\varphi-1$	& 5	& 0	\\
 	& 	1	& 40	& 14	\\
 	& 	1.1552	& 2	& 2	\\ 
 	& 	1.4068	& 2	& 2	\\ 
    & 	$\sqrt{2}$	& 14 & 4 \\
\hline
Yeast PPI &$-\sqrt{2}$ & 2 & 0\\ 
     & $-1$ & 28 & 9\\
      & 0 & 564 & 154\\
       & 1 & 9 & 2\\
        & $\sqrt{2}$& 2 & 0\\
\end{tabular} 
\end{ruledtabular}
\end{center}
\caption{\label{spectratable}\small{\textbf{High multiplicity eigenvalues in empirical networks}. All high-multiplicity eigenvalues $\lambda$ of the networks of FIG.~\ref{spectra} are given along with their multiplicity $m_\mathcal{P}$ in the parent network and multiplicity $m_\mathcal{Q}$ in the quotient network. The redundant multiplicity $m_\mathcal{P} - m_\mathcal{Q}$ is explained by the symmetry on the network, as described in the main text. Observe that all redundant eigenvalues $(m_\mathcal{P} - m_\mathcal{Q} > 0)$ are in the set $\textup{RSpec}_3$ except $\pm\sqrt{5}$, which is due to the complex motif in FIG.~\ref{motifsqrt5}. Note that the redundant eigenvalues may nevertheless come from different BSMs or even from complex motifs in some cases.}}
\end{table}

\section{\label{sec:conclusions} Conclusions}
Due to the forces which form and shape them, many real-world empirical networks contain a significant ammount of structural redundancy.  Since structurally redundant elements may be permuted without altering network structure, redundancy may be formally investigated by examining network automorphism groups. By considering the relationship between network topology and automorphism group structure, we have shown how specific automorphism subgroups may be associated with specific network motifs. Furthermore, we have shown that certain network eigenvalues may be directly associated with these symmetric motifs. Thus, we have explained how the presence of a variety of local network structures may be seen in network spectra and have shown that the portion of a network's spectrum associated with symmetric motifs is precisely the part of the spectrum due to redundancy. In addition we have computed the redundant spectrum of the most common symmetric motifs up to 3 orbits and any number of vertices and demonstrated their presence in a variety of real-world empirical networks.

Although the theoretical details are somewhat involved, in practice it is extremely easy to find the redundant portion of a networks spectrum and its associated symmetric motifs, even for large networks, using the \texttt{nauty} algorithm \cite{mckay} and a computational group theory package such as GAP \cite{gap}.

In summary, the symmetry approach we have outlined in this paper confirms previous results connecting network spectra with simple local network structures. Additionally, since symmetry can take many forms, this approach also extends these results by providing a general means to relate network eigenvalues to a variety of disparate network structures in a simple, flexible algebraic manner. However, our results are limited by the very nature of the automorphism group: only global symmetries are taken into account, and they fail to measure other internal symmetries (as opposed to the purely combinatorial motifs of Milo and coworkers \cite{milo}), since they are very sensitive to the addition of new vertices. It would be interesting to relax the group notion to that of a groupoid \cite{ramsay} to see if these results can be extended in this more general setting.

\section{Acknowledgments}
This work was funded by the EPSRC and by a London Mathematical Society scheme 6 grant. 

\appendix

\section{} \label{appendixA}
Let $\mathcal{G}$ be a graph with $n$ vertices and adjacency matrix $A = (a_{ij})$. Suppose that the action of $G=\textup{Aut}(\mathcal{G})$ on $\mathcal{G}$ has $m$ $G$-orbits. We show that there is an orthogonal basis of eigenvectors $v_1, \ldots, v_n$ such that $v_1,\ldots, v_m$ are constant on each $G$-orbit and $v_{m+1},\ldots,v_n$ are redundant (the sum of the coordinates at each $G$-orbit is zero). 

The proof follows is a consequence of well-known results in graph theory (see for instance Chapters 8 and 9 in \cite{godsil}). A partition $\pi = \{ C_1, \ldots, C_r\}$ of the vertex set of $\mathcal{G}$ is called \emph{equitable} if the number of neighbours in $C_j$ of \emph{any} vertex in $C_i$ is a constant $b_{ij}$. For example, the orbits of any subgroup of $\textup{Aut}(\mathcal{G})$ gives an equitable partition. The \emph{quotient} of $\mathcal{G}$ by an equitable partition $\pi$, denoted $\mathcal{G}/\pi$, is the directed multigraph with $r$ vertices and adjacency matrix $B = (b_{ij})$. The \emph{characteristic matrix} of a partition $\pi$ is the $n \times r$ matrix $P =  (p_{ij})$ such that $p_{ij}=1$ if the $i$th vertex of $\mathcal{G}$ belong to $C_j$ and 0 otherwise. That is, in column notation $P = (w_1|\ldots|w_r)$ with $w_j$ the vector with 1's in the vertices of $C_j$ and 0 elsewhere. We have that $P$ is the characteristic matrix of an equitable partition if and only if 
\begin{equation}
  \label{eq:1}
  AP = PB,
\end{equation}
since the $(i,j)$-entry of either matrix is the number of neighbours of the $i$th vertex in $C_j$.
A subspace $U$ is called \emph{$A$-invariant} if $Au \in U$ for all $u \in U$. Note that (\ref{eq:1}) is equivalent to saying that the space $W$ spanned by the columns of $P$ is $A$-invariant. One can show \cite{godsil} that every non-zero $A$-invariant subspace has an orthogonal basis of eigenvectors. Furthermore, the orthogonal complement of an $A$-invariant subspace if also $A$-invariant \cite[8.4.3]{godsil}. Consequently, we can write $\mathbb{R}^n = W \oplus W^\bot$ and find an orthogonal basis of eigenvectors of $W$ and another for $W^\bot$. Finally note that:

\begin{itemize}
  \item [(1)] $\textup{dim}(W)=r$.
  \item [(2)] $u \in W \Leftrightarrow$ $u$ is constant on each $C_j$.
  \item [(3)] $u \in W^\bot \Leftrightarrow$ the sum of the coordinates of $u$ on each cell is zero.
\end{itemize}

\section{} \label{appendixB}
Suppose that $\mathcal{G}$ is a graph with $n$ vertices and $m$ $G$-orbits, where $G=\textup{Aut}(\mathcal{G})$. Consider the associated geometric decomposition, 
$G = H_1 \times H_2 \times \ldots H_k$ and corresponding symmetric motifs $\mathcal{M}_1, \ldots, \mathcal{M}_k$. Suppose that $\mathcal{M}_i$ has $n_i$ vertices and $m_i$ $H_i$-orbits (which then coincide with the $G$-orbits). Call $n_0$ to the number of fixed points in $\mathcal{G}$. Then we have 
$$
   n_1 + \ldots + n_k + n_0 = n \textup{ and } m_1 + \ldots +m_k + n_0 = m\,.
$$
For each motif $\mathcal{M}_i$ we can apply the result in Appendix \ref{appendixA} to find an orthogonal basis of eigenvectors such that $n_i-m_i$ of them, say $\{\mathbf{v}^i_j\}$, are redundant. Hence they give $\mathcal{M}_i$-local eigenvectors $\{\overline{\mathbf{v}}^i_j\}$ of $\mathcal{G}$, for each $1\le i \le k$. Note that the $\overline{\mathbf{v}}^i_j$'s are pairwise orthogonal and hence in particular are linearly independent. 

Now choose an orthogonal basis of eigenvectors of the quotient, $\{\mathbf{w}_1, \ldots, \mathbf{w}_m\}$. Each $\widehat{\mathbf{w}}_i$ is an eigenvector of $\mathcal{G}$, constant on each orbit. Then $\{\widehat{\mathbf{w}}_1, \ldots, \widehat{\mathbf{w}}_m\} \cup \{ \overline{\mathbf{v}}^i_j\}$ is an orthogonal system of $m + (n_1-m_1) + \ldots (n_k - m_k) = m + n - n_0 - m + n_0 = n$ vectors, that is, an orthogonal basis of eigenvectors of $\mathcal{G}$.\\

\section{} \label{appendixC} 
Let $\mathcal{M}$ be a graph with an orbit of $n$ vertices $x_1, \ldots, x_n$ such that all $n!$ permutations of the vertices are automorphisms of $\mathcal{M}$. We demonstrate that there is a redundant eigenvalue $\lambda$ of redundant multiplicity at least $n-1$. 

We can assume that $n \ge 2$. Let $(\lambda, \mathbf{v})$ be a redundant eigenpair (there is at least one, by Appendix \ref{appendixA}). Suppose that $v_1, \ldots, v_n$ are the entries of $\mathbf{v}$ at $x_1, \ldots, x_n$. Recall that any permutation of the $v_i$'s (fixing the other entries) gives an eigenvector of the same eigenvalue. Since $\mathbf{v}$ is redundant, it cannot be constant on the orbit, thus we can assume without loss of generality that $v_1 \neq v_2$. Let $\sigma$ be a permutation interchanging the first and second coordinates while fixing the other $n-2$ entries in the orbit. Thus $v - \sigma v$ is a multiple of the vector with values $(1, -1, 0, \ldots, 0)$ on the $x_i$'s. Further permuting the coordinates gives $n-1$ linearly independent eigenvectors of $\lambda$, as required.

\bibliographystyle{amsplain}

\end{document}